\documentclass[5p]{elsarticle}

\usepackage{amsmath,amssymb,amsfonts}
\usepackage{url}
\usepackage{algorithmic}
\usepackage{graphicx}
\usepackage{textcomp}
\usepackage{hyperref}
\usepackage{multicol}
\usepackage{multirow}
\usepackage[table,xcdraw]{xcolor}
\usepackage{xcolor}
\usepackage{adjustbox}
\usepackage{ragged2e}
\usepackage{array}
\usepackage{mathtools}
\usepackage{rotating}
\usepackage{lipsum}
\usepackage{subfigure}
\usepackage[utf8]{inputenc}

\journal{Journal of Systems and Software}

\begin{document}

\begin{frontmatter}



\title{\textit{EvaluateXAI}: A Framework to Evaluate the Reliability and Consistency of Rule-based XAI Techniques for Software Analytics Tasks}


\author[inst1]{Md Abdul Awal} 
\author[inst1]{Chanchal K. Roy} 

\affiliation[inst1]{
            organization={Department of Computer Science, University of Saskatchewan},
            city={Saskatoon}, 
            state={Saskatchewan},
            country={Canada}}




\begin{abstract}

The advancement of machine learning (ML) models has led to the development of ML-based approaches to improve numerous software engineering tasks in software maintenance and evolution. Nevertheless, research indicates that despite their potential successes, ML models may not be employed in real-world scenarios because they often remain a black box to practitioners, lacking explainability in their reasoning. Recently, various rule-based model-agnostic Explainable AI (XAI) techniques, such as PyExplainer and LIME, have been employed to explain the predictions of ML models in software analytics tasks. In this paper, we assess the ability of these techniques, particularly the state-of-the-art PyExplainer and LIME, to generate reliable and consistent explanations for ML models across various software analytics tasks, including Just-in-Time (JIT) defect prediction, clone detection, and the classification of useful code review comments. Our manual investigations find inconsistencies and anomalies in the explanations generated by these techniques. Therefore, we design a novel framework: Evaluation of Explainable AI (\textit{EvaluateXAI}), along with granular-level evaluation metrics, to automatically assess the effectiveness of rule-based XAI techniques in generating reliable and consistent explanations for ML models in software analytics tasks. After conducting in-depth experiments involving seven state-of-the-art ML models trained on five datasets and six evaluation metrics, we find that none of the evaluation metrics reached 100\%, indicating the unreliability of the explanations generated by XAI techniques. Additionally, PyExplainer and LIME failed to provide consistent explanations for 86.11\% and 77.78\% of the experimental combinations, respectively. Therefore, our experimental findings emphasize the necessity for further research in XAI to produce reliable and consistent explanations for ML models in software analytics tasks.

\end{abstract}


\end{frontmatter}



\section{Introduction}
In software maintenance and evolution, many software engineering tasks such as method name prediction \cite{alon2019code2vec, alon2018code2seq}, code summarization \cite{fernandes2018structured}, code comment generation \cite{hu2018deep}, code review \cite{bacchelli2013expectations}, code clone detection \cite{roy2007survey, nafi2019clcdsa}, and Just-in-Time (JIT) defect prediction \cite{catolino2019cross, kamei2012large, yatish2019mining, choi2022empirical} have been widely applied to improve the software development and maintenance process. Thanks to the development of various machine learning (ML) models, researchers have proposed many ML-based approaches to facilitate those software engineering tasks in software maintenance and evolution. For example, in modern code review, it is unlikely to maintain the quality and perform exhaustive code review activities for all the incoming commits with limited Software Quality Assurance (SQA) resources \cite{pornprasit2021pyexplainer}. Therefore, in this regard, the JIT defect prediction model can be used to predict defect-introducing commits before performing any commit operations. Thus, it assists software developers in prioritizing their limited SQA resources by focusing on the riskiest commits during code review \cite{catolino2019cross, kamei2012large}. It also helps create proactive strategies for improving software quality to avoid past mistakes, which could result in software defects in future releases \cite{mcintosh2018fix}.


Despite their potential usefulness in various software analytics tasks, ML models may not be employed in real-world scenarios because ML practitioners often need more understanding of why specific predictions are made \cite{jiarpakdee2020empirical}. For example, in the case of JIT defect prediction, they usually ask why a commit is being predicted as defect-introducing, as JIT defect models solely focus on prediction without explaining their decisions \cite{pornprasit2021pyexplainer, jiarpakdee2020empirical, jiarpakdee2021practitioners}. Therefore, due to the lack of explainability to understand the reasoning behind the decisions of the JIT defect models, practitioners may encounter difficulties in enhancing software quality with limited SQA resources, potentially leading to less effective SQA practices. As a result, many XAI methods such as SHapley Additive exPlanation (SHAP) \cite{lundberg2017unified}, Local Interpretable Model-agnostic Explanations (LIME) \cite{ribeiro2016should}, Anchors \cite{ribeiro2018anchors}, BreakDown \cite{gosiewska2019ibreakdown}, SQAPlanner \cite{rajapaksha2021sqaplanner}, Integrated Gradient (IG) \cite{sundararajan2017axiomatic}, PyExplainer \cite{pornprasit2021pyexplainer} and so on have been developed to explain the outcome of the ML models.

\begin{figure}[htbp]
\centering     
\subfigure[]{\label{motivation_a}\includegraphics[width=6.5cm, height=2.15cm]{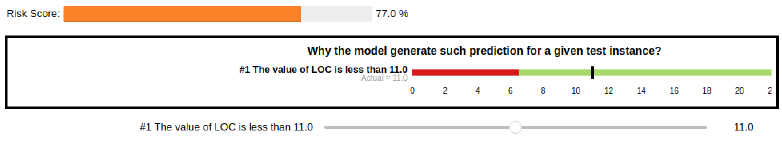}}
\vspace{-0.5em}
\subfigure[]{\label{motivation_d}\includegraphics[width=6.5cm, height=1.75cm]{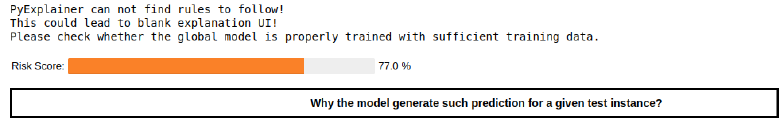}}
\vspace{-0.5em}
\subfigure[]{\label{motivation_b}\includegraphics[width=6.5cm, height=3cm]{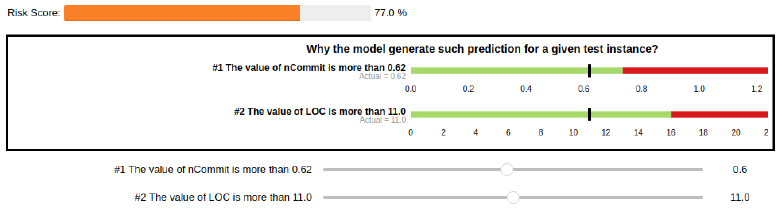}}
\vspace{-0.5em}
\subfigure[]{\label{motivation_c}\includegraphics[width=6.5cm, height=3cm]{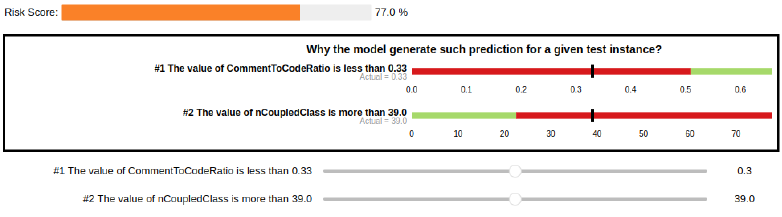}}

\caption{Explanations generated by PyExplainer for a single instance reveal that the model predicts the instance as a defect-introducing commit with a 77\% probability, as indicated by the Risk Score.}
\label{pyexp explanation}
\vspace{-1.25em}
\end{figure}
\begin{figure}[htbp]
\centering     
\subfigure[]{\label{lime_a}\includegraphics[width=6.5cm, height=2.75cm]{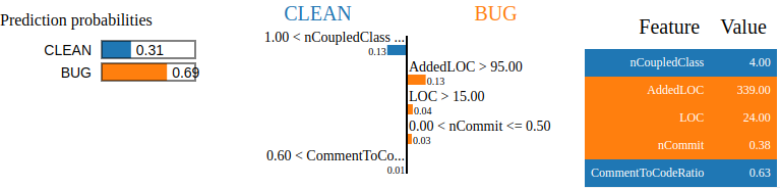}}
\subfigure[]{\label{lime_b}\includegraphics[width=6.5cm, height=2.75cm]{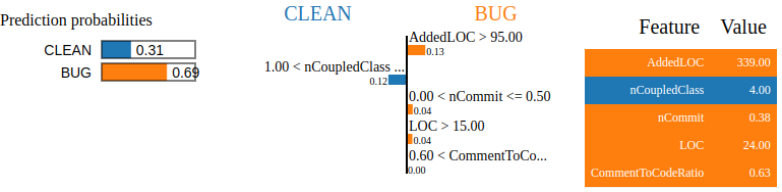}}
\subfigure[]{\label{lime_c}\includegraphics[width=6.5cm, height=2.75cm]{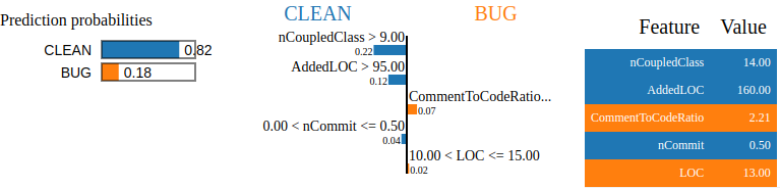}} 
\caption{Figures \ref{lime_a} and \ref{lime_b} demonstrate that LIME produces different explanations when regenerated for the same instance. These two figures collectively visualize why the given instance is predicted as a buggy commit. Figure \ref{lime_c} visually explains an instance predicted as a clean commit.}
\label{lime_anomaly}
\end{figure}

The rule-based XAI methods have been used in software analytics very recently \cite{jiarpakdee2020empirical, rajbahadur2021impact, roy2022don}. For example, the state-of-the-art model-agnostic\footnote{Model-agnostic refers to techniques that are universally applicable to ML models regardless of their specific type or architectural design.} XAI technique called LIME \cite{ribeiro2016should} was adopted in software analytics to explain the predictions of the JIT defect models at line-level and file-level \cite{jiarpakdee2020empirical, pornprasit2021jitline}. Research shows that the synthetic neighbourhood generation process significantly impacts how well LIME produces its explanations \cite{pornprasit2021pyexplainer}. In addition, LIME fails to generate consistent explanations for the same instance if we apply it multiple times in a row \cite{jiarpakdee2020empirical}. Recently, a local rule-based model-agnostic XAI technique called PyExplainer was proposed to overcome the shortcomings of LIME in explaining the outcome of the JIT defect prediction models. However, our manual investigations find that they produce inconsistent explanations, as shown in Figures \ref{pyexp explanation} and \ref{lime_anomaly} for the same instance if we rerun it. In addition, PyExplainer sometimes fails to generate explanations and shows a blank UI to its practitioners, as depicted in Figure \ref{motivation_d}. Thus, the inconsistent explanations generated by XAI techniques may confuse and discourage practitioners from using XAI techniques in real-world applications \cite{roy2022don}.

Lundberg et al. \cite{lundberg2017unified} argued that instance explanation generation must remain consistent upon regeneration for the same instance. However, our manual investigations find that PyExplainer and LIME do not adhere to this statement when rerunning on the same instance. Moreover, they exhibit anomalous behaviours, as illustrated in Section \ref{MExam}, raising concerns about the reliability of the explanations they generate for ML models, such as JIT defect prediction models. Therefore, it is essential to systematically evaluate the reliability and consistency of the explanations produced by rule-based XAI techniques for various software analytics tasks, including JIT defect prediction, clone detection, and the classification of useful code review comments.

To the best of our knowledge, no framework or granular-level evaluation metrics are currently available for systematically assessing the ability of rule-based XAI techniques to generate reliable and consistent explanations for ML models in software analytics tasks. Therefore, in this paper, we introduce a novel framework: Evaluation of Explainable AI (\textit{EvaluateXAI}), along with six granular-level evaluation metrics, to evaluate the reliability and consistency of the explanations generated by rule-based XAI techniques. Specifically, we use \textit{EvaluateXAI} to assess PyExplainer and LIME in the context of seven machine learning models trained on five commonly used datasets in software analytics tasks. Thus, the key contributions of our study are listed below:
\begin{itemize}

    \item We design a novel framework called \textit{EvaluateXAI} to assess the reliability and consistency of rule-based XAI techniques in generating explanations for ML models in software analytics tasks.
    \item We adopt two metrics from Pornprasit et al. \cite{pornprasit2021pyexplainer} and develop four additional granular-level evaluation metrics within \textit{EvaluateXAI}.
    \item We demonstrate that \textit{EvaluateXAI} is applicable to any rule-based XAI techniques, enabling the evaluation of their reliability and consistency in generating explanations for ML models. 
    \item Our code and the datasets used in this study are publicly available to advance the research in XAI. The replication packages can be accessed through this link\footnote{\href{https://zenodo.org/doi/10.5281/zenodo.7869326}{Replication-packages}}.

\end{itemize}


    
    



\section{Motivating Examples}
\label{MExam}
This study aims to evaluate the reliability and consistency of the explanations generated by rule-based XAI techniques by designing a novel framework and granular-level evaluation metrics. To achieve this, we consider state-of-the-art rule-based model-agnostic XAI techniques, such as PyExplainer and LIME, as candidates for designing the framework and evaluation metrics. We deliberately selected PyExplainer and LIME due to their extensive study in software analytics tasks, particularly for JIT defect predictions \cite{pornprasit2021pyexplainer, jiarpakdee2020empirical, roy2022don}. Additionally, we emphasize JIT defect prediction as an illustrative example to demonstrate the challenges and complexities associated with applying rule-based XAI techniques to practical software analytics tasks.

PyExplainer was introduced to address the limitations of LIME in explaining the outcomes of JIT defect prediction models. However, while it has shown effectiveness in providing explanations for JIT defect prediction models, our manual investigations reveal inconsistencies and anomalies in its explanation generation. To illustrate the shortcomings of PyExplainer, let us consider a ML (e.g., Random Forest) model trained on JIT defect-introducing commit data. We follow the official ``PART A - Quick Start" tutorial provided by the PyExplainer\footnote{\href{https://github.com/awsm-research/PyExplainer/blob/master/TUTORIAL.ipynb}{``PART A - Quick Start" Tutorial}}. We want to know why PyExplainer says a commit introduces a bug before performing the commit operation and how consistent the generated explanations are by PyExplainer. It is important to note that to find the inconsistencies and anomalous behaviours of PyExplainer and LIME; we rerun them on the same instance 10 times in a row. Table \ref{instance} shows the instance on which we apply PyExplainer and LIME to generate explanations, and all the generated explanations are shown in Figures \ref{pyexp explanation}, \ref{lime_a}, and \ref{lime_b}.

\begin{table}[htbp]
\tiny
\centering
\caption{An instance to be explained using PyExplainer and LIME.}
\resizebox{\columnwidth}{!}{\begin{tabular}{c|c|c|c|c}
\hline
nCommit  & AddedLOC & nCoupledClass & LOC  & CommentToCodeRatio \\ \hline
0.615385 & 246.0    & 39.0          & 11.0 & 0.33               \\ \hline
\end{tabular}}
\vspace{-1.25em}
\label{instance}
\end{table}

Lundberg et al. \cite{lundberg2017unified} argued that instance explanation generation must remain consistent upon regeneration for the same instance. However, Figure \ref{pyexp explanation} depicts that PyExplainer generates four different explanations for the same instance upon re-execution. Additionally, we observe that among the five features (as shown in Table \ref{instance}), four of them (e.g., \textit{LOC}, \textit{nCommit}, \textit{CommentToCodeRatio}, and \textit{nCoupleedClass}) have been used to generate explanations for different executions. Furthermore, in Figure \ref{motivation_a}, PyExplainer generates the explanation where the feature value \textit{LOC} is less than $11.0$. On the other hand, in Figure \ref{motivation_b}, PyExplainer generates the explanation where the feature value \textit{LOC} is more than $11.0$. Finally, Figure \ref{motivation_d} represents the scenario where PyExplainer fails to generate an explanation. We observe a similar inconsistency for LIME as depicted in Figures \ref{lime_a} and \ref{lime_b}. These inconsistent explanations may confuse practitioners and discourage them from using PyExplainer and LIME to explain the outcomes of JIT defect prediction models \cite{roy2022don}.

\begin{figure}[htbp]
\centering     
\subfigure[]{\label{interpret_a}\includegraphics[width=6.5cm, height=2.15cm]{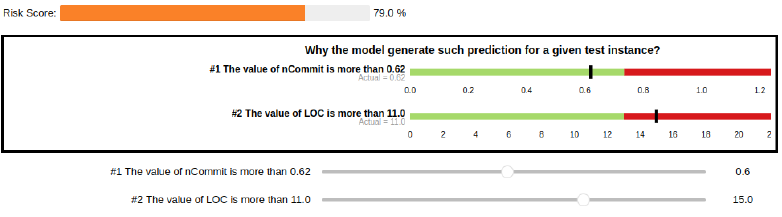}}
\vspace{-0.5em}
\subfigure[]{\label{interpret_b}\includegraphics[width=6.5cm, height=3cm]{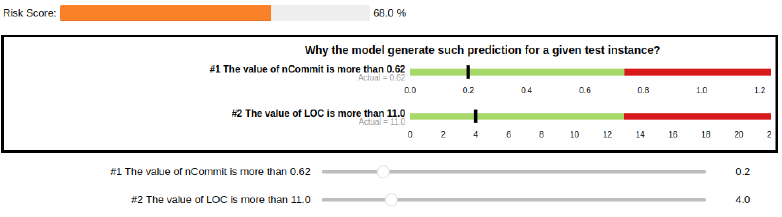}} \\
\vspace{-0.5em}
\subfigure[]{\label{interpret_c}\includegraphics[width=6.5cm, height=2.15cm]{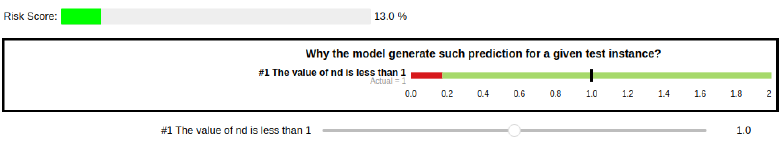}}
\vspace{-0.5em}
\subfigure[]{\label{interpret_d}\includegraphics[width=6.5cm, height=2.15cm]{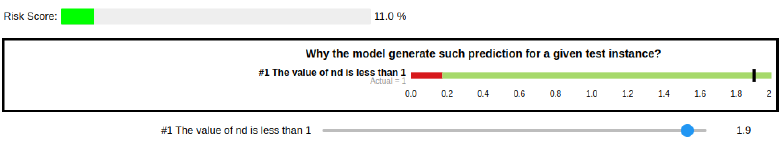}} \\
\vspace{-0.5em}
\subfigure[]{\label{interpret_e}\includegraphics[width=6.5cm, height=2.15cm]{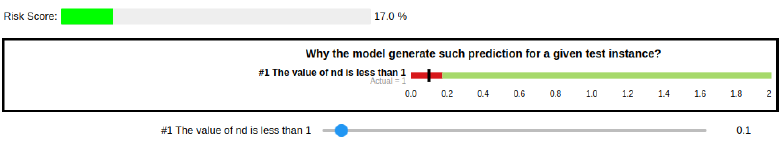}}
\vspace{-0.5em}
\caption{Explanations generated by PyExplainer for a single instance reveal that the model predicts the instance as a defect-introducing commit with a 77\% probability (as indicated by the Risk Score) as shown in Figures \ref{interpret_a} and \ref{interpret_b}. On the other hand, Figures \ref{interpret_c}, \ref{interpret_d}, and \ref{interpret_e} depict the generated explanations when the ML model predicts the given instance as a clean commit.}
\label{interpret visualization}
\vspace{-1.25em}
\end{figure}

\begin{figure*}[t]
\centering     
\subfigure[]{\label{anomaly_a}\includegraphics[width=6.5cm, height=2.15cm]{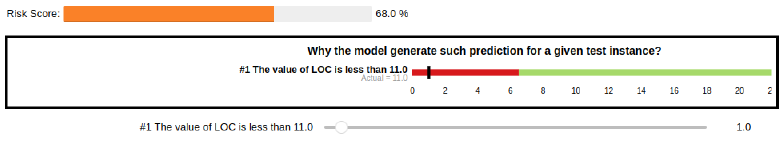}}
\vspace{-0.5em}
\subfigure[]{\label{anomaly_b}\includegraphics[width=6.5cm, height=2.15cm]{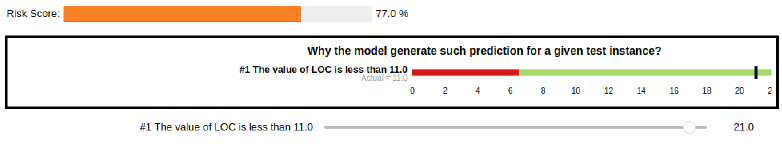}} \\
\vspace{-0.5em}
\subfigure[]{\label{anomaly_c}\includegraphics[width=6.5cm, height=3cm]{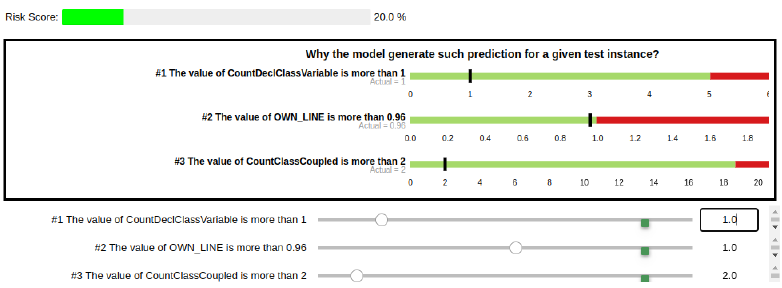}}
\vspace{-0.5em}
\subfigure[]{\label{anomaly_d}\includegraphics[width=6.5cm, height=3cm]{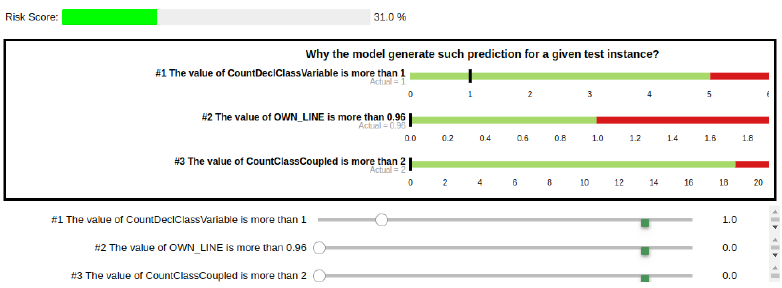}} \\
\vspace{-0.5em}
\subfigure[]{\label{anomaly_e}\includegraphics[width=6.5cm, height=3cm]{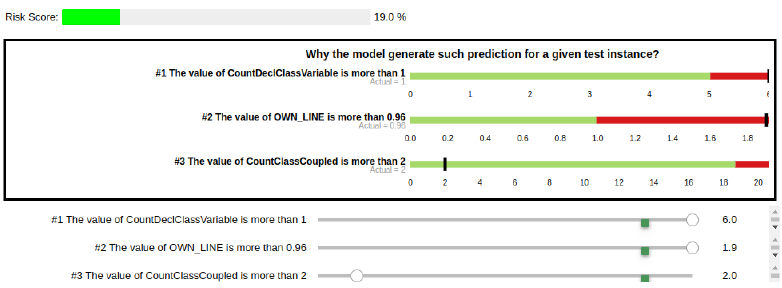}}
\vspace{-0.5em}
\caption{While changing the feature values in the guided directions, PyExplainer fails to function correctly, which we refer to as the anomalous behaviour of PyExplainer.}
\label{pyexp anomaly}
\end{figure*}

Another essential aspect of PyExplainer is whether it generates quality explanations and whether we can rely on them. To illustrate this, let us consider a scenario where a ML model predicts the given instance (e.g., as shown in Table \ref{instance}) as a buggy commit with a risk score of $77\%$, as shown in Figure \ref{pyexp explanation}. The visual interpretation of PyExplainer (details in Section \ref{back}) illustrates that if we change the feature values in the \textit{green} zone, the risk score must be decreased, and changing feature values in the \textit{red} zone increases the risk score. Now, if we change the feature values in the \textit{green} zone, the risk score decreases to $68\%$ from $77\%$, considering Figures \ref{motivation_b} and \ref{interpret_b}. However, Figure \ref{anomaly_b} shows that this is not always true. Figure \ref{motivation_a} shows that when the value of $LOC$ is 11, the ML model predicts the given instance as defective commits with a risk score of 77\%. Despite changing the feature value of \textit{LOC} from $11.0$ to $21.0$ in the \textit{green} zone, it remains the same instead of decreasing, as shown in Figure \ref{anomaly_b}.

Figure \ref{interpret_a} shows that changing feature values in the \textit{red} zone increases the risk score from $77\%$ (as shown in Figure \ref{motivation_b}) to $79\%$. However, Figure \ref{anomaly_a} shows that this is not always true. For example, when considering Figures \ref{motivation_a} and \ref{anomaly_a}, after changing the feature value of \textit{LOC} from $11.0$ to $1.0$, the risk score decreased from $77\%$ to $68\%$ instead of increasing, demonstrating the anomalous behaviour of PyExplainer. Based on our manual investigations, it is evident that LIME produces more consistent explanations than PyExplainer, although it should be noted that the explanations are not entirely consistent. For example, LIME generates two different explanations, whereas PyExplainer generates four different explanations in 10 consecutive runs on the same instance. Therefore, these inconsistent and anomalous explanation generations prompt us to conduct in-depth investigations and design a novel framework with different evaluation metrics at a granular level to assess the reliability and consistency of the explanations generated by rule-based XAI techniques for software analytics tasks.

\section{Background}
\label{back}
We introduce \textit{EvaluateXAI} and propose granular-level evaluation metrics using PyExplainer as a case study. Therefore, understanding the working principles and interpretation of visual explanations generated by PyExplainer is essential for comprehending \textit{EvaluateXAI} and the granular-level metrics. Thus, this section provides a brief description of how PyExplainer works and how we interpret the explanations generated by PyExplainer for JIT defect prediction models, along with some basic terminologies related to the study.

\textbf{Explainable Artificial Intelligence (XAI)}: It refers to a set of techniques, tools, and methods used in artificial intelligence (AI) and machine learning (ML) to make the decisions and outputs of AI and ML systems understandable and interpretable by humans. XAI provides transparency and visibility into how AI or ML models make decisions.

\textbf{Instance}: In machine learning, an `instance' typically refers to a single occurrence or example of data used in training, testing, or prediction. An instance is a specific set of input features that the machine learning algorithm processes to make predictions or learn patterns. For example, in the context of just-in-time (JIT) defect prediction, an instance would be a unit of code or a software component, and the features associated with that instance would include various metrics or characteristics related to the code, such as lines of code, complexity, historical defect information, and so on.

\textbf{Risk Score}: The risk score defines the probability that an instance is predicted as a predefined class. To clarify, let us consider a scenario where a ML model was trained on the JIT defect prediction dataset, such as cross-project mobile apps. Figure \ref{motivation_a} shows that the model predicts the selected instance as a defect-introducing commit with 77\% probability. Therefore, we can say that the risk score for that instance is 77\%. In this paper, we use the terms `prediction probability for positive class' and `risk score' interchangeably, denoting the same thing.

\textbf{Simulated Instance:} It represents the instance we generate by changing the feature values in the guided direction as per the PyExplainer explanation. Subsection \ref{sim ins gen} provides a detailed overview of the simulated instance generation process.


\textbf{PyExplainer Overview}: PyExplainer is a model-agnostic method that uses local rules to explain predictions made by JIT defect models. It consists of four steps: in step 1, PyExplainer constructs synthetic neighbours around the instance to be explained using crossover and mutation approaches. In step 2, PyExplainer obtains the predictions of the synthetic neighbours from the global model to mimic the global model's behaviour. In step 3, PyExplainer creates a local model based on the logistic regression technique called RuleFit. RuleFit aggregates ensembles trees and linear models to interpret the model while making sense of the logical reasons learned from the rule features. Finally, in step 4, PyExplainer generates an explanation from the local rule-based model and visualizes the explanation as shown in Figure \ref{pyexp explanation}.

\textbf{Interpret PyExplainer Visual Explanation}:
The PyExplainer tool represents rule-based explanations through bullet plot visualization and written descriptions. For example, the visual illustration shown in Figure \ref{motivation_b} tells us three key pieces of information: (1) the rules that describe why a commit is predicted as defect-introducing; (2) the vertical black bars that denote the actual feature values; and (3) the risk score is associated with a set of feature values that fall within a specific range. In addition, the feature values that fall within a risky range are represented by \textit{red} shades, while \textit{green} shades denote the non-risky range of feature values. Figure \ref{motivation_b} shows two textual descriptions: (1) ``\textit{the value of nCommit is more than 0.62}", and (2) ``\textit{the value of LOC is more than 11.0}". Finally, the risk score is 77\% which denotes that the instance is predicted as a defect-introducing commit with $77\%$ probability by the JIT defect model. 

Ideally, if we increase the value of both features in the direction of the \textit{red} zone, the risk score must be increased, as shown in Figure \ref{interpret_a}. On the other hand, if we decrease the value of both features in the direction of the \textit{green} zone, the risk score must be reduced, as shown in Figure \ref{interpret_b}. Thus, we can change the feature values to alter the prediction of buggy commits to clean commits and vice-versa. In addition, Figure \ref{motivation_b} demonstrates that the features \textit{nCommit} and \textit{LOC} influence the JIT defect model to predict the given instance as a buggy commit.

Figure \ref{interpret_c} describes that the selected instance (e.g., an instance of the cross-project mobile apps dataset) is predicted as clean commits with $13\%$ probability. Ideally, if we change the feature value of \textit{nd} (number of modified directories) in the \textit{green} zone, the probability score must be decreased, as shown in Figure \ref{interpret_d}. But, on the other hand, if we change the feature value of \textit{nd} in the \textit{red} zone, then the probability score must be increased, as shown in Figure \ref{interpret_e}.  Finally, we observe that any changes in the feature values affect the prediction of the JIT defect model. Thus, PyExplainer generates visual explanations to explain the predictions of the JIT defect model.

\section{Framework Design}
\label{method}
\begin{figure}[htbp] 
\centering
\includegraphics[width=8.5cm, height = 5cm]{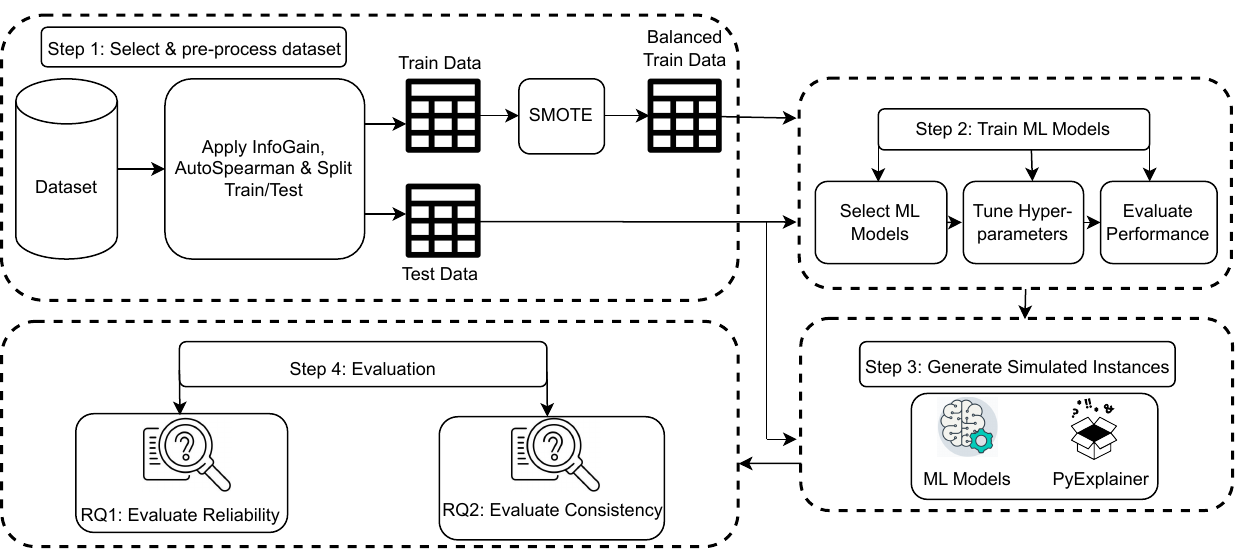}
\caption{Our proposed framework for evaluating the reliability and consistency of rule-based XAI techniques. Step 1 includes the dataset selection and pre-processing methods. Step 2 involves the training of ML models, while Step 3 involves the generation of simulated instances. Step 4 investigates the effectiveness of PyExplainer and LIME in generating reliable and consistent explanations for ML models in software analytics tasks.} 
\label{workflow} 
\vspace{-1.25em}
\end{figure}

In this section, we discuss the key components of our proposed framework called \textit{EvaluateXAI}, which is designed to evaluate the reliability and consistency of the explanations generated by rule-based XAI techniques through an extensive study. The framework consists of four key elements, each denoted as a step, as illustrated in Figure \ref{workflow}. Below, we provide a detailed description of each of these steps.

\subsection{Dataset Selection and Pre-processing}
\label{DC}
The dataset selection process is crucial because the quality and suitability of the dataset directly impact the validity and generalizability of ML models' results. Additionally, dataset pre-processing is critical in the data analysis and ML pipeline. Therefore, in this study, we carefully select datasets and pre-process them in the following way:

\subsubsection{Dataset Selection}
This study evaluates the reliability and consistency of the explanations generated by XAI techniques for ML models trained on tabular data for various software analytics tasks, such as defect prediction, cross-language clone detection (CLCDSA), and code review comment classification. We utilize five different publicly available datasets: three JIT defect prediction datasets, one clone detection dataset, and one code review comment classification dataset from related studies. For JIT defect prediction, we employ three datasets: a dataset from cross-project mobile apps (Catolino et al. \cite{catolino2019cross}) and Java project datasets containing data from 10 different Java projects (Yatish et al. \cite{kamei2012large}). In this category, we utilize a carefully selected cross-project mobile apps dataset with 14 features and approximately 30,000 data points. Various Java projects contribute datasets with 65 distinct features, comprising about 25,000 data points. Additionally, we select the Postgres dataset, containing 14 features and approximately 20,000 data points, from a pool of six subject systems for Desktop applications, as introduced by Kamei et al. \cite{yatish2019mining}.

The code review comment dataset \cite{rahman2017predicting}, featuring 15 distinct attributes and consisting of 1,100 data points, aids in identifying valuable comments. We also extend our analysis to include the cross-language clone (CLCDSA) dataset by Nafi et al. \cite{nafi2019clcdsa}. The CLCDSA dataset, designed for cross-language clone detection, comprises 18 distinctive features and roughly 30,000 data points. By leveraging these curated datasets, each with its unique attributes, our study offers comprehensive insights into the reliability and consistency of XAI techniques when generating explanations for ML models across various software analytics tasks.

\subsubsection{Feature Engineering}
In our methodology, we employ feature engineering strategies to enhance the performance of our chosen ML models. We draw inspiration from prior research endeavours to ascertain the significance of specific attributes. For instance, Catolino et al. \cite{catolino2019cross} employed the \textit{InfoGain} method for feature selection, identifying 6 relevant features out of the original 14. These selected attributes are then integrated into our model for training. Similarly, Nafi et al. \cite{nafi2019clcdsa} conducted feature selection within their code fragments, resulting in the retention of 9 features from an initial pool of 24. As a result, each clone pair in their dataset featured 18 feature values and a single class label, forming the basis for our ML model training.

When dealing with the Java project dataset, we confront the issue of collinearity among its features. To overcome this challenge, we employ the \textit{AutoSpearman} technique, a method previously utilized in research endeavours by Yatish et al. \cite{yatish2019mining}, Pornprasit et al. \cite{pornprasit2021pyexplainer}, Catolino et al. \cite{catolino2019cross}, and Roy et al. \cite{roy2022don}. This technique enables us to identify and retain 27 relevant features from the original set of 65 in our experiment on the Java project dataset. In examining the Postgres dataset, we also encounter the challenge of multicollinearity. To effectively address this issue, we implement the same feature selection and normalization methodologies outlined in the study by Kamei et al. \cite{kamei2012large}. This approach ensures that our model is well-equipped to handle the complexities inherent in the dataset while preserving the integrity of the selected features.

\subsubsection{Dataset Balancing}
The task of JIT defect prediction suffers from an imbalanced binary classification problem, where one class (e.g., clean commit) has a disproportionately large number of examples compared to the other class (e.g., bug-introducing commit). To address this issue, we apply the Synthetic Minority Over-sampling Technique (SMOTE) \cite{chawla2002smote}, which has been used in previous studies \cite{catolino2019cross, roy2022don, pornprasit2021pyexplainer} on the same dataset. Similarly, we use the same technique to address the imbalanced dataset problem in the cross-language clone dataset. It is important to note that we exclusively apply SMOTE to the training dataset, leaving the test dataset unaltered \cite{jiarpakdee2020empirical}.

\subsection{Machine Learning Model Selection and Training}
ML model selection and training are critical to developing a successful ML project. These steps involve choosing an appropriate algorithm or model architecture for the specific task and training the selected model on the pre-processed dataset. Additionally, finding the best combination of hyperparameter values that optimize the model's performance ensures that the model generalizes well to new, unseen data and avoids overfitting the training data.

\subsubsection{Machine Learning Models}
This study evaluates the reliability and consistency of rule-based XAI techniques for generating explanations for classical ML models. Therefore, we target classical ML models extensively used in different software analytics tasks and XAI research. For example, in software defect prediction, Logistic Regression and Random Forest are two of the most commonly utilized classification techniques \cite{shihab2012exploration, hall2011systematic, nadim2023utilizing, nadim2022leveraging}. Additionally, although PyExplainer is a model-agnostic XAI technique, its current implementation only supports Sklearn's supervised classification models\footnote{https://pyexplainer.readthedocs.io/en/latest/usage.html}. Therefore, we consider this criterion when selecting classical ML models. Thus, for our experiments, we select three classical ML models (Logistic Regression (LR), Multi-Layer Perceptron (MLP), and Decision Tree (DT)) and four ensemble methods (Gradient Boosting Classifier (GBC), Random Forest (RF), AdaBoost (ADA), and Bagging (BAG)). These models have been used in previous software analytics tasks and XAI studies \cite{catolino2019cross, kamei2012large, pornprasit2021pyexplainer, roy2022don, huang2024aligning, brughmans2024disagreement, begum2023software, mehta2021improved, sharma2023ensemble, singh2024improved, malhotra2023empirical, ali2023analysis, kumar2023software, ali2024software, zhang2023comparative, abdou2024severity, rao2023study, dewangan2023severity, shrimankar2022software, saha2022optimized, wang2023towards, xiao2004mirrorfair, feng2024machine}; hence, we select all of them.



\subsubsection{Hyperparameters Tuning}
\label{tune_param}
The main objective of hyperparameter tuning is to enhance the model's performance on a validation dataset, ensuring its ability to effectively generalize to new and unseen data while avoiding overfitting the training data. By exploring various combinations of hyperparameters, tuning methods aim to identify the most suitable set of values that yield the highest model performance. In our study, we follow commonly used hyperparameter settings to train ML models \cite{catolino2019cross, kamei2012large, pornprasit2021pyexplainer, roy2022don, tantithamthavorn2016automated, tantithamthavorn2016empirical, tantithamthavorn2018impact, yang2020hyperparameter}. To tune the hyperparameters, we apply grid search \cite{liashchynskyi2019grid}, random search \cite{bergstra2012random}, and Bayesian optimization \cite{wu2019hyperparameter}. After obtaining the results from each technique, we compare the AUC values to identify the best-performing hyperparameter combinations. For example, Table \ref{hyper_param_tuning} presents the three sets of hyperparameters after tuning the LR model on the cross-project mobile app dataset. This table shows that the random search provides us with the best hyperparameter combinations regarding the AUC value. This ensures that our final selection leverages the strengths of each optimization method, leading to the most effective hyperparameter tuning. Thus, we select the best hyperparameter combinations from the three search techniques. The thorough optimization process is fundamental to our ML model training, ensuring that our models are precisely tuned for optimal performance.

\begin{table}[htbp]
\centering
\caption{Optimized hyperparameter combinations selected from different search techniques}
\resizebox{\linewidth}{!}{
\begin{tabular}{l|l|l}
\hline
Search technique      & AUC value & Parameter value                                                                                    \\ \hline
Grid search           & \cellcolor[HTML]{C0C0C0}\textbf{0.6979}    & C: 2.78, dual: True, max iter: 130, penalty: `l2', solver:   `liblinear'                           \\ \hline
Random search         & 0.6675    & solver: `lbfgs', penalty: `l2', max\_iter: 110, dual: False,   C: 166.81 \\ \hline
Bayesian optimization & 0.6765    & solver: `lbfgs', penalty: `l2', max\_iter: 130, dual: False,   C: 1291.55 \\ \hline
\end{tabular}}
\label{hyper_param_tuning}
\end{table}

\subsubsection{Evaluate Model Performance}
Validation of ML models is a crucial step during model development, aimed at evaluating the model's capability to perform well on new and unseen data. Therefore, to ensure that PyExplainer and LIME provide explanations based on well-trained models, we evaluate the performance of the studied classification techniques. Since PyExplainer often takes excessive time to generate an explanation for a single instance, we split the dataset into training (90\%) and testing (10\%) to speed up the overall process. To validate ML models using the 90\% training data, we apply a 10-fold cross-validation technique like other defect prediction studies \cite{d2010extensive, wahono2015systematic}.

We assess the model's performance based on accuracy and F1-score. Additionally, we employ the Area Under the Receiver Operating Characteristic Curve (AUC-ROC) to assess the discriminatory power of our models, following the recommendation of recent research \cite{catolino2019cross, jiarpakdee2020empirical, lyu2021towards, ghotra2015revisiting, lessmann2008benchmarking, rahman2013and}, and the 0.75 test-AUC value is the minimum threshold for ensuring the reliability of explanations \cite{roy2022don, lyu2021towards}. The AUC curve measures the true negative rate (coverage of the negative class) on the x-axis and the true positive rate (coverage of the positive class) on the y-axis. As a threshold-independent performance measure, the AUC evaluates the models' ability to discriminate between positive (e.g., defective commits) and negative (e.g., clean commits) instances. The AUC values range from 0 (worst) to 0.5 (no better than random guessing) up to 1 (best) \cite{hanley1982meaning}. Thus, the AUC-ROC is a comprehensive evaluation technique for binary classification models, gauging their effectiveness in distinguishing between positive and negative instances.

\subsection{Simulated Instance Generation}
\label{sim ins gen}
We want to evaluate the reliability and consistency of the explanations generated by rule-based XAI techniques (e.g., PyExplainer and LIME) based on the simulated instances (e.g., a single commit) using \textit{EvaluateXAI} and granular-level evaluation metrics described in \ref{eval}. Pornprasit et al. \cite{pornprasit2021pyexplainer} changed the feature values by one standard deviation (STD) from the threshold values found in the explanation. For example, we want to alter the prediction of defect-introducing commit to clean commit, as shown in Figure \ref{motivation_a}. From training data, we calculate the STD value of each feature, and let us assume the STD value of the feature \textit{LOC} is $1.9$. From Figure \ref{motivation_a}, we also see the threshold value of \textit{LOC} is approximately $6.35$. As we want to change the prediction from defect to clean, we must add the STD value to the threshold value to increase the value in the \textit{green} zone. Therefore, the simulated feature value is set to $1.90 + 6.35=8.25$. In the case of Figure \ref{motivation_b}, we must subtract one STD value from the threshold value of the \textit{nCommit} to move in the \textit{green} zone.

PyExplainer always creates two types of rules while generating explanations for a given instance: (1) the \textit{less than } rule, which denotes that the value of a particular feature is always less than a threshold value (e.g., \textbf{The value of LOC is less than 11.0}, as shown in \ref{motivation_a}) impacting the model's prediction, and (2) the \textit{more than} rule, which denotes that the value of a particular feature is always greater than a threshold value (e.g., \textbf{The value of nCommit is more than 0.62}, as shown in \ref{motivation_b}) impacting the model's prediction. Considering Figures \ref{motivation_a} and \ref{motivation_b}, we observe two rules and if we want to change the prediction of the JIT defect models from defect to clean: (1) for \textit{less than} rule, we add one STD value with the threshold to generate simulated instances; (2) in contrary, for \textit{more than} rule, we subtract one STD value from the threshold to generate simulated instances. Furthermore, if we want to change the prediction of the JIT defect model from clean to defect, we have to do the exact opposite, as described above. Thus, we generate simulated instances for our experiments. We define the generation of a single simulated instance by the following equation:

\begin{equation}
\small
\label{sim eqn}
sim(x') = \forall f \in F, |x[f] \pm (\alpha * std)|
\end{equation}

Where $x$ denotes a single instance of a dataset $X$, $x'$ represents the generated simulated instance, and $f$ is a single feature selected from the set of features $F$. In Equation \ref{sim eqn}, $\alpha$ acts as the controlling parameter, determining the amount of standard deviation to be added or subtracted from each feature value (e.g., the threshold obtained from explanation) of the original instance in generating a simulated instance. Since LIME provides explanations for tabular data in a rule-based format similar to PyExplainer, we can generate simulated instances for LIME using the process outlined above. Below, we describe the detailed process of generating simulated instances for LIME.

The visual explanation provided by PyExplainer illustrates that altering feature values within the \textit{red} zone increases the probability (e.g., risk score) of the given instance being predicted as a positive class (e.g., bug-inducing commit, cross-language clone, and useful code review comments). Conversely, adjusting feature values within the \textit{green} zone reduces the probability. In Figure \ref{motivation_a}, it is demonstrated that the given instance is predicted as a buggy commit with a probability of $77\%$, and the associated rule is ``the value of LOC is less than 11.0 ($LOC<11.0$)." Thus, in alignment with the visual explanation, adhering to this rule and decreasing the value of \textit{LOC} in the guided direction (e.g., satisfying the rule $LOC<11.0$) increases the probability, and vice versa.

A similar procedure can be applied to generate simulated instances using LIME. Figure \ref{lime_a} indicates that the instance is predicted as a buggy commit with a probability of $69\%$. In this context, the \textit{orange} colour represents the \textit{red} zone. Therefore, adjusting feature values within the \textit{orange} zone to satisfy the rules increases the probability of the given instance being predicted as a positive class (e.g., a buggy commit).

Figure \ref{lime_a} illustrates the influence of three features on the ML model's prediction of the instance as a buggy commit. For example, the actual value of the \textit{AddedLOC} feature is $339.00$. In the generated rule (e.g., $AddedLOC > 95.00$), any \textit{AddedLOC} value surpassing $95.00$ (by altering feature values within the \textit{orange} zone in accordance with the rules, referred to as the guided direction) raises the probability of the instance being predicted as a positive class. Conversely, modifying feature values counter to the guided direction diminishes the probability of the given instance being predicted as a positive class. For instance, decreasing the feature value (e.g., $AddedLOC < 95.00$) results in a reduced probability of the instance being predicted as a positive class. Thus, this process allows us to generate a simulated instance for any given instance that is predicted as a positive class (e.g., a buggy commit).

We follow similar steps as illustrated above to generate simulated instances for the negative class (e.g., clean commit). Figure \ref{lime_c} indicates that the given instance is predicted as a clean commit with a probability of $82\%$. In this case, the \textit{blue} colour corresponds to the \textit{green} zone. Therefore, adjusting feature values within the \textit{blue} zone to adhere to the rules increases the probability of the given instance being predicted as a negative class (e.g., a clean commit).

Figure \ref{lime_c} illustrates the influence of three features on the ML model's prediction of the instance as a clean commit. For instance, the actual value of the \textit{nCoupledClass} feature is $14.00$. In the generated rule (e.g., $nCoupledClass > 9.00$), any \textit{AddedLOC} value exceeding $9.00$ (by altering feature values within the \textit{green} zone in accordance with the rules, referred to as the guided direction) increases the probability of the instance being predicted as a negative class. Conversely, modifying feature values counter to the guided direction decreases the probability of the given instance being predicted as a negative class. For instance, reducing the feature value (e.g., $nCoupledClass < 9.00$) leads to a decreased probability of the instance being predicted as a negative class. Thus, this process enables us to generate a simulated instance for any given instance predicted as a negative class (e.g., a clean commit).

\subsection{Evaluation}
\label{eval}
We aim to evaluate the reliability and consistency of the explanations generated by rule-based XAI techniques. In this section, we introduce a set of evaluation metrics at different granularity levels to assess the reliability and consistency of these explanations. Since PyExplainer was designed to address the shortcomings of LIME in explanation generation, we use PyExplainer as a case study to develop the evaluation metrics. We adopt two evaluation metrics from Pornprasit et al. \cite{pornprasit2021pyexplainer} for reliability and consistency assessment and design four additional granular-level metrics, as detailed below.

\textbf{Positive class}: We consider defective commits, cloned-pair, and useful comments in code review as the positive class.

\textbf{Negative class}: We consider clean commit, not clone-pair, and not useful comments in code review as negative class.

\textbf{\textit{\%Reversed}}: $\%Reversed$ assesses the reliability of the explanations generated by PyExplainer. Ideally, the simulated instances generated based on the explanations should cause the model's predictions to flip on those instances. We generate simulated instances as depicted in Figure \ref{interpret visualization} and described in Subsection \ref{sim ins gen}. Subsequently, we calculate the proportion of simulated instances that successfully flip the predictions of the ML models. A higher $\%Reversed$ value indicates the effectiveness of PyExplainer in generating reliable explanations for ML models. The $\%Reversed$ metric is defined as:

\begin{equation}
\small
\label{reversed equn}
\%Reversed = \frac{|\{x|x \in X \wedge M(x') \neq M(x)\}|}{|X|}
\end{equation}

where $X$ is a dataset, $x \epsilon X$ is the original instance, $x'$ is the generated simulated instance, and $M$ is the ML model. We separately calculate the $\%Reversed$ metric value for correct and wrong predictions. For instance, consider an ML model trained on a dataset containing 1000 samples with a reported accuracy of $90\%$. We then determine, out of the $900$ correct predictions, how many instances experience a prediction flip. Similarly, we assess how many samples undergo prediction reversal among the $100$ wrong predictions. As a result, the combined sum of $\%Reversed$ metrics for correct and wrong predictions may not equate to $100\%$. 

\textbf{\textit{\%Prob\_diff}}: $\%Prob\_diff$ denotes the difference between the probability of the original and the simulated instances. Ideally, the probability difference between them should be as high as possible. A higher $\%Prob\_diff$ value demonstrates PyExplainer's effectiveness in producing reliable explanations for ML models. The $\%Prob\_diff$ is defined as:

\begin{equation}
\small
\label{prob diff equn}
\%Prob\_diff = \forall x \in X, |P(M(x)) - P(M(x'))|
\end{equation}

In equation \ref{prob diff equn}, $P(M(x))$ denotes the prediction probability of the original instance, $P(M(x'))$ denotes the prediction probability of the simulated instance, and $M$ denotes the ML model. For the positive class, we calculate \textit{\%Prob\_diff} by subtracting the simulated instance's prediction probability from the original instance's prediction probability. Therefore, in the ideal case, the \textit{\%Prob\_diff} value must never be negative for each test instance of the positive class. Similarly, for the negative class, we calculate \textit{\%Prob\_diff} by subtracting the original instance's prediction probability from the simulated instance's prediction probability. Again, the \textit{\%Prob\_diff} value must never be negative for each test instance of the negative class.

In addition to $\%Reversed$ and $\%Prob\_diff$ metrics, we design the following four granular-level evaluation metrics:

\textbf{\textit{Positive Class Probability Decrease (PCPD)}}: If we change the feature values of an instance of the positive class in the \textit{green} zone, the risk score must decrease. For instance, Figure \ref{motivation_b} shows that when the actual values of $nCommit = 0.6$ and $LOC = 11$, the ML model predicts the given instance as defective commit with a risk score of 77\%. Now, after changing the feature values to $nCommit = 0.2$ and $LOC = 4$ in the \textit{green} zone, as demonstrated in Fig \ref{interpret_b}, the risk score decreases to 68\%. However, Figure \ref{anomaly_b} illustrates that this is not always the case. For example, Figure \ref{motivation_a} shows that when the value of $LOC=11$, the ML model predicts the given instance as defective commits with a risk score of 77\%. Despite changing the feature value of \textit{LOC} from $11.0$ to $21.0$, it remains the same instead of decreasing, as shown in Figure \ref{anomaly_b}. To detect this anomalous behaviour, we define \textit{PCPD} as follows:

\begin{equation}
\small
\label{PCPD}
PCPD = \frac{\sum\limits_{{x,x'} \epsilon X_p} 1 : \mathit{if} R(M(x)) > R(M(x'))}{|X_p|}
\end{equation}


\textbf{\textit{Positive Class Probability Increase (PCPI)}}: If we change the feature values of an instance of the positive class in the \textit{red} zone, the risk score must increase. For instance, Figure \ref{motivation_b} shows that when the actual values of $nCommit = 0.6$ and $LOC = 11$, the ML model predicts the given instance as defective commit with a risk score of 77\%. Now, after changing the feature value to $LOC = 15$ in the \textit{red} zone, as shown in Figure \ref{interpret_a}, the risk score increases to 79\%. However, Figure \ref{anomaly_a} shows this is not always true. For example, considering Figures \ref{motivation_a} and \ref{anomaly_a}, after changing the feature values of \textit{LOC} from $11.0$ to $1.0$, the risk score decreased from $77\%$ to $68\%$ instead of increasing. To detect this anomalous behaviour, we define \textit{PCPI} as follows:

\begin{equation}
\small
\label{PCPI}
PCPI = \frac{\sum\limits_{{x,x'} \epsilon X_p} 1 : \mathit{if} R(M(x')) > R(M(x))}{|X_p|}
\end{equation}

\textbf{\textit{Negative Class Probability Decrease (NCPD)}}: Keeping similarity with the $PCPD$, if we change the feature values of an instance of the negative class in the \textit{green} zone, the risk score must decrease. For instance, Figure \ref{interpret_c} shows that when the actual value of $nd$ is 1, the ML model predicts the given instance as a clean commit with a risk score of 13\%. Now, after changing the feature values to $nd = 1.9$ in the \textit{green} zone, the risk score decreases to 11\%, as shown in Figure \ref{interpret_d}. However, Figure \ref{anomaly_d} shows this is not always true. For example, considering Figures \ref{anomaly_c} and \ref{anomaly_d}, after changing the feature values of $OWN\_LINE = 0$ and $CountClassCoupled = 0$, the risk score increases to $31\%$ from $20\%$ instead of decreasing. To detect this anomalous behaviour, we define \textit{NCPD} as follows:

\begin{equation}
\small
\label{NCPD}
NCPD = \frac{\sum\limits_{{x,x'} \epsilon X_c} 1 : \mathit{if} R(M(x)) > R(M(x'))}{|X_c|}
\end{equation}


\textbf{\textit{Negative Class Probability Increase (NCPI)}}: Keeping similarity with the $PCPI$, if we change the feature values of an instance of the negative class in the \textit{red} zone, the risk score must increase. For instance, Figure \ref{interpret_c} shows that when the actual value of $nd$ is 1, the ML model predicts the given instance as a clean commit with a risk score of 13\%. Now, after changing the feature value to $nd = 0.1$ in the \textit{red} zone, the risk score increases to 17\%, as shown in Figure \ref{interpret_e}. However, Figure \ref{anomaly_e} shows that this is not always true. For example, considering Figures \ref{anomaly_c} and \ref{anomaly_e}, after changing the feature values of $OWN\_LINE$ to 1.9 and $CountDeclClassVariable$ to 6.0, the risk score decreases to $19\%$ from $20\%$ instead of increasing. To detect this anomalous behaviour, we define \textit{NCPI} as follows:

\begin{equation}
\small
\label{NCPI}
NCPI = \frac{\sum\limits_{{x,x'} \epsilon X_c} 1 : \mathit{if} R(M(x')) > R(M(x))}{|X_c|}
\end{equation}

In equations \ref{PCPD} and \ref{PCPI}, $X_p$ is the subset of the dataset $X$ containing only the positive class instances, in equations \ref{NCPD} and \ref{NCPI}, $X_c$ is the subset of the dataset $X$ containing only the negative class instances, $R(M(x))$ denotes the risk score of the original instance, $R(M(x'))$ denotes the risk score of the simulated instance.

We evaluate the reliability and consistency of the explanations generated by PyExplainer and LIME against all four granular-level evaluation metrics for each test instance. For example, for each instance, we increment the value of \textit{PCPD} by one when PyExplainer or LIME satisfies that specific metric. We apply the same rule to all other granular-level evaluation metrics. Ideally, we anticipate an increase in the value of each metric for every test instance. For instance, a \textit{PCPD} value approaching $100\%$ indicates the effectiveness of XAI techniques (e.g., PyExplainer, LIME) in producing reliable explanations for ML models and vice-versa. Finally, the increasing controlling parameter $\alpha$ indicates that a higher STD value will be added or subtracted from the original feature values during the generation of a simulated instance. Therefore, as the value of $\alpha$ increases, the values of each evaluation metric also increase, demonstrating the effectiveness of XAI techniques in generating reliable explanations.

Lundberg et al. \cite{lundberg2017unified} stated that instance explanation generation must remain the same upon regeneration for the same instance. Therefore, we evaluate the consistency of XAI techniques in generating explanations by running them on the same instance multiple times using \textit{EvaluateXAI}, as well as considering the various evaluation metrics described above. Any discrepancies among the multiple runs indicate the inconsistency of XAI techniques in explanation generation.

\section{Experimental Results}
\label{exp}
In our study, we utilize scikit-learn\footnote{https://scikit-learn.org/stable/}, LIME\footnote{https://github.com/marcotcr/lime}, and PyExplainer libraries\footnote{https://pyexplainer.readthedocs.io/en/latest/index.html} to conduct our experiments. We consider five different datasets: three for JIT defect prediction, one for clone detection, and one for code review comment classification and seven ML models previously used in other studies, resulting in a total of $5 \times 7 = 35$ experimental combinations. While Pornprasit et al. \cite{pornprasit2021pyexplainer} only considered the proportion of correctly predicted instances in the test dataset for \textit{What-if} analysis, our study analyzes both the correctly and wrongly predicted instances to enable extensive experimental analysis. For example, the RF model trained on the CLCDSA dataset correctly predicted 2821 samples and wrongly predicted 280 samples out of 3101 in total. Our study considers all 3101 instances. Thus, we first analyze the performance of our selected ML models, and then, we present the results of our study by addressing the following two research questions:

\textbf{RQ1: Do rule-based XAI techniques generate reliable explanations for machine learning models in software analytics tasks?}
    
    

\textbf{RQ2: Do rule-based XAI techniques maintain consistency in generating explanations for machine learning models in software analytics tasks?}


\subsection{Analyze Model Performance}
Section \ref{tune_param} outlines the process of fine-tuning the hyperparameters for our ML models during training. The table \ref{hyper_val} displays the hyperparameters and their optimized values for the ML models trained on the cross-project mobile apps dataset. For hyperparameter settings pertaining to other datasets, please refer to our replication package. Following the identification of these optimized hyperparameters, we proceed to train our ML models.

\begin{table}[htbp]
\centering
\caption{Optimized hyperparameters for different ML models trained on the cross-project mobile apps dataset after tuning.}
\resizebox{\linewidth}{!}{
\begin{tabular}{l|l}
\hline
\multicolumn{1}{c|}{ML Models} & \multicolumn{1}{c}{Hyperparameters}                                                                                                      \\ \hline
LR                            & C: 2.78, dual: True, max\_iter: 130, penalty: `l2', solver: `liblinear'                                                       \\ \hline
DT                            & min\_samples\_split: 10, min\_samples\_leaf: 15, max\_depth: 15,   criterion: `gini'                                              \\ \hline
RF                            & bootstrap: False, max\_depth: None, max\_features: `sqrt',   min\_samples\_leaf: 2, min\_samples\_split: 2, n\_estimators: 50 \\ \hline
MLP                           & solver: `adam', learning\_rate: `constant', hidden\_layer\_sizes:   (10, 30, 10), alpha: 0.0001, activation: `tanh'             \\ \hline
ADA                           & n\_estimators: 20, learning\_rate: 1.04, algorithm: `SAMME.R'                                                                       \\ \hline
BAG                           & max\_features: 5, max\_samples: 100, n\_estimators: 1200                                                                            \\ \hline
GBC                           & learning\_rate: 1, max\_depth: 7, min\_samples\_leaf: 0.1,   min\_samples\_split: 0.1, n\_estimators: 200                       \\ \hline
\end{tabular}}
\label{hyper_val}
\end{table}

Table \ref{dataset} represents the accuracy, F1-score, and AUC values of ML models trained on the different datasets. This table shows that the LR model exhibits the lowest AUC value ($0.73$), whereas the BAG model shows the highest AUC value ($0.85$) when trained on the cross-project mobile apps dataset. The ML models trained on the CLCDSA dataset show the highest AUC values, ranging from $0.82$ to $0.96$. Additionally, Table \ref{dataset} demonstrates comparatively low AUC values for different ML models trained on the code review dataset. By examining Table \ref{dataset}, we can conclude that all the ML models demonstrate acceptable accuracy, F1-score, and AUC values, indicating their high accuracy and non-overfitting nature. Furthermore, all test-AUCs (except two cases and code review datasets) surpass 0.75, a threshold advocated by prior research as the minimum value necessary for ensuring the reliability of explanations \cite{roy2022don, lyu2021towards}. Figure \ref{AUC_ROC} depicts the AUC–ROC curves for all the ML models trained on the selected datasets, as presented in \ref{AUC_ROC_Appendix}.

\begin{table}[htbp]
\footnotesize
\centering
\caption{Performance of the ML models on different datasets considering different evaluation metrics such as Accuracy (Acc), F1-score (F-1), and AUC values.}
\resizebox{\linewidth}{!}{
\begin{tabular}{l|lllllllllllllll}
\hline
\multirow{3}{*}{\textbf{\begin{tabular}[c]{@{}l@{}}ML\\ Model\end{tabular}}} & \multicolumn{15}{c}{\textbf{Dataset}}                                                                                                                                                                                                                                                                                                                                                                       \\ \cline{2-16} 
                                                                             & \multicolumn{3}{c|}{\textbf{Cross Project}}                                       & \multicolumn{3}{c|}{\textbf{Java Project}}                                        & \multicolumn{3}{c|}{\textbf{Postgres}}                                            & \multicolumn{3}{c|}{\textbf{CLCDSA}}                                              & \multicolumn{3}{c}{\textbf{Code Review}}                    \\ \cline{2-16} 
                                                                             & \multicolumn{1}{l|}{Acc}  & \multicolumn{1}{l|}{F-1}  & \multicolumn{1}{l|}{AUC}  & \multicolumn{1}{l|}{Acc}  & \multicolumn{1}{l|}{F-1}  & \multicolumn{1}{l|}{AUC}  & \multicolumn{1}{l|}{Acc}  & \multicolumn{1}{l|}{F-1}  & \multicolumn{1}{l|}{AUC}  & \multicolumn{1}{l|}{Acc}  & \multicolumn{1}{l|}{F-1}  & \multicolumn{1}{l|}{AUC}  & \multicolumn{1}{l|}{Acc}  & \multicolumn{1}{l|}{F-1}  & AUC  \\ \hline
LR                                                                           & \multicolumn{1}{l|}{0.68} & \multicolumn{1}{l|}{0.49} & \multicolumn{1}{l|}{0.73} & \multicolumn{1}{l|}{0.75} & \multicolumn{1}{l|}{0.39} & \multicolumn{1}{l|}{0.76} & \multicolumn{1}{l|}{0.75} & \multicolumn{1}{l|}{0.57} & \multicolumn{1}{l|}{0.76} & \multicolumn{1}{l|}{0.76} & \multicolumn{1}{l|}{0.81} & \multicolumn{1}{l|}{0.82} & \multicolumn{1}{l|}{0.62} & \multicolumn{1}{l|}{0.69} & 0.63 \\ \hline
DT                                                                           & \multicolumn{1}{l|}{0.79} & \multicolumn{1}{l|}{0.60} & \multicolumn{1}{l|}{0.83} & \multicolumn{1}{l|}{0.80} & \multicolumn{1}{l|}{0.49} & \multicolumn{1}{l|}{0.80} & \multicolumn{1}{l|}{0.78} & \multicolumn{1}{l|}{0.59} & \multicolumn{1}{l|}{0.79} & \multicolumn{1}{l|}{0.86} & \multicolumn{1}{l|}{0.88} & \multicolumn{1}{l|}{0.93} & \multicolumn{1}{l|}{0.67} & \multicolumn{1}{l|}{0.73} & 0.68 \\ \hline
RF                                                                           & \multicolumn{1}{l|}{0.81} & \multicolumn{1}{l|}{0.58} & \multicolumn{1}{l|}{0.84} & \multicolumn{1}{l|}{0.88} & \multicolumn{1}{l|}{0.55} & \multicolumn{1}{l|}{0.85} & \multicolumn{1}{l|}{0.80} & \multicolumn{1}{l|}{0.59} & \multicolumn{1}{l|}{0.82} & \multicolumn{1}{l|}{0.91} & \multicolumn{1}{l|}{0.93} & \multicolumn{1}{l|}{0.96} & \multicolumn{1}{l|}{0.71} & \multicolumn{1}{l|}{0.76} & 0.72 \\ \hline
MLP                                                                          & \multicolumn{1}{l|}{0.75} & \multicolumn{1}{l|}{0.57} & \multicolumn{1}{l|}{0.83} & \multicolumn{1}{l|}{0.80} & \multicolumn{1}{l|}{0.48} & \multicolumn{1}{l|}{0.81} & \multicolumn{1}{l|}{0.71} & \multicolumn{1}{l|}{0.51} & \multicolumn{1}{l|}{0.73} & \multicolumn{1}{l|}{0.83} & \multicolumn{1}{l|}{0.86} & \multicolumn{1}{l|}{0.92} & \multicolumn{1}{l|}{0.60} & \multicolumn{1}{l|}{0.68} & 0.63 \\ \hline
ADA                                                                          & \multicolumn{1}{l|}{0.73} & \multicolumn{1}{l|}{0.56} & \multicolumn{1}{l|}{0.81} & \multicolumn{1}{l|}{0.79} & \multicolumn{1}{l|}{0.47} & \multicolumn{1}{l|}{0.81} & \multicolumn{1}{l|}{0.78} & \multicolumn{1}{l|}{0.58} & \multicolumn{1}{l|}{0.79} & \multicolumn{1}{l|}{0.81} & \multicolumn{1}{l|}{0.77} & \multicolumn{1}{l|}{0.85} & \multicolumn{1}{l|}{0.63} & \multicolumn{1}{l|}{0.65} & 0.67 \\ \hline
BAG                                                                          & \multicolumn{1}{l|}{0.77} & \multicolumn{1}{l|}{0.59} & \multicolumn{1}{l|}{0.85} & \multicolumn{1}{l|}{0.78} & \multicolumn{1}{l|}{0.50} & \multicolumn{1}{l|}{0.83} & \multicolumn{1}{l|}{0.78} & \multicolumn{1}{l|}{0.61} & \multicolumn{1}{l|}{0.82} & \multicolumn{1}{l|}{0.81} & \multicolumn{1}{l|}{0.84} & \multicolumn{1}{l|}{0.89} & \multicolumn{1}{l|}{0.65} & \multicolumn{1}{l|}{0.73} & 0.71 \\ \hline
GBC                                                                          & \multicolumn{1}{l|}{0.81} & \multicolumn{1}{l|}{0.59} & \multicolumn{1}{l|}{0.84} & \multicolumn{1}{l|}{0.81} & \multicolumn{1}{l|}{0.48} & \multicolumn{1}{l|}{0.80} & \multicolumn{1}{l|}{0.78} & \multicolumn{1}{l|}{0.57} & \multicolumn{1}{l|}{0.80} & \multicolumn{1}{l|}{0.85} & \multicolumn{1}{l|}{0.89} & \multicolumn{1}{l|}{0.94} & \multicolumn{1}{l|}{0.63} & \multicolumn{1}{l|}{0.69} & 0.65 \\ \hline
\end{tabular}}
\label{dataset}
\vspace{-1.5em}
\end{table}

\subsection{Present Experimental Results}
\label{present_exp_resutls}
We present the results of our experimental study in relation to our two research questions. The detailed analysis and outcomes are elaborated in the following sections by addressing each research question individually.

\textbf{RQ1: Do rule-based XAI techniques generate reliable explanations for machine learning models in software analytics tasks?} 

\textbf{Motivation}: The quality and reliability of the explanations generated by the XAI techniques are crucial. In a scenario where a ML model predicts a defective commit with a 77\% risk score as shown in Figure \ref{pyexp explanation}, PyExplainer's visual interpretation suggests that changing feature values in the \textit{green} zone should decrease the risk score and in the \textit{red} zone, it should increase it. However, Figure \ref{anomaly_b} contradicts this. For example, when the $LOC$ value is 11, the risk score remains unchanged despite changing it to 21 in the \textit{green} zone.

Similarly, in Figure \ref{interpret_a}, changing feature values in the \textit{red} zone increases the risk score from 77\% to 79\% (Figure \ref{motivation_b}). However, Figure \ref{anomaly_a} reveals an exception to this. Changing $LOC$ from 11 to 1 reduces the risk score from 77\% to 68\%, which is an anomaly in PyExplainer's explanations. We found similar anomalies in the explanations generated by LIME. These anomalies call for further investigations to evaluate the reliability of XAI's generated explanations.

\textbf{Approach}:
We address this research question using \textit{EvaluateXAI}, six evaluation metrics, and seven ML models trained on five datasets. For each testing sample of each dataset, we generate simulated instances. Then, we calculate the values of the six evaluation metrics. Finally, we report the experimental results using various tables, figures, and boxplots.

\begin{figure}[htbp]
\centering     
\subfigure[Cross-project (correct predictions)]{\label{mob_rev_pyexp_true}\includegraphics[width=4.25cm, height=3cm]{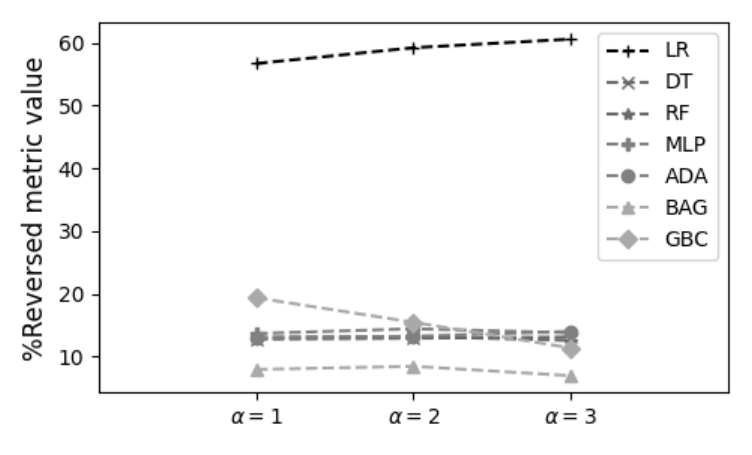}} 
\subfigure[Cross-project (wrong predictions)]{\label{mob_rev_pyexp_false}\includegraphics[width=4.25cm, height=3cm]{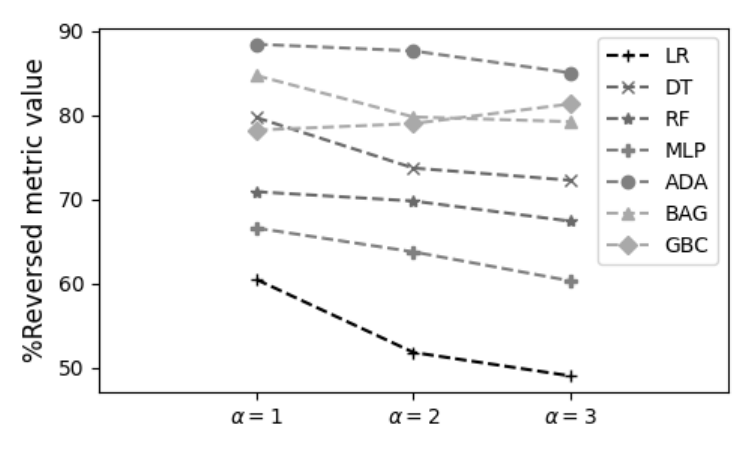}}
\subfigure[Java (correct predictions)]{\label{java_rev_pyexp_true}\includegraphics[width=4.25cm, height=3cm]{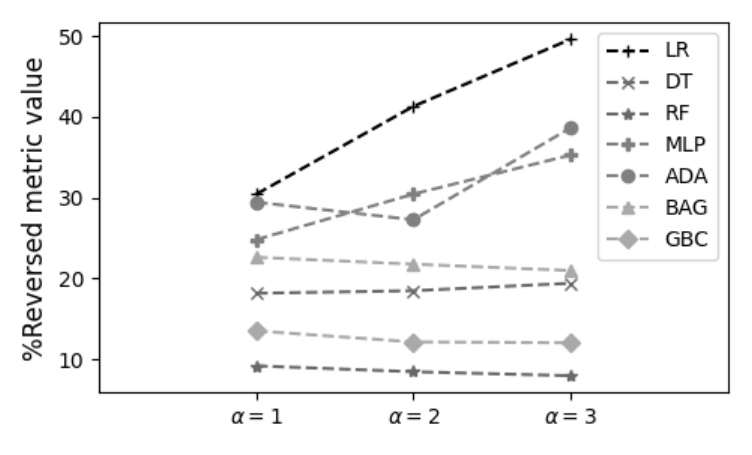}} 
\subfigure[Java (wrong predictions)]{\label{java_rev_pyexp_false}\includegraphics[width=4.25cm, height=3cm]{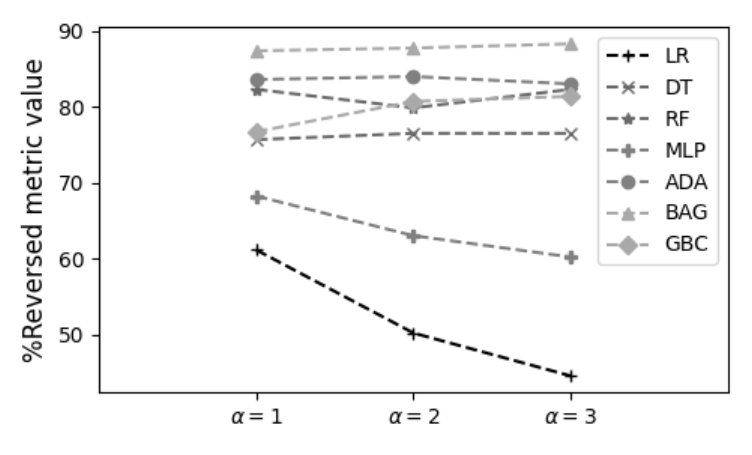}}
\subfigure[Postgres (correct predictions)]{\label{postgres_rev_pyexp_true}\includegraphics[width=4.25cm, height=3cm]{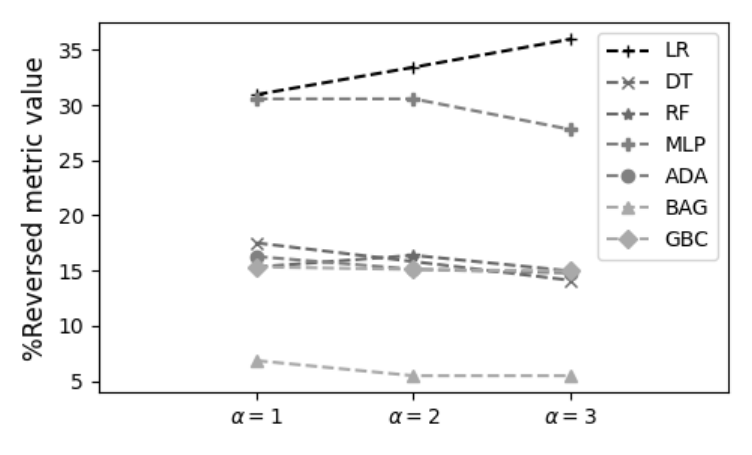}}
\subfigure[Postgres (wrong predictions)]{\label{postgres_rev_pyexp_false}\includegraphics[width=4.25cm, height=3cm]{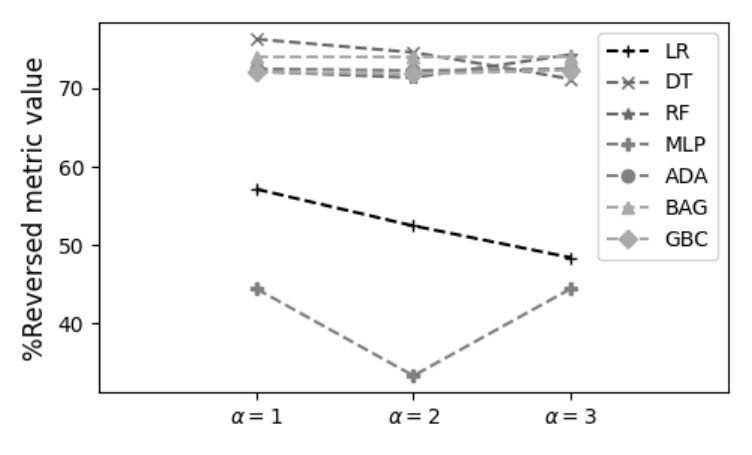}}
\subfigure[CLCDSA (correct predictions)]{\label{clcdsa_rev_pyexp_true}\includegraphics[width=4.25cm, height=3cm]{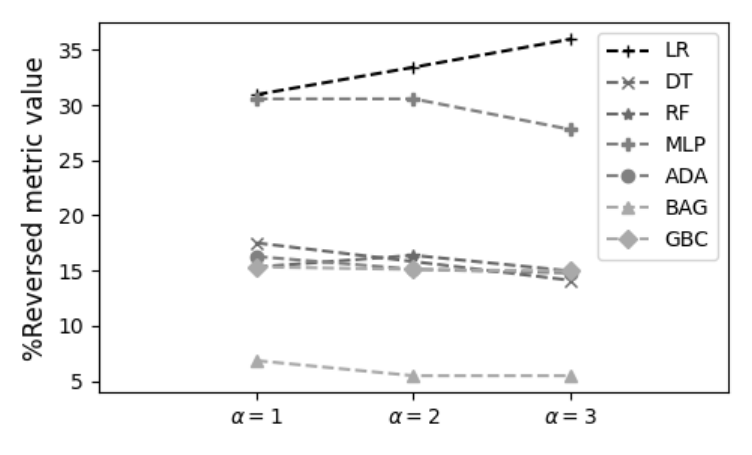}}
\subfigure[CLCDSA (wrong predictions)]{\label{clcdsa_rev_pyexp_false}\includegraphics[width=4.25cm, height=3cm]{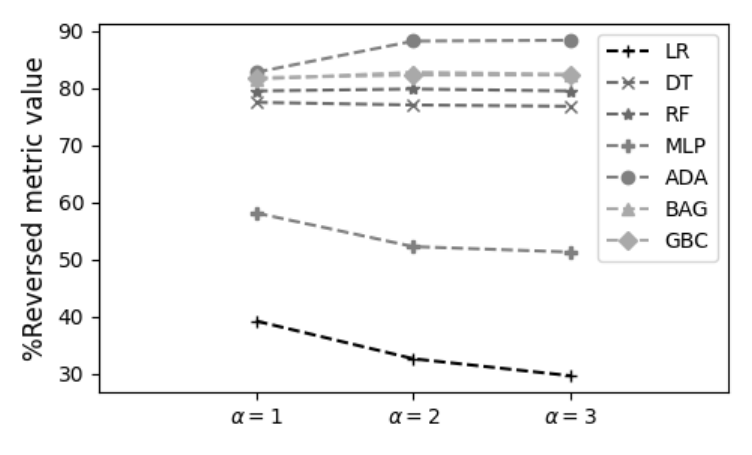}}
\subfigure[Code review (correct predictions)]{\label{code_rev_pyexp_true}\includegraphics[width=4.25cm, height=3cm]{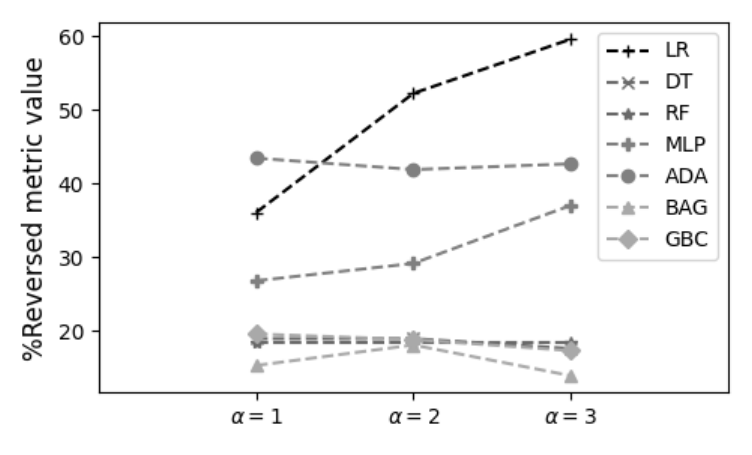}}
\subfigure[Code review (wrong predictions)]{\label{code_rev_pyexp_false}\includegraphics[width=4.25cm, height=3cm]{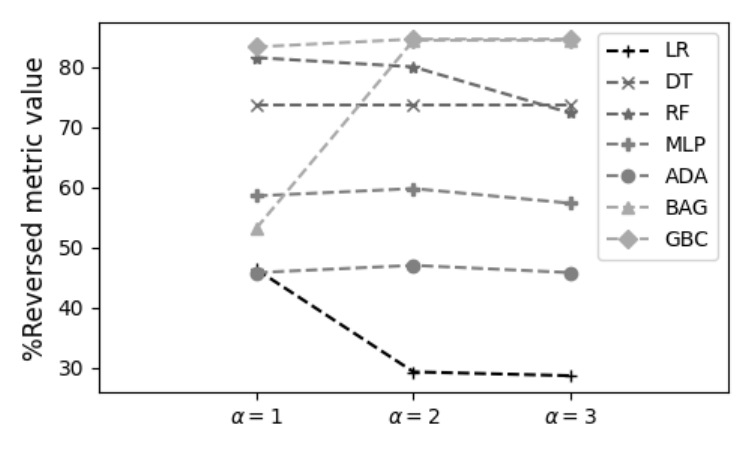}}
\caption{The characteristics of $\%Reversed$ metric values with different $\alpha$ when we apply PyExplainer to the simulated instances of the selected datasets. For example, Figure \ref{mob_rev_pyexp_true} represents the variation in $\%Reversed$ metric values for different $\alpha$ when ML models make correct predictions.}
\label{rev_pyexp_all_dataset}

\end{figure}

\begin{figure}[htbp]
\centering     
\subfigure[Cross-project (correct predictions)]{\label{mob_rev_lime_true}\includegraphics[width=4.25cm, height=3cm]{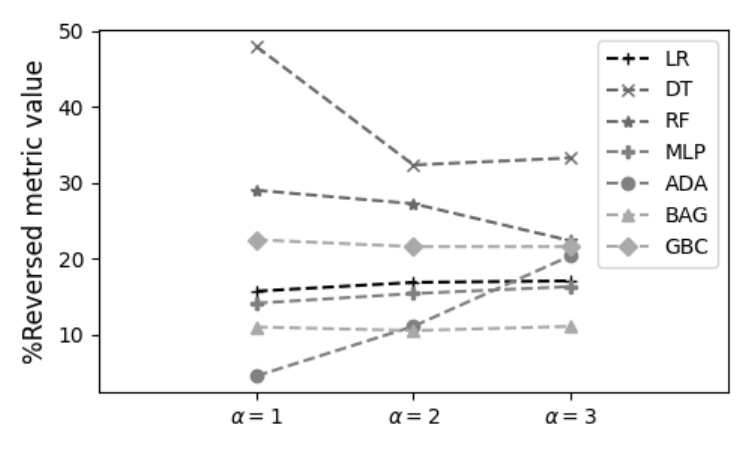}} 
\subfigure[Cross-project (wrong predictions)]{\label{mob_rev_lime_false}\includegraphics[width=4.25cm, height=3cm]{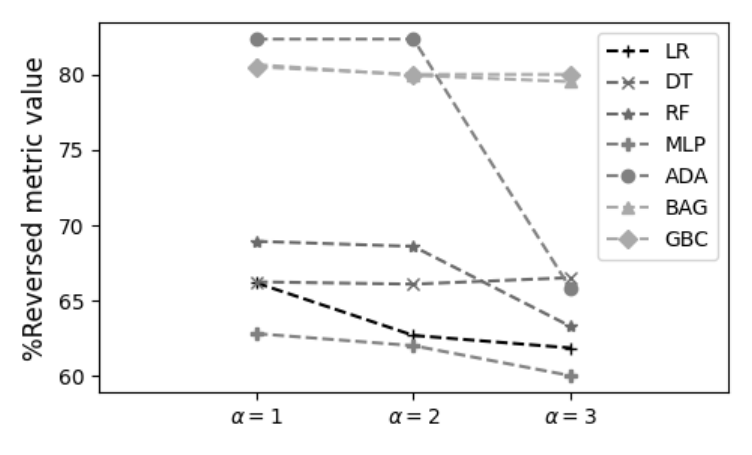}}
\subfigure[Java (correct predictions)]{\label{java_rev_lime_true}\includegraphics[width=4.25cm, height=3cm]{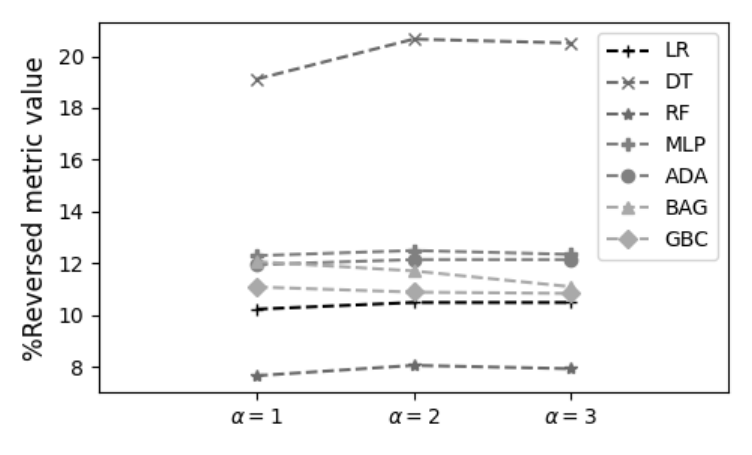}}
\subfigure[Java (wrong predictions)]{\label{java_rev_lime_false}\includegraphics[width=4.25cm, height=3cm]{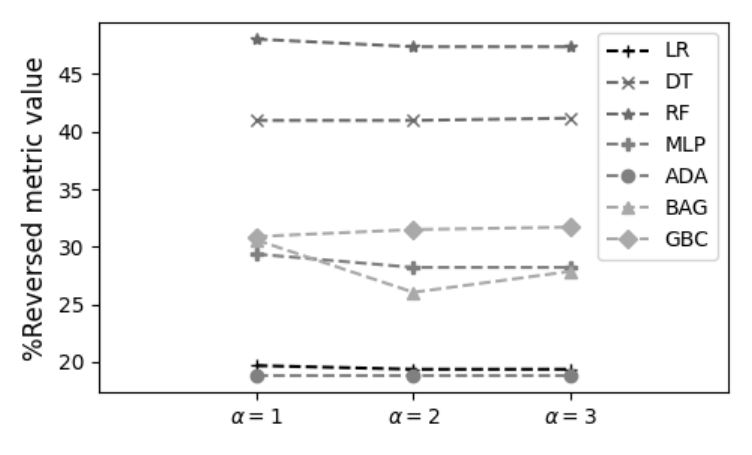}}
\subfigure[Postgres (correct predictions)]{\label{postgres_rev_lime_true}\includegraphics[width=4.25cm, height=3cm]{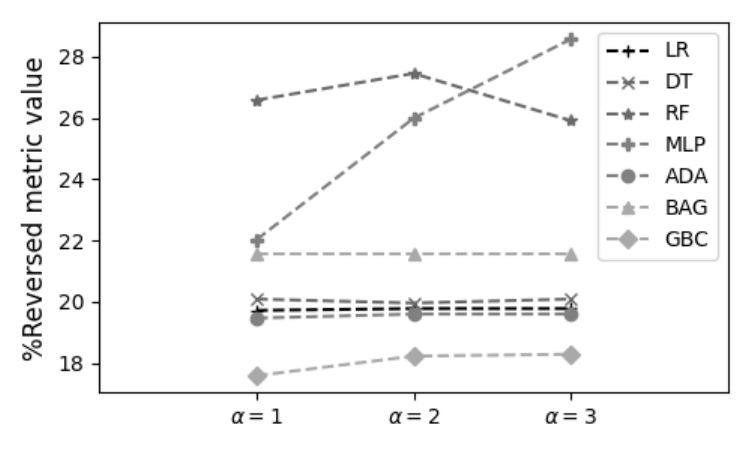}}
\subfigure[Postgres (wrong predictions)]{\label{postgres_rev_lime_false}\includegraphics[width=4.25cm, height=3cm]{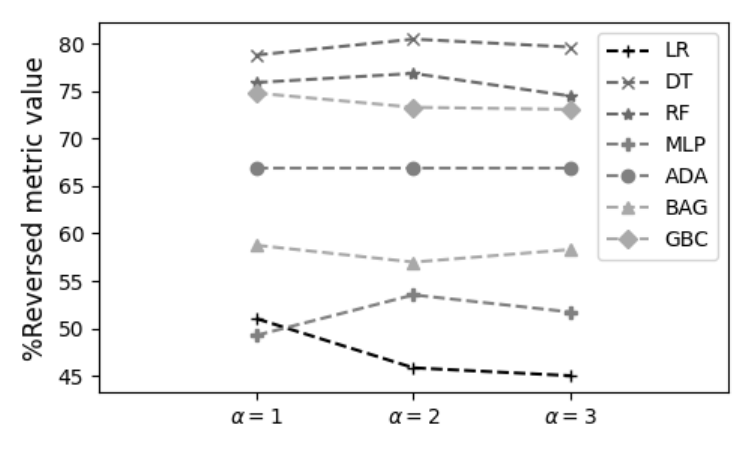}}
\subfigure[CLCDSA (correct predictions)]{\label{clcdsa_rev_lime_true}\includegraphics[width=4.25cm, height=3cm]{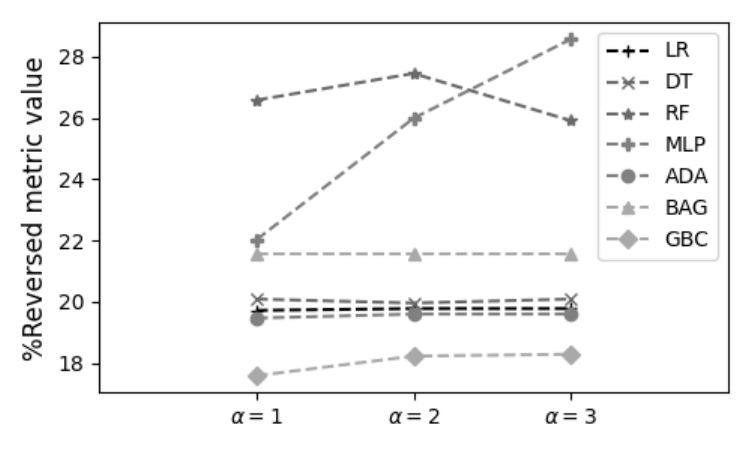}}
\subfigure[CLCDSA (wrong predictions)]{\label{clcdsa_rev_lime_false}\includegraphics[width=4.25cm, height=3cm]{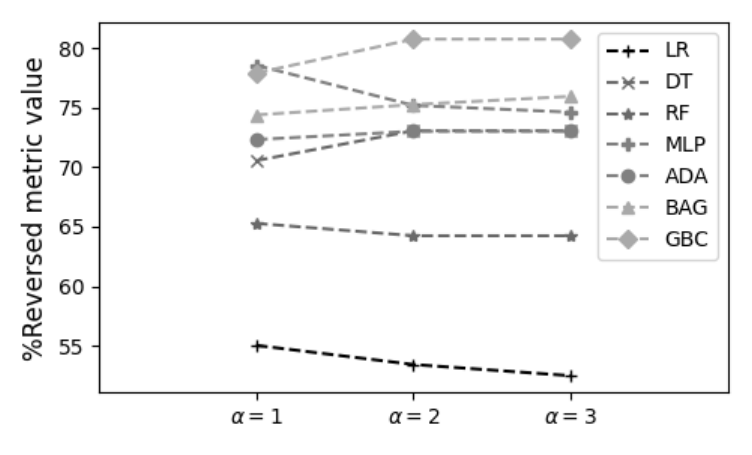}}
\subfigure[Code review (correct predictions)]{\label{code_rev_lime_true}\includegraphics[width=4.25cm, height=3cm]{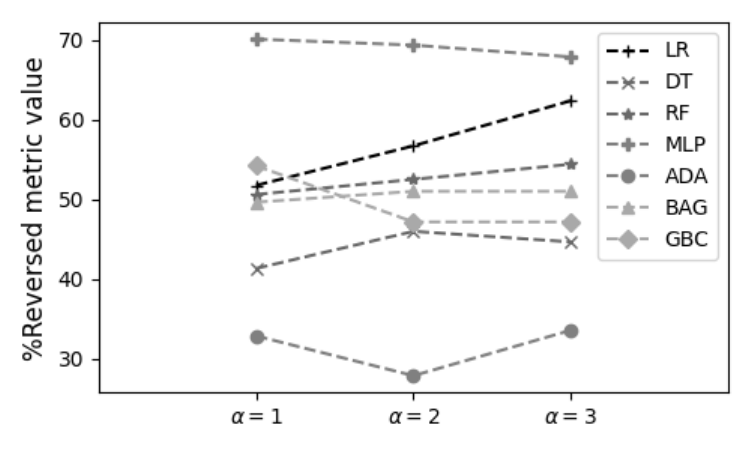}}
\subfigure[Code review (wrong predictions)]{\label{code_rev_lime_false}\includegraphics[width=4.25cm, height=3cm]{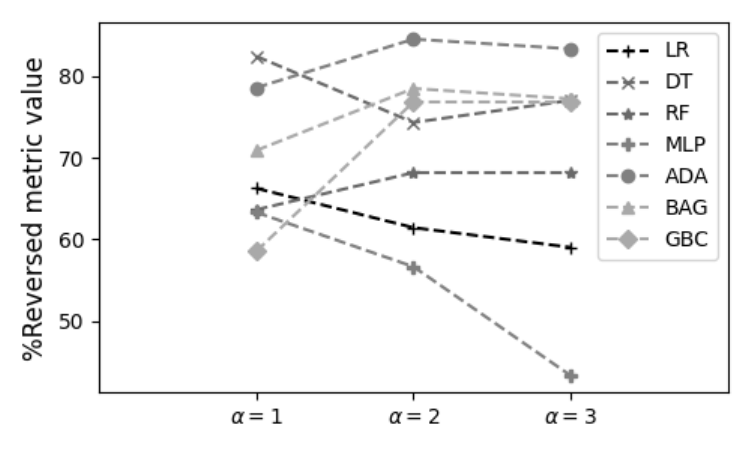}}
\caption{The characteristics of $\%Reversed$ metric values with different $\alpha$ when we apply LIME to the simulated instances of the selected datasets. For example, Figure \ref{mob_rev_pyexp_true} represents the variation in $\%Reversed$ metric values for different $\alpha$ when ML models make correct predictions.}
\label{rev_lime_all_dataset}
\end{figure}

\textbf{Results}:
Figures \ref{rev_pyexp_all_dataset} and \ref{rev_lime_all_dataset} display the values of the $\%Reversed$ metric when PyExplainer and LIME are applied to simulated instances of the cross-project mobile apps, Java project, Postgres, CLCDSA, and code review datasets, respectively. The experiments are conducted with varying $\alpha$ values.

We start by evaluating the reliability of the explanations generated by PyExplainer and LIME for ML models focusing on correct predictions. From Figure \ref{mob_rev_pyexp_true}, we observe the highest $\%Reversed$ metric values for different $\alpha$ values in the case of the LR model when PyExplainer is run on simulated instances of the cross-project mobile apps dataset. In the case of other ML models, we observe a relatively low $\%Reversed$ metric value (e.g., $<<20\%$). Similar to PyExplainer, LIME exhibits a very similar trend. For example, we observe the highest $\%Reversed$ metric values for different $\alpha$ values in the case of the DT model when LIME is run on simulated instances of the cross-project mobile apps dataset. In the case of other ML models, we observe a relatively low $\%Reversed$ metric value (e.g., $<<29\%$). 

Moving to wrong predictions, we observe improved $\%Reversed$ metric values for all ML models when PyExplainer and LIME are applied to simulated instances from the cross-project mobile apps dataset. For instance, with $\alpha=1$, we find that the $\%Reversed$ metric values vary between 60.51\% and 88.4\% for PyExplainer and between 62.82\% and 82.34\% for LIME. Similar results are observed with the other four datasets. Overall, when considering the $\%Reversed$ metric, PyExplainer and LIME generate relatively reliable explanations for the ML models when they make wrong predictions. However, their performance is less impressive when ML models make correct predictions, with only a few exceptions.

Another important aspect of PyExplainer and LIME is that, as we increase the $\alpha$ values during the generation of simulated instances, we anticipate an increase in the $\%Reversed$ metric value. However, the $\%Reversed$ metric values projected with the increase of $\alpha$ values, as shown in Figures \ref{rev_pyexp_all_dataset} and \ref{rev_lime_all_dataset}, contradict this expectation. For example, in the case of the GBC model trained on the cross-project mobile apps dataset, PyExplainer exhibits a downward trend in the $\%Reversed$ metric. Additionally, in the case of the RF model trained on the CLCDSA dataset, LIME exhibits a downward trend in the $\%Reversed$ metric. For some ML models, the $\%Reversed$ metric value increases from $\alpha=1$ to $\alpha=2$, and then decreases when $\alpha=3$. Overall, we observe similar results for all the ML models trained on the other datasets. Therefore, the reliability of the explanation generated by PyExplainer and LIME is in question.

We further investigate the variations in prediction probabilities between the original and simulated instances using the $\%Prob\_diff$ metric. We present the experimental results for LR and RF models as these two models were extensively studied for JIT defect predictions \cite{pornprasit2021pyexplainer, jiarpakdee2020empirical, roy2022don}. Additionally, we select two datasets, CLCDSA and code review, because, for these two datasets, our selected ML models show the highest and lowest accuracy and AUC values, respectively. 

\begin{figure}[htbp]
\centering     
\subfigure[LR + CLCDSA + PyExplainer]{\label{LR_fig_cort_CLCDSA}\includegraphics[width=4.25cm, height=3.5cm]{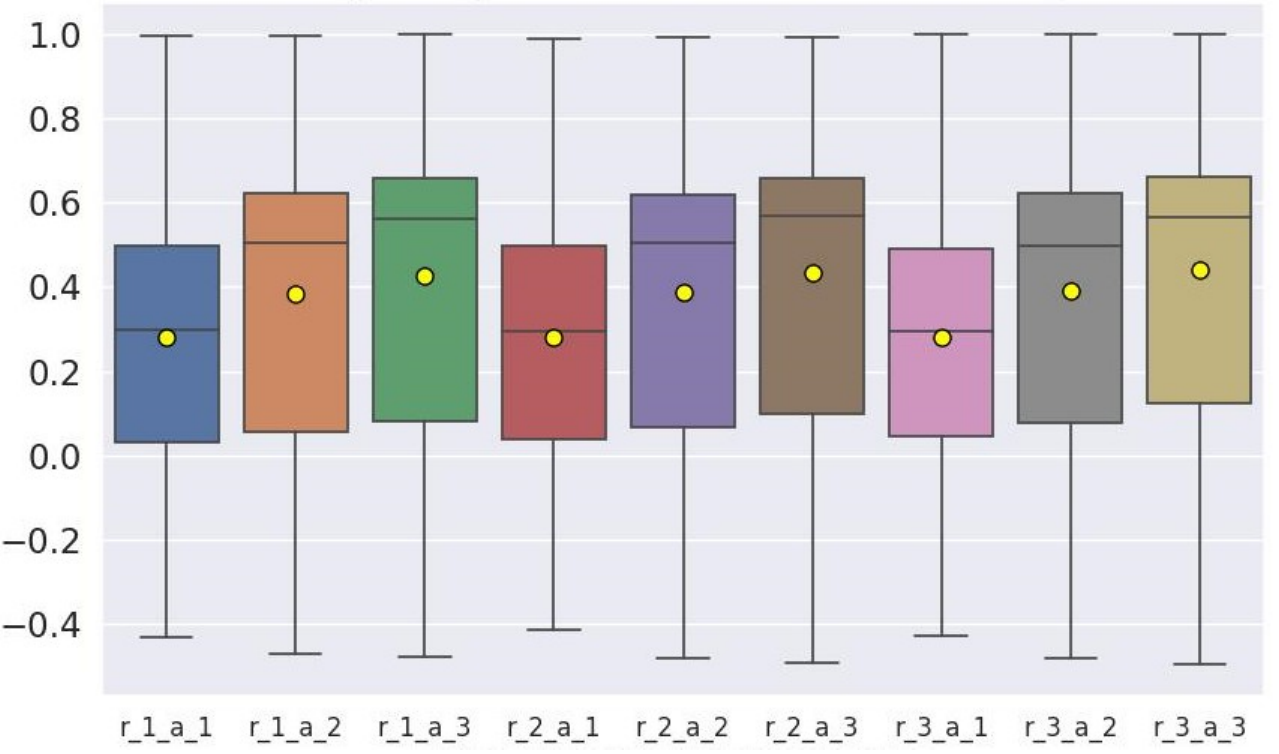}} 
\subfigure[LR + code review + PyExplainer]{\label{LR_fig_cort_CR}\includegraphics[width=4.25cm, height=3.5cm]{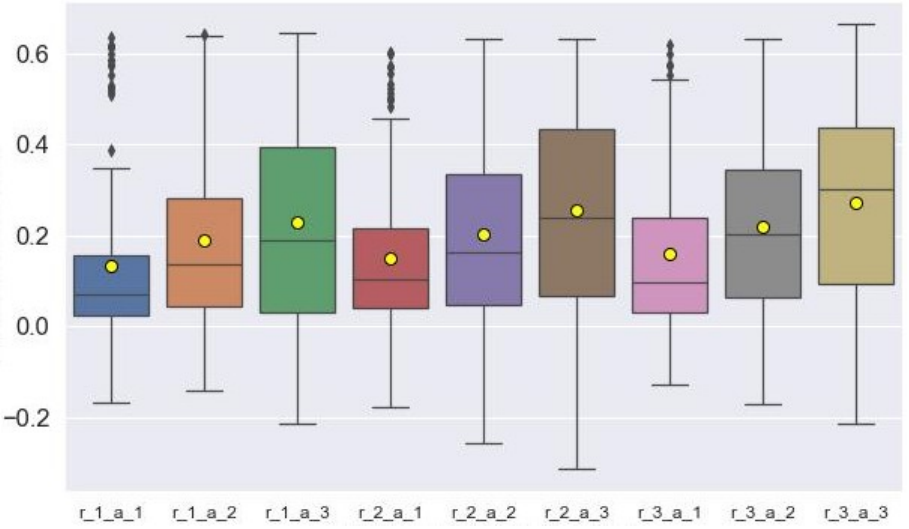}}
\subfigure[LR + CLCDSA + LIME]{\label{LR_LIME_CLCDSA_Correct}\includegraphics[width=4.25cm, height=3.5cm]{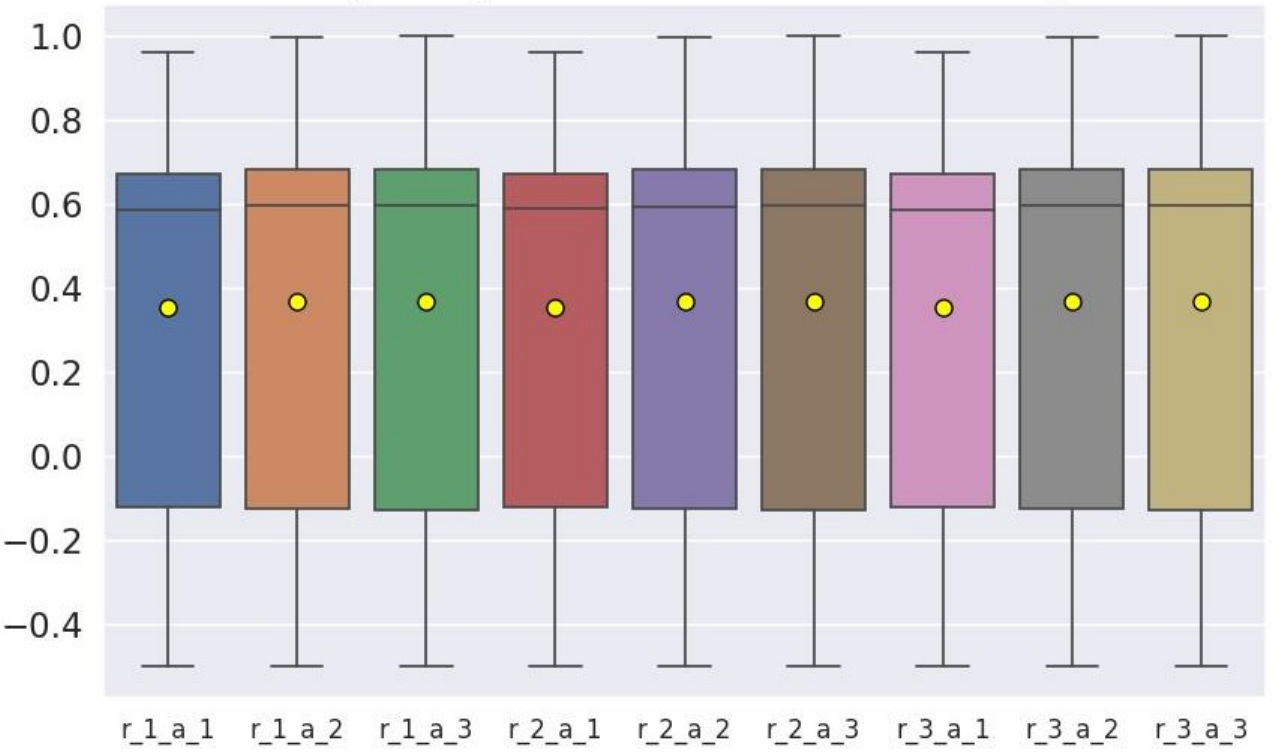}}
\vspace{-0.5em}
\subfigure[LR + code review + LIME]{\label{LR_LIME_Code_Correct}\includegraphics[width=4.25cm, height=3.5cm]{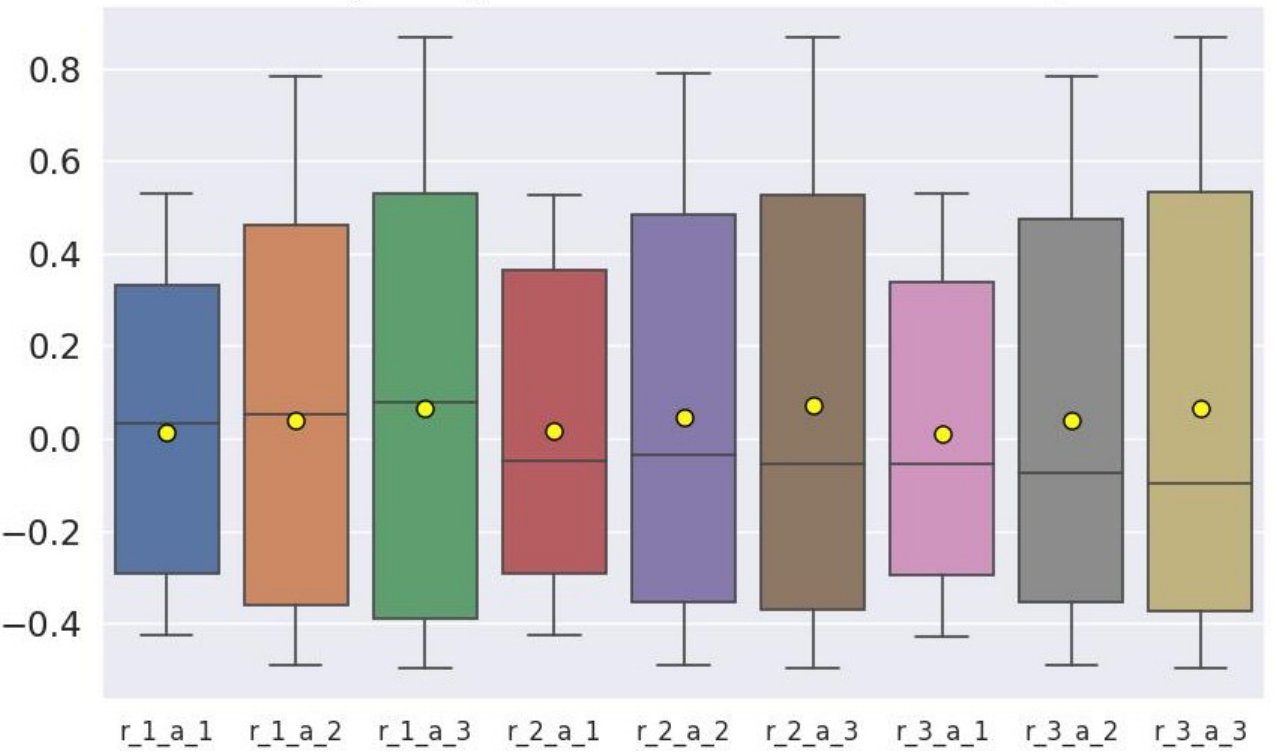}}
\subfigure[RF + CLCDSA + PyExplainer]{\label{RF_fig_cort_CLCDSA}\includegraphics[width=4.25cm, height=3.5cm]{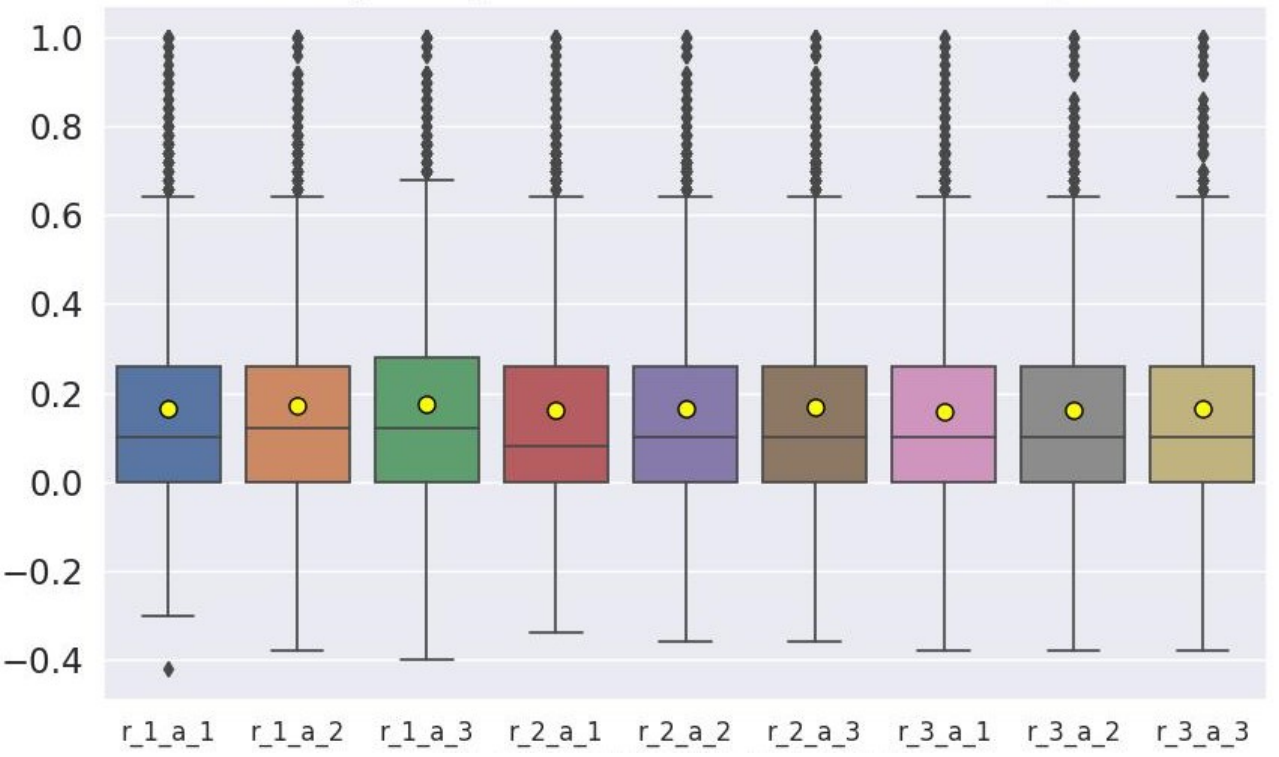}} 
\subfigure[RF + code review + PyExplainer]{\label{RF_fig_cort_CR}\includegraphics[width=4.25cm, height=3.5cm]{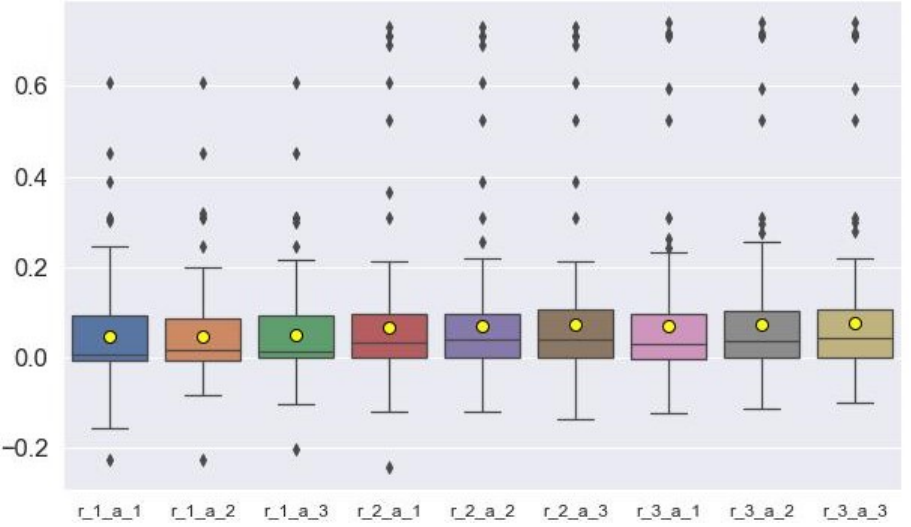}}
\subfigure[RF + CLCDSA + LIME]{\label{RF_LIME_CLCDSA_Correct}\includegraphics[width=4.25cm, height=3.5cm]{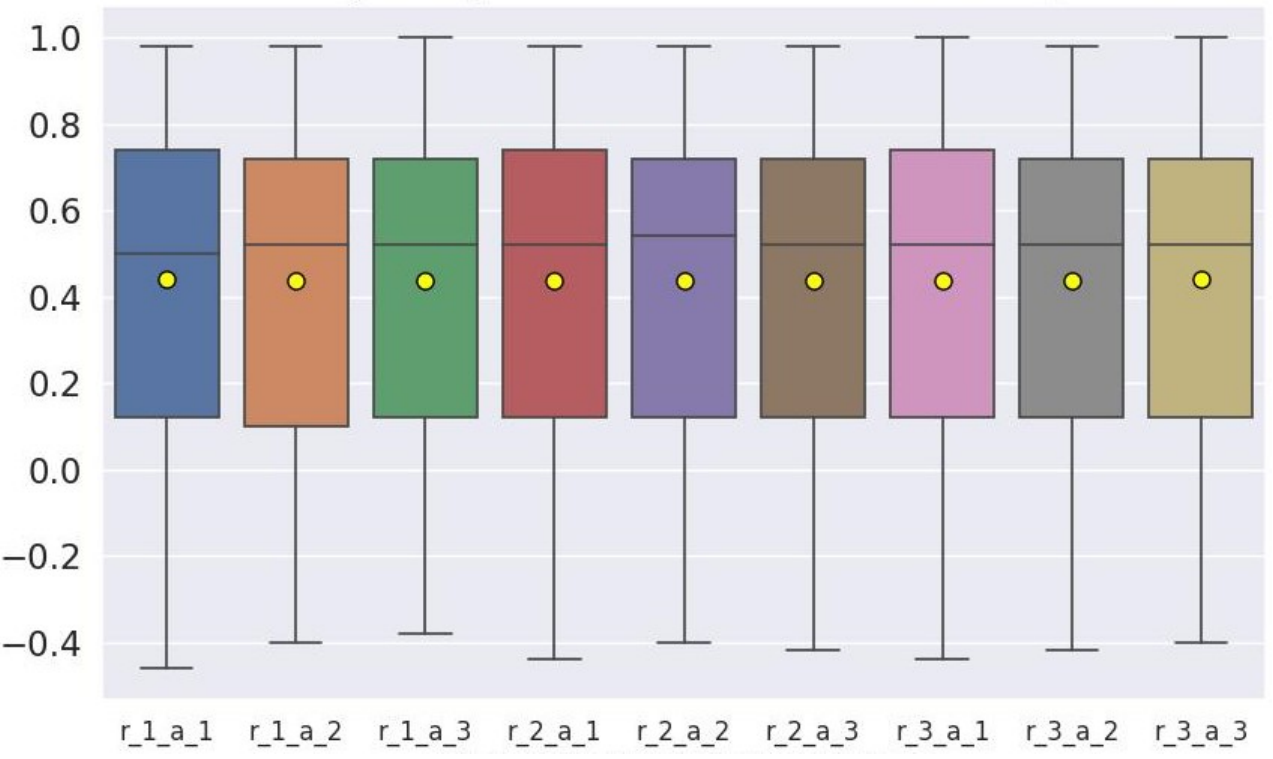}}
\vspace{-0.5em}
\subfigure[RF + code review + LIME]{\label{RF_LIME_Code_Correct}\includegraphics[width=4.25cm, height=3.5cm]{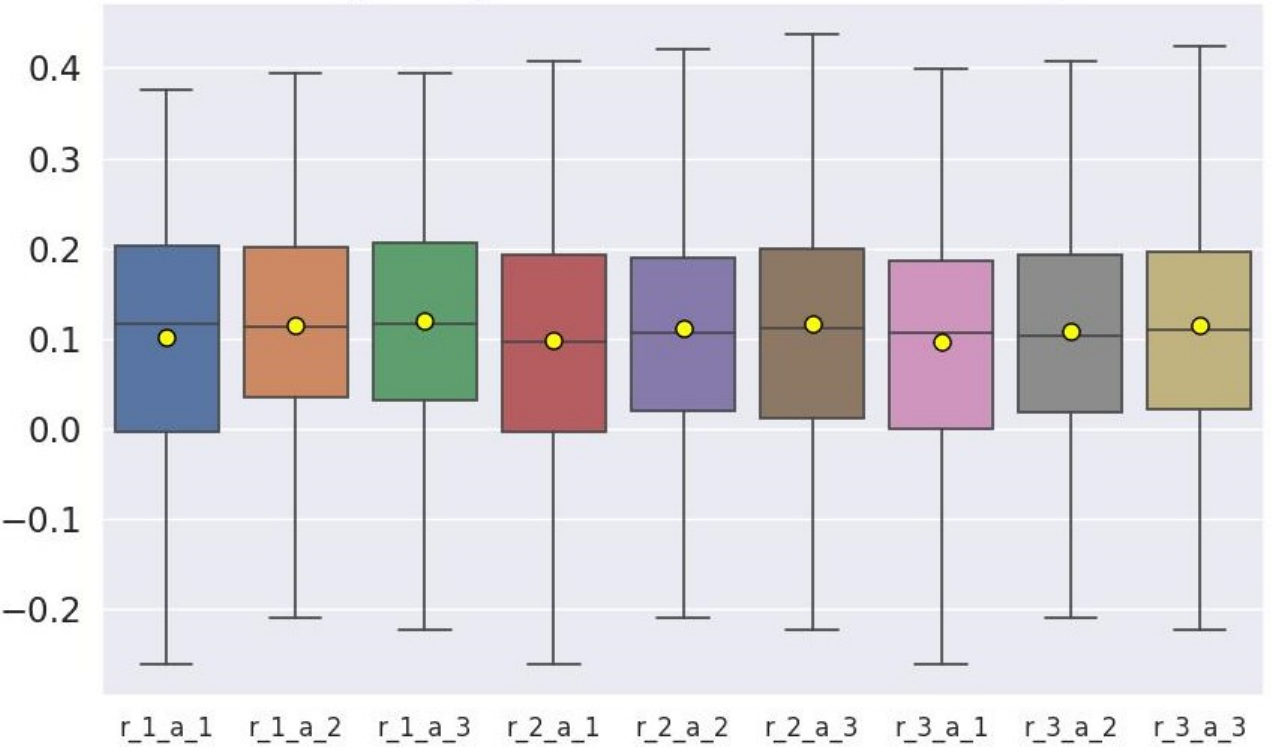}}
\caption{Distribution of the probability difference between the original and the simulated instances. Figures \ref{LR_fig_cort_CLCDSA} and \ref{LR_fig_cort_CR} show the probability difference distribution for the LR model trained on CLCDSA and code review datasets when we use PyExplainer for explanation generation. Figures \ref{LR_LIME_CLCDSA_Correct} and \ref{LR_LIME_Code_Correct} show the probability difference distribution for the LR model trained on the CLCDSA and code review datasets when we use LIME for explanation generation. Figures \ref{RF_fig_cort_CLCDSA} and \ref{RF_fig_cort_CR} show the probability difference distribution for the RF model trained on CLCDSA and code review datasets when we use PyExplainer for explanation generation. Figures \ref{RF_LIME_CLCDSA_Correct} and \ref{RF_LIME_Code_Correct} show the probability difference distribution for the RF model trained on the CLCDSA and code review datasets when we use LIME for explanation generation. Here, r\_i\_a\_j denotes the $i^{th}$ execution with $\alpha=j$, and LR + CLCDSA + PyExplainer denotes that PyExplainer is applied to explain the outcome of the LR model trained on the CLCDSA dataset.}
\label{prob_diff boxplot}
\vspace{-1.25em}
\end{figure}

First, we evaluate the effectiveness of PyExplainer in generating reliable explanations using the $\%Prob\_diff$ metric. Figures \ref{LR_fig_cort_CLCDSA} and \ref{LR_fig_cort_CR} demonstrate that, with the increase in $\alpha$ values, the mean value of $\%Prob\_diff$ consistently increases for the LR model trained on both the CLCDSA and code review datasets. For example, in the first execution with $\alpha=1$ (e.g., r\_1\_a\_1), the mean value is approximately $0.29$ for correct predictions. In the first execution with $\alpha=2$ (e.g., r\_1\_a\_2), the mean value increases to approximately $0.37$, and in the first execution with $\alpha=3$ (e.g., r\_1\_a\_3), the mean value further rises to approximately $0.43$ for correct predictions. In this case, the mean values indicate a comparative difference between the original and the simulated instances. In contrast to the LR model, Figures \ref{RF_fig_cort_CLCDSA} and \ref{RF_fig_cort_CR} demonstrate that, with the increase in $\alpha$ values, the mean value of $\%Prob\_diff$ remains almost the same for the RF model trained on both the CLCDSA and code review datasets. Furthermore, Figure \ref{RF_fig_cort_CR} clearly demonstrates that for some generated simulated instances, the $\%Prob\_diff$ values are negative, which contradicts the characteristic of the $\%Prob\_diff$ evaluation metric as described in Section \ref{eval}.

Regarding LIME, when we examine the LR and RF models trained on the CLCDSA dataset, we notice a substantial difference in the prediction probabilities between the original and simulated instances, as depicted in Figures \ref{LR_LIME_CLCDSA_Correct} and \ref{RF_LIME_CLCDSA_Correct}. However, as we increase the $\alpha$ values, the $\%Prob\_diff$ metric remains relatively constant. In the case of the LR model trained on the code review dataset, we observe an increase in $\%Prob\_diff$ metric values with higher $\alpha$ values. Interestingly, in this case, we encounter some simulated instances with negative $\%Prob\_diff$ values, which contradicts the expected characteristics of the $\%Prob\_diff$ evaluation metric as described in Section \ref{eval}. Finally, Figure \ref{RF_LIME_Code_Correct} illustrates that, for the RF model trained on the code review dataset, the $\%Prob\_diff$ metric value does not show a consistent increase with higher $\alpha$ values. Additionally, the probability difference between the original and simulated instances is insignificant. We observe similar experimental results for the LR and RF models, making wrong predictions. The details of the results can be found in Figure \ref{prob_diff_wrng_appendix} in \ref{prob_diff_appendix}.

\begin{figure}[htbp]
\centering     
\subfigure[PCPD (correct predictions)]{\label{mob_PCPD_pyexp_rev_true}\includegraphics[width=4.25cm, height=3cm]{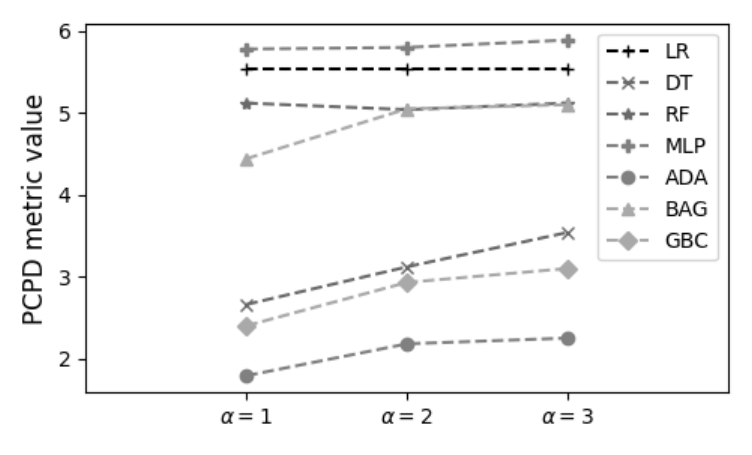}}
\subfigure[PCPD (wrong predictions)]{\label{mob_PCPD_pyexp_rev_false}\includegraphics[width=4.25cm, height=3cm]{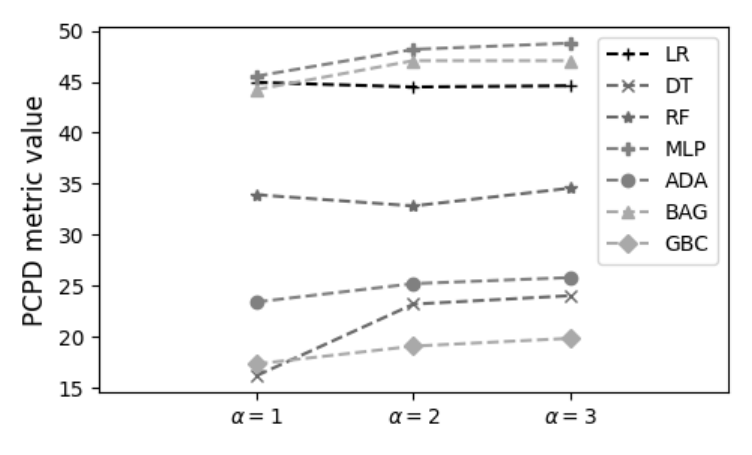}}
\subfigure[PCPI (correct predictions)]{\label{mob_PCPI_pyexp_rev_true}\includegraphics[width=4.25cm, height=3cm]{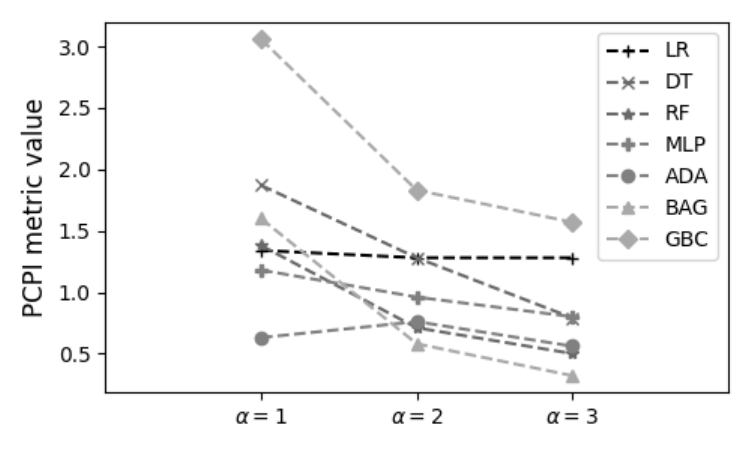}}
\subfigure[PCPI (wrong predictions)]{\label{mob_PCPI_pyexp_rev_false}\includegraphics[width=4.25cm, height=3cm]{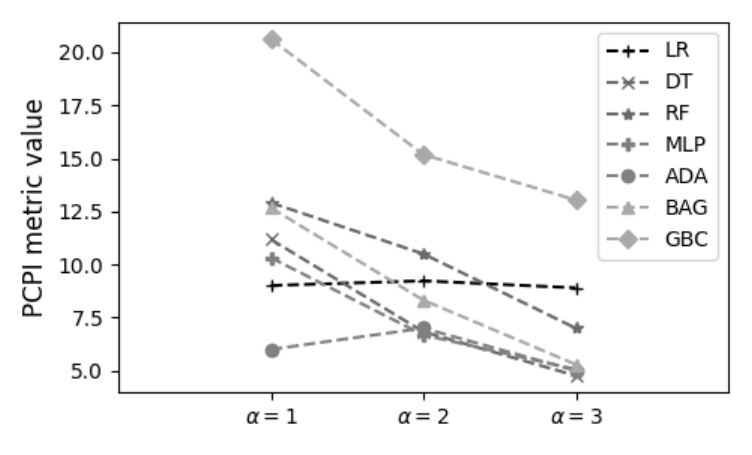}}
\subfigure[NCPD (correct predictions)]{\label{mob_NCPD_pyexp_rev_true}\includegraphics[width=4.25cm, height=3cm]{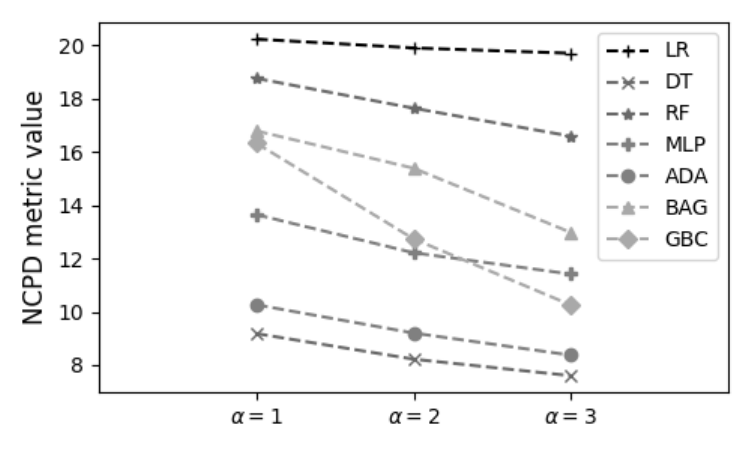}} 
\subfigure[NCPD (wrong predictions)]{\label{mob_NCPD_pyexp_rev_false}\includegraphics[width=4.25cm, height=3cm]{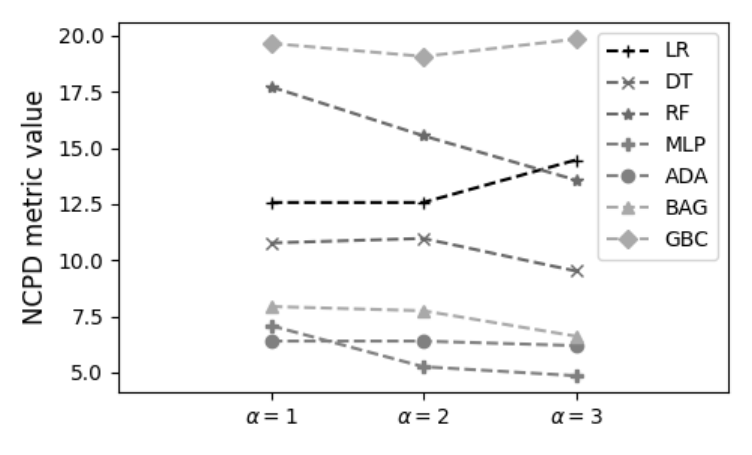}}
\subfigure[NCPI (correct predictions)]{\label{mob_NCPI_pyexp_rev_true}\includegraphics[width=4.25cm, height=3cm]{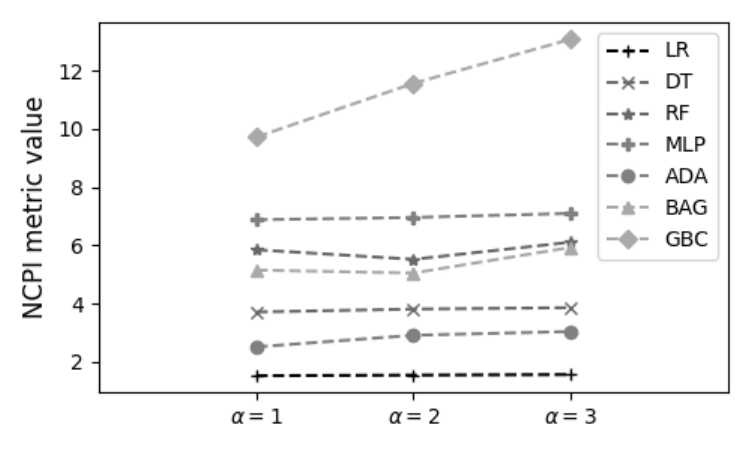}}
\subfigure[NCPI (wrong predictions)]{\label{mob_NCPI_pyexp_rev_false}\includegraphics[width=4.25cm, height=3cm]{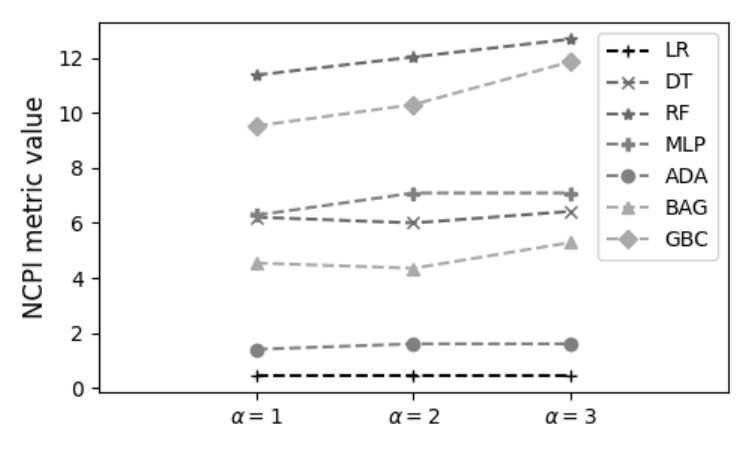}}
\caption{The values of granular-level evaluation metrics when PyExplainer is applied to the simulated instances of the cross-project mobile apps dataset.}
\label{granular_mob_pyexp}
\end{figure}

\begin{figure}[htbp]
\centering     
\subfigure[PCPD (correct predictions)]{\label{mob_PCPD_lime_rev_true}\includegraphics[width=4.25cm, height=3cm]{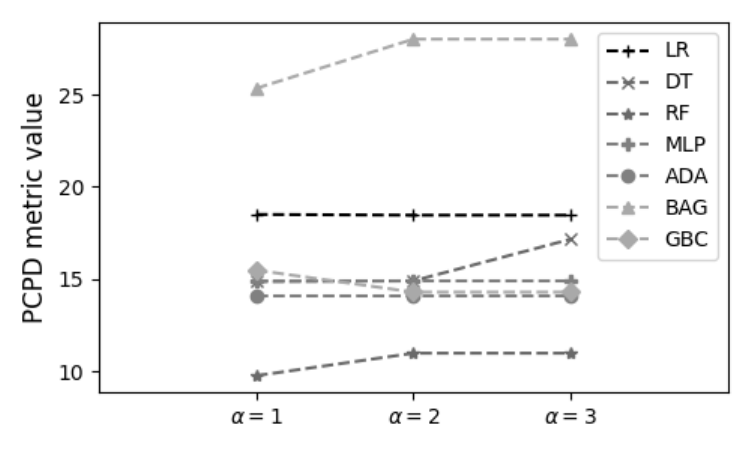}}
\subfigure[PCPD (wrong predictions)]{\label{mob_PCPD_lime_rev_false}\includegraphics[width=4.25cm, height=3cm]{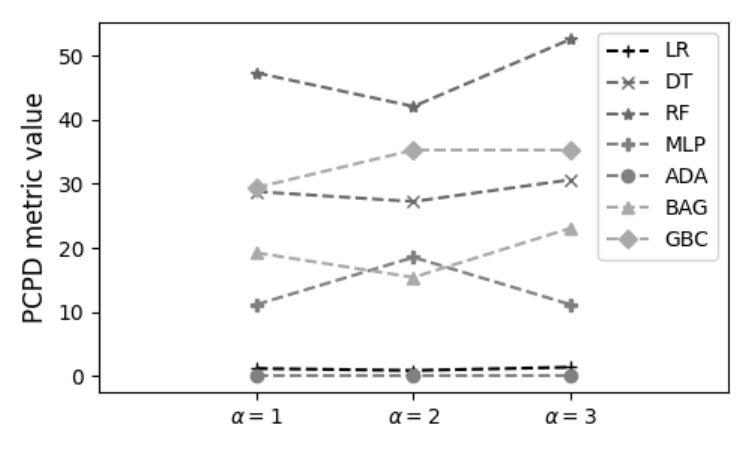}}
\subfigure[PCPI (correct predictions)]{\label{mob_PCPI_lime_rev_true}\includegraphics[width=4.25cm, height=3cm]{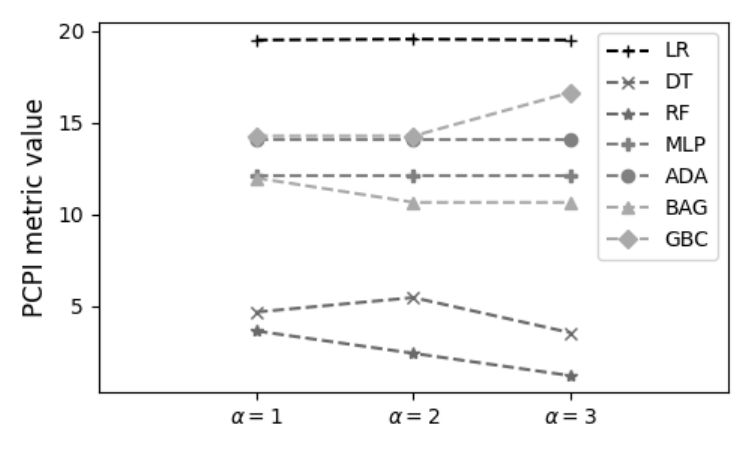}}
\subfigure[PCPI (wrong predictions)]{\label{mob_PCPI_lime_rev_false}\includegraphics[width=4.25cm, height=3cm]{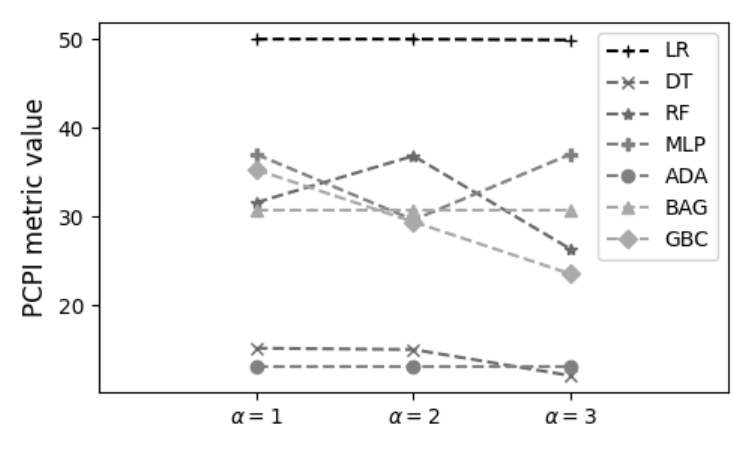}}
\subfigure[NCPD (correct predictions)]{\label{mob_NCPD_lime_rev_true}\includegraphics[width=4.25cm, height=3cm]{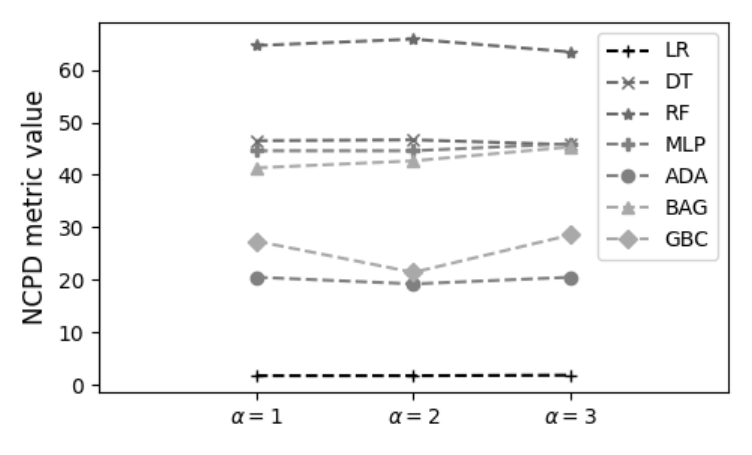}} 
\subfigure[NCPD (wrong predictions)]{\label{mob_NCPD_lime_rev_false}\includegraphics[width=4.25cm, height=3cm]{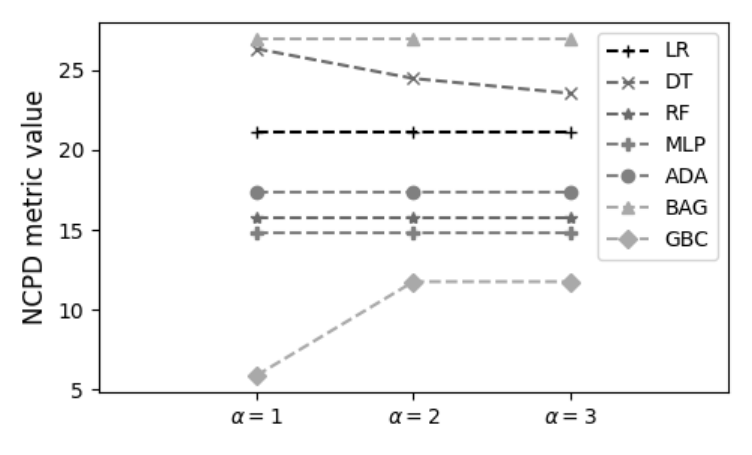}}
\subfigure[NCPI (correct predictions)]{\label{mob_NCPI_lime_rev_true}\includegraphics[width=4.25cm, height=3cm]{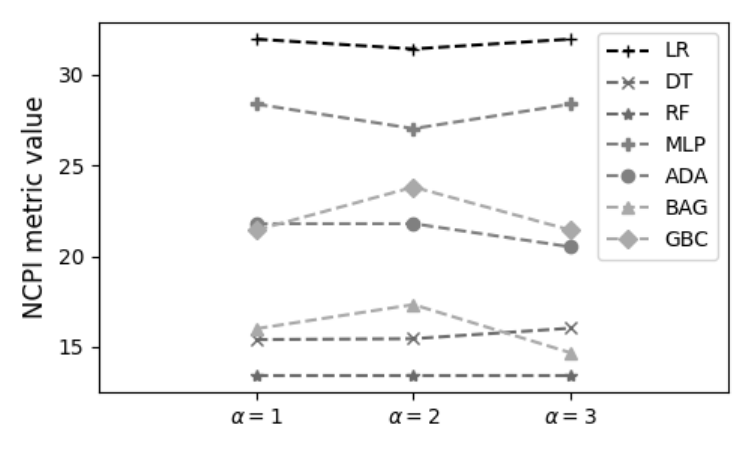}}
\subfigure[NCPI (wrong predictions)]{\label{mob_NCPI_lime_rev_false}\includegraphics[width=4.25cm, height=3cm]{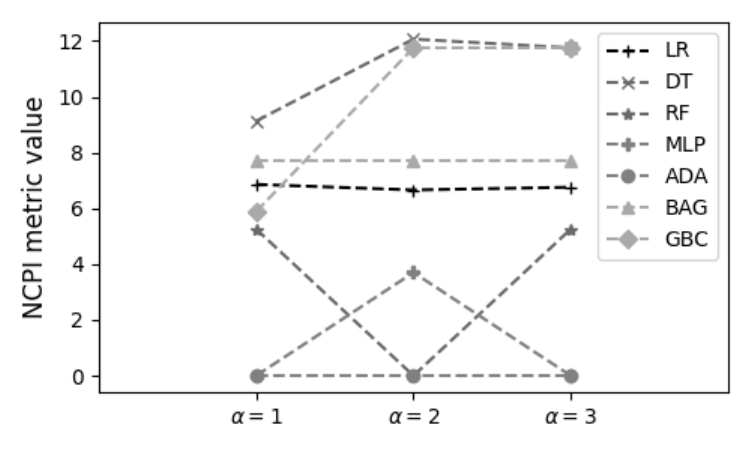}}
\caption{The values of granular-level evaluation metrics when LIME is applied to the simulated instances of the cross-project mobile apps dataset.}
\label{granular_mob_lime}
\end{figure}

Figures \ref{granular_mob_pyexp} and \ref{granular_mob_lime} demonstrate the experimental results of four granular-level evaluation metrics (i.e., PCPD, PCPI, NCPD, and NCPI) for PyExplainer and LIME applied to ML models trained on the cross-project mobile apps dataset, both for correct and wrong predictions. In an ideal scenario, we expect an increase in each evaluation metric for each simulated instance. However, Figure \ref{granular_mob_pyexp} clearly shows that, with PyExplainer, we observe very low values for PCPD, PCPI, NCPD, and NCPI metrics in the case of correct predictions, with only a few exceptions. For instance, when examining the LR model, we find a PCPD value of 24.59 and an NCPI value of 20.22. Conversely, we consistently observe very low PCPI and NCPD values for correct predictions when PyExplainer is used. Turning our attention to LIME, we observe similar results for the ML models making correct predictions.

For wrong predictions, we observe granular-level evaluation metric values that are relatively higher or similar, except for the NCPI metric, when we apply LIME to the simulated instances for all the ML models compared to the correct predictions. From Figures \ref{granular_mob_pyexp} and \ref{granular_mob_lime}, it is evident that considering both correct and wrong predictions, the four granular-level evaluation metrics often yield low values, raising concerns about the reliability of the explanations generated by PyExplainer and LIME.

Similar to the $\%Reversed$ and $\%Prob\_diff$ metrics, we can assess whether the values of granular-level evaluation metrics such as PCPD, PCPI, NCPD, and NCPI increase with the rise of $\alpha$ when PyExplainer and LIME are applied to the simulated instances. Figures \ref{granular_mob_pyexp} and \ref{granular_mob_lime} demonstrate that, as $\alpha$ increases, the granular-level evaluation metrics do not consistently increase for all the ML models. For example, in the case of the NCPD metric, we observe either a downward trend or a steady trend with the increase of $\alpha$ values. Therefore, the reliability of the explanation generated by PyExplainer and LIME is in question. Please refer to Tables \ref{granular_PyExp_appendix} and \ref{granular_LIME_appendix} in \ref{granular_level_appendix} for the detailed results of granular-level metrics for different ML models trained on the Java project, Postgres, CLCDSA, and code review datasets.

Table \ref{dataset} reveals that, for certain datasets and ML models, F1-scores are comparatively low due to data imbalance. For example, the Java project test dataset (e.g., after the train-test split) contains 2191 samples of the negative class (e.g., clean commits) and 323 samples of the positive class (e.g., bug-inducing commits). Therefore, we perform additional experiments by balancing the test set to better assess the reliability and consistency of the explanations generated by rule-based XAI techniques (e.g., PyExplainer) using EvaluateXAI. Table \ref{balance_dataset} reveals that after balancing the test set, F1-scores improve for different ML models trained on the Java project dataset.

\begin{table}[htbp]
\centering
\caption{Performance of the ML models on Java project dataset considering different evaluation metrics such as Accuracy (Acc), F1-score (F-1), and AUC values.}
\resizebox{\columnwidth}{!}{
\begin{tabular}{l|cc|cc|cc}
\hline
\multirow{2}{*}{\begin{tabular}[c]{@{}l@{}}ML\\ Model\end{tabular}} & \multicolumn{2}{c|}{Acc}                                                                                                                    & \multicolumn{2}{c|}{F-1}                                                                                                                    & \multicolumn{2}{c}{AUC}                                                                                                                    \\ \cline{2-7} 
                                                                    & \multicolumn{1}{c|}{\begin{tabular}[c]{@{}c@{}}Before\\ balancing\end{tabular}} & \begin{tabular}[c]{@{}c@{}}After\\ balancing\end{tabular} & \multicolumn{1}{c|}{\begin{tabular}[c]{@{}c@{}}Before\\ balancing\end{tabular}} & \begin{tabular}[c]{@{}c@{}}After\\ balancing\end{tabular} & \multicolumn{1}{c|}{\begin{tabular}[c]{@{}c@{}}Before\\ balancing\end{tabular}} & \begin{tabular}[c]{@{}c@{}}After\\ balancing\end{tabular} \\ \hline
LR                                                                  & \multicolumn{1}{c|}{0.75}                                                       & 0.69                                                      & \multicolumn{1}{c|}{0.37}                                                       & 0.67                                                      & \multicolumn{1}{c|}{0.76}                                                       & 0.76                                                      \\ \hline
DT                                                                  & \multicolumn{1}{c|}{0.80}                                                       & 0.74                                                      & \multicolumn{1}{c|}{0.49}                                                       & 0.72                                                      & \multicolumn{1}{c|}{0.80}                                                       & 0.79                                                      \\ \hline
RF                                                                  & \multicolumn{1}{c|}{0.88}                                                       & 0.76                                                      & \multicolumn{1}{c|}{0.55}                                                       & 0.70                                                      & \multicolumn{1}{c|}{0.85}                                                       & 0.87                                                      \\ \hline
MLP                                                                 & \multicolumn{1}{c|}{0.80}                                                       & 0.73                                                      & \multicolumn{1}{c|}{0.48}                                                       & 0.70                                                      & \multicolumn{1}{c|}{0.81}                                                       & 0.80                                                      \\ \hline
ADA                                                                 & \multicolumn{1}{c|}{0.79}                                                       & 0.76                                                      & \multicolumn{1}{c|}{0.47}                                                       & 0.75                                                      & \multicolumn{1}{c|}{0.81}                                                       & 0.81                                                      \\ \hline
BAG                                                                 & \multicolumn{1}{c|}{0.78}                                                       & 0.75                                                      & \multicolumn{1}{c|}{0.50}                                                       & 0.75                                                      & \multicolumn{1}{c|}{0.83}                                                       & 0.82                                                      \\ \hline
GBC                                                                 & \multicolumn{1}{c|}{0.81}                                                       & 0.72                                                      & \multicolumn{1}{c|}{0.48}                                                       & 0.68                                                      & \multicolumn{1}{c|}{0.80}                                                       & 0.79                                                      \\ \hline
\end{tabular}}
\label{balance_dataset}
\end{table}

\begin{figure}[htbp]
\centering     
\subfigure[$\%Reversed$ (correct predictions)]{\label{balance_pyexp_reverse_true}\includegraphics[width=4.25cm, height=3cm]{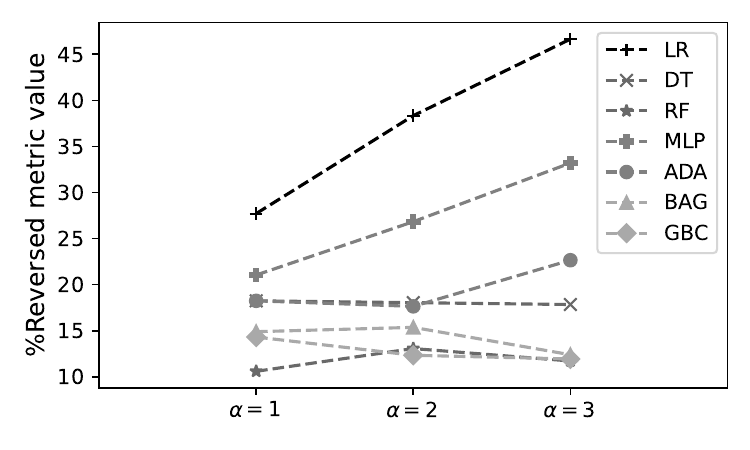}} 
\subfigure[$\%Reversed$ (wrong predictions)]{\label{balance_pyexp_reverse_false}\includegraphics[width=4.25cm, height=3cm]{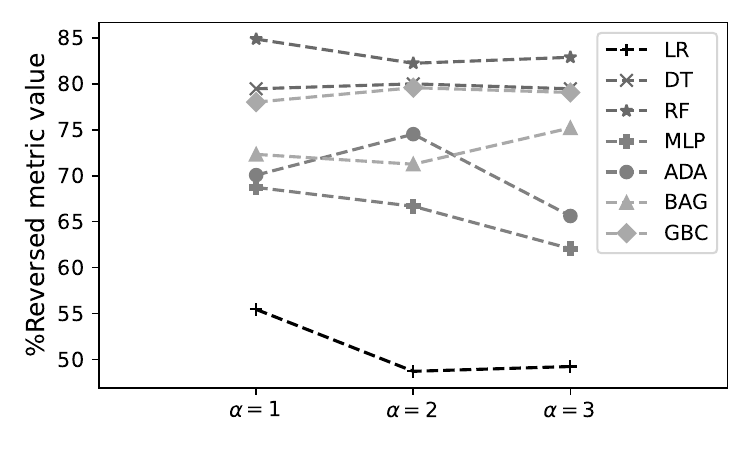}}
\subfigure[PCPD (correct predictions)]{\label{balance_PCPD_pyexp_true}\includegraphics[width=4.25cm, height=3cm]{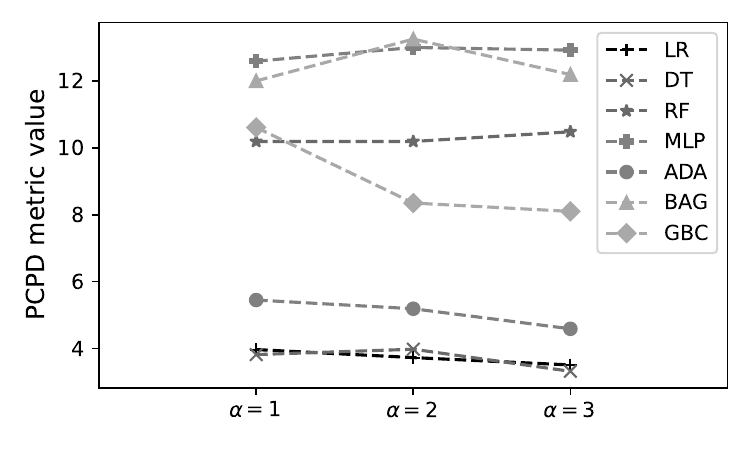}} 
\subfigure[PCPD (wrong predictions)]{\label{balance_PCPD_pyexp_false}\includegraphics[width=4.25cm, height=3cm]{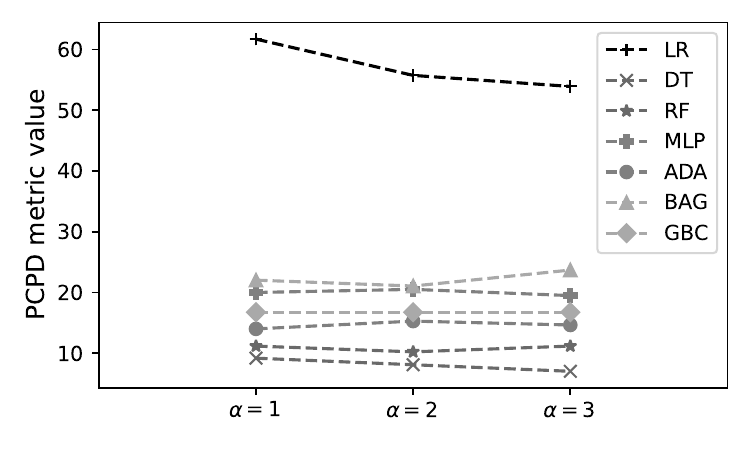}}
\subfigure[PCPI (correct predictions)]{\label{balance_PCPI_pyexp_true}\includegraphics[width=4.25cm, height=3cm]{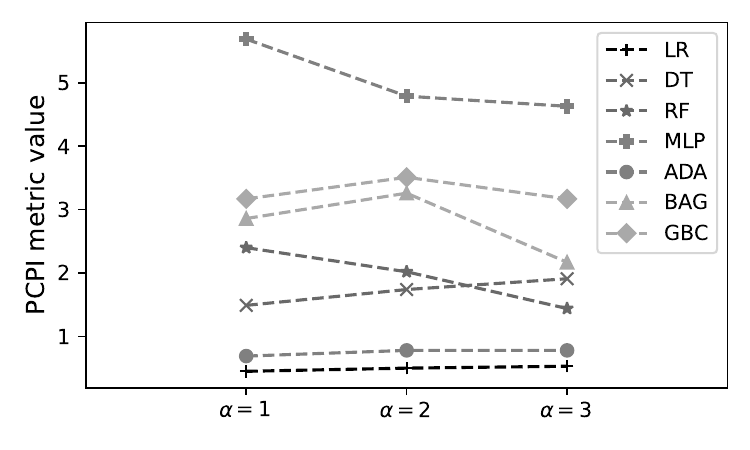}}
\subfigure[PCPI (wrong predictions)]{\label{balance_PCPI_pyexp_false}\includegraphics[width=4.25cm, height=3cm]{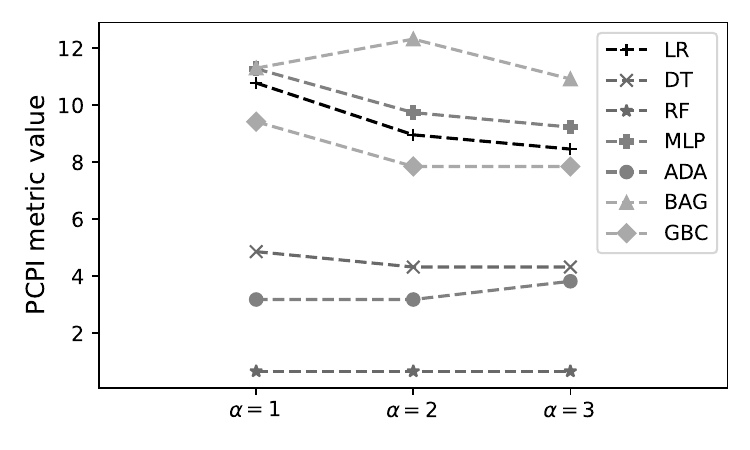}}
\subfigure[NCPD (correct predictions)]{\label{balance_NCPD_pyexp_true}\includegraphics[width=4.25cm, height=3cm]{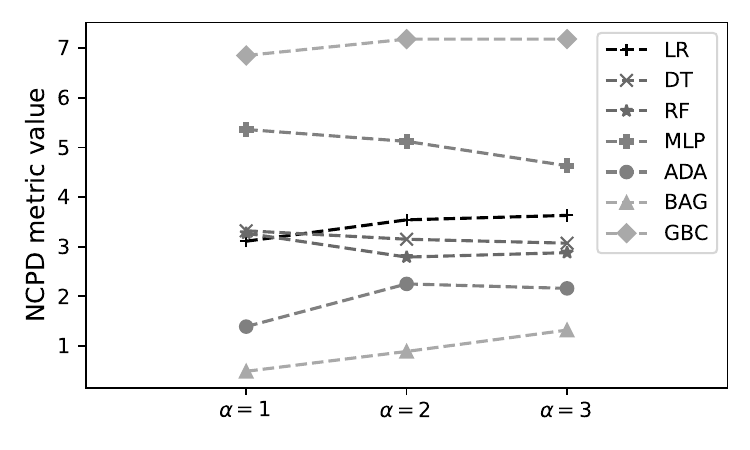}}
\subfigure[NCPD (wrong predictions)]{\label{balance_NCPD_pyexp_false}\includegraphics[width=4.25cm, height=3cm]{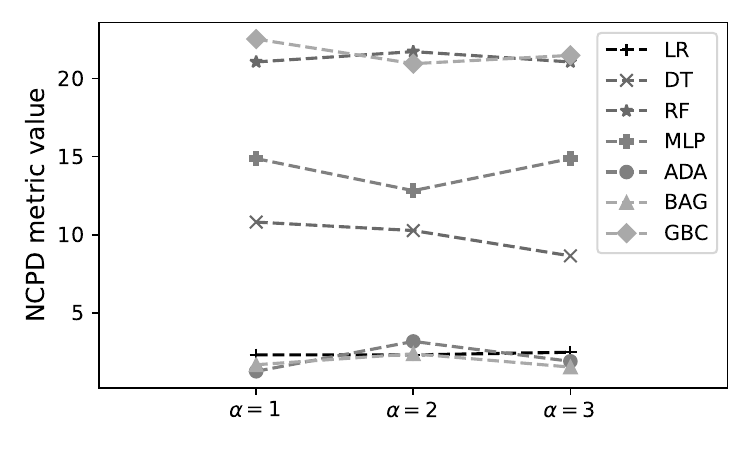}}
\subfigure[NCPI (correct predictions)]{\label{balance_NCPI_pyexp_true}\includegraphics[width=4.25cm, height=3cm]{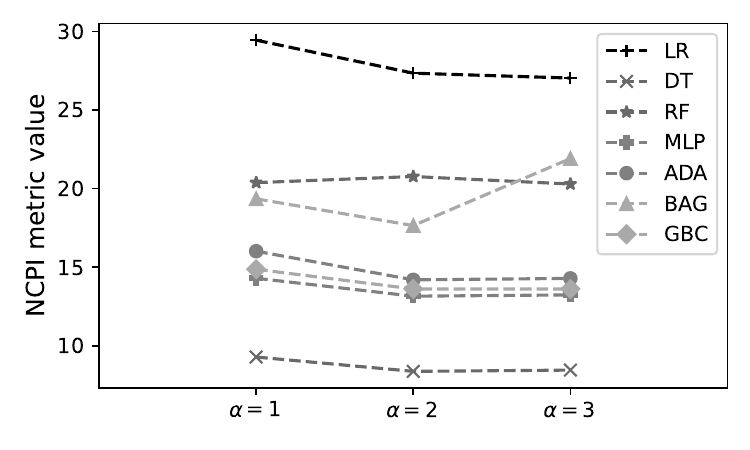}}
\subfigure[NCPI (wrong predictions)]{\label{balance_NCPI_pyexp_false}\includegraphics[width=4.25cm, height=3cm]{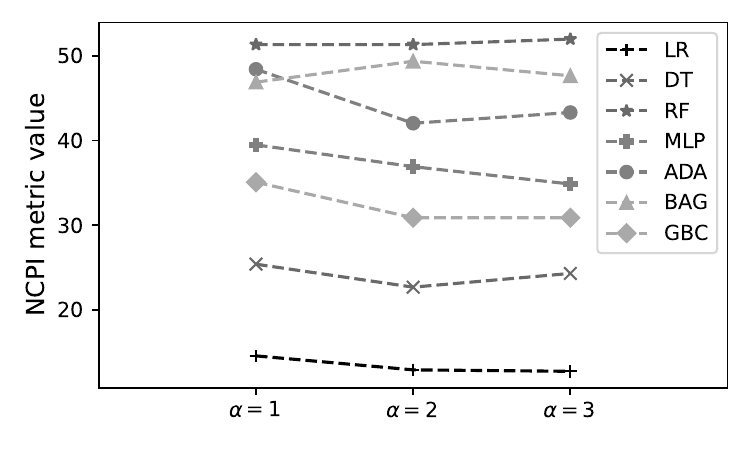}}
\caption{The characteristics of different metric values with different $\alpha$ when we apply PyExplainer to the simulated instances of the Java project dataset. For example, Figure \ref{balance_pyexp_reverse_true} represents the variation in $\%Reversed$ metric values for different $\alpha$ when ML models make correct predictions.}
\label{balance_pyexp_all_metric}

\end{figure}

Figure \ref{balance_pyexp_all_metric} shows that, in the case of PyExplainer, none of the evaluation metrics reach 100\% when considering all examples of the balanced test set of the Java project dataset. Additionally, as the $\alpha$ value increases, the metric values do not increase accordingly. Therefore, after balancing the test set, we still obtain similar results to those without balancing the test set of the Java project dataset. We also observe similar results for the $\%Prob\_diff$ metric with the balanced test set of the Java project dataset. The detailed results can be found in our replication package, following the instructions in the README.md file.

Ideally, we expect an increase in the value of each evaluation metric for every test instance. For instance, when the $\%Reversed$ metric value approaches 100\%, it indicates the effectiveness of PyExplainer and LIME in producing reliable explanations for ML models. However, the relatively low values of each evaluation metric in our experiments demonstrate that PyExplainer and LIME often fail to generate reliable explanations for ML models in software analytics tasks.

In this study, we introduce a novel framework that employs six evaluation metrics to evaluate the reliability of explanations generated by XAI techniques. Our experiments reveal that XAI techniques do not consistently provide reliable explanations for ML models. This underscores the need for further research to enhance the effectiveness of XAI techniques in generating reliable explanations for ML models in software analytics tasks.


\vspace{1em}
\noindent
\fbox{\begin{minipage}{24em}
\textbf{Result RQ1}: Our extensive study, employing \textit{EvaluateXAI} on five datasets and seven ML models, concludes that none of the evaluation metrics reach 100\% when considering all the testing instances for explanation generation in our experiments. This indicates that rule-based XAI techniques, such as PyExplainer and LIME, often fail to generate reliable explanations for ML models in software analytics tasks.
\end{minipage}}
\vspace{1em}

\textbf{RQ2: Do rule-based XAI techniques maintain consistency in generating explanations for machine learning models in software analytics tasks?}

\textbf{Motivation}: Lundberg et al. \cite{lundberg2017unified} argued that instance explanation generation must remain consistent upon regeneration for the same instance. However, Figure \ref{pyexp explanation} shows that PyExplainer (while PyExplainer involves randomization in the neighbour generation process, Pornprasit et al. \cite{pornprasit2021pyexplainer} claimed that \textbf{``we repeated the experiment five times, the conclusion of our paper remains the same"}) provides four different explanations for the same instance upon re-execution. On the other hand, Figures \ref{lime_a} and \ref{lime_b} demonstrate that LIME (which involves random perturbation) provides two different explanations for the same instance upon re-execution. Furthermore, we notice that among the five features, PyExplainer utilizes four of them (e.g., Line of Code (\textit{LOC}), \textit{nCommit}, \textit{CommentToCodeRatio}, and \textit{nCoupledClass}) to generate explanations for various executions. Additionally, in Figure \ref{motivation_a}, PyExplainer generates an explanation for cases where the feature value \textit{LOC} is less than $11.0$, while in Figure \ref{motivation_b}, it generates explanations for instances where the feature value \textit{LOC} is greater than $11.0$. Lastly, Figure \ref{motivation_d} illustrates a scenario in which PyExplainer fails to generate an explanation. These inconsistent explanations may confuse practitioners and discourage them from using XAI techniques to explain the results of ML models in software analytics tasks \cite{roy2022don}. Therefore, further research is warranted to evaluate the consistency of the explanations generated by XAI techniques.

\textbf{Approach}: 
Jiarpakdee et al. \cite{jiarpakdee2020empirical} assessed inconsistency by running 100 consecutive iterations on a single instance and measuring the rank difference of each metric. We employ a similar approach but consider 500 instances instead of a single one. Thus, we address this research question by conducting the same experimental analysis performed for the first research question (\textbf{RQ1}) multiple times. We compare the results from each execution using six different evaluation metrics. For the evaluation metric $\%Prob\_diff$, to measure whether the distributions of the $\%Prob\_diff$ between two executions are statistically different, we apply the Wilcoxon Signed-Rank Test \cite{wilcoxon1992individual}. Additionally, we utilize Cliff’s $|\delta|$ effect size \cite{cliff1993dominance} to quantify the extent of the differences. Finally, we report the experimental results using several tables and boxplots.

\textbf{Results}:
We evaluate the consistency of PyExplainer and LIME by examining the results of six evaluation metrics. We expect that when we run PyExplainer and LIME multiple times on the same test instances with the same experimental settings, we should obtain consistent results across different evaluation metrics. Any discrepancies among successive executions indicate the inconsistency of PyExplainer and LIME in generating explanations for ML models.

\begin{figure}[htbp]
\centering     
\subfigure[LR]{\label{Mob_LR_100}\includegraphics[width=4.25cm, height=3.25cm]{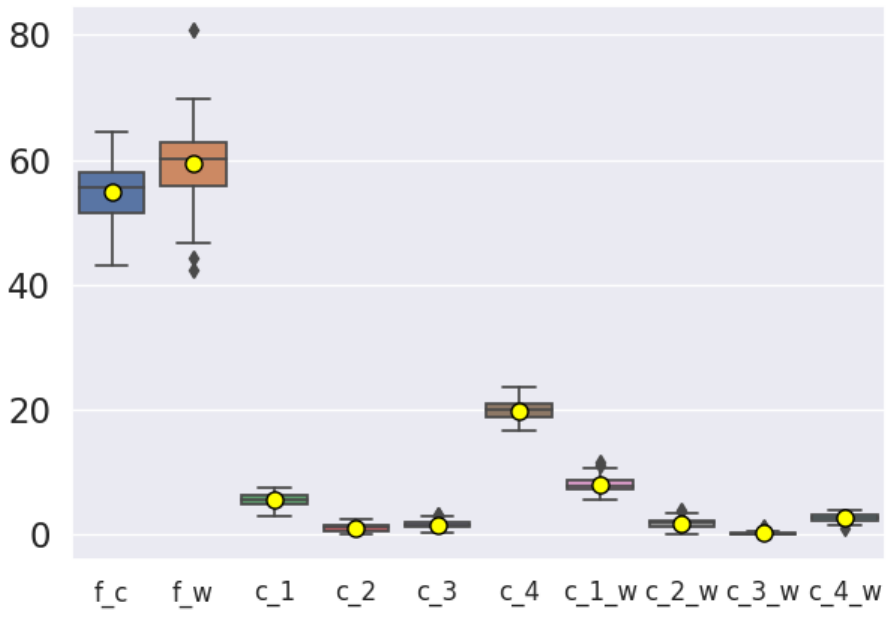}} 
\subfigure[DT]{\label{Mob_DT_100}\includegraphics[width=4.25cm, height=3.25cm]{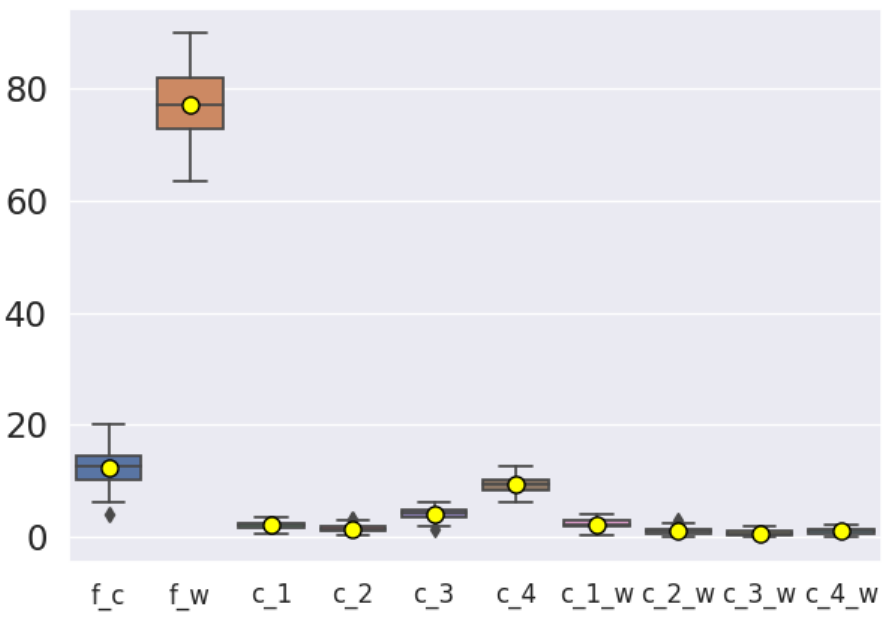}}
\subfigure[RF]{\label{Mob_RF_100}\includegraphics[width=4.25cm, height=3.25cm]{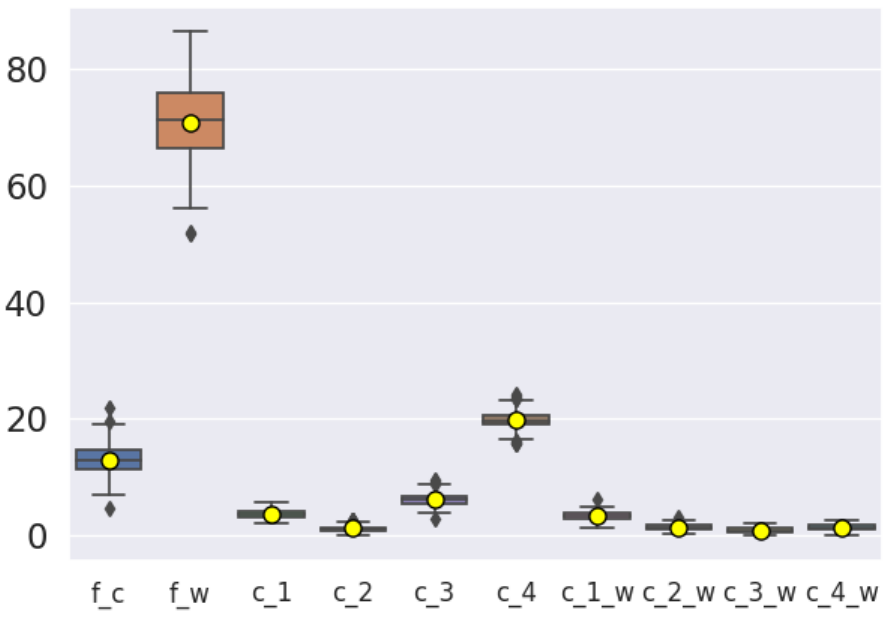}}
\vspace{-0.5em}
\subfigure[MLP]{\label{Mob_MLP_100}\includegraphics[width=4.25cm, height=3.25cm]{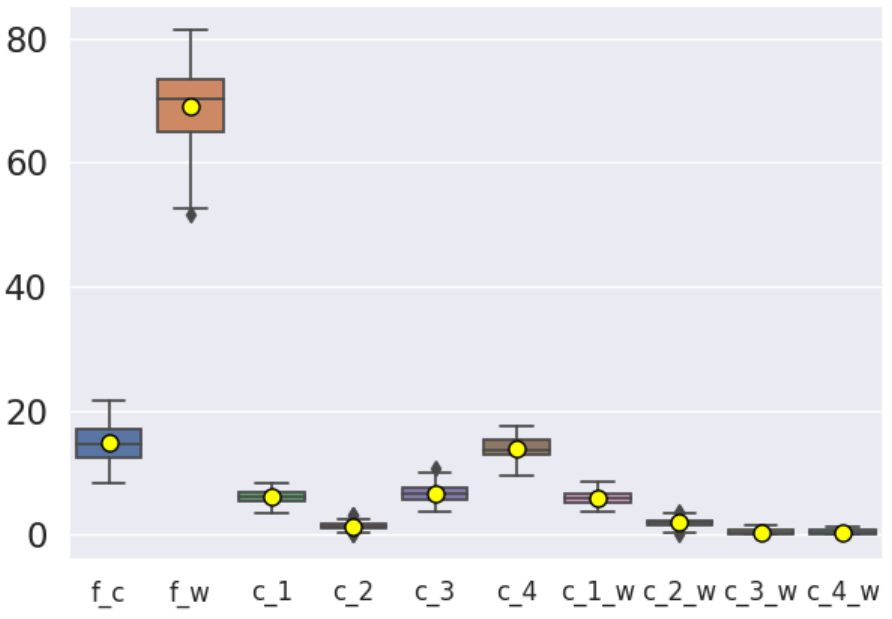}}
\subfigure[ADA]{\label{Mob_ADA_100}\includegraphics[width=4.25cm, height=3.25cm]{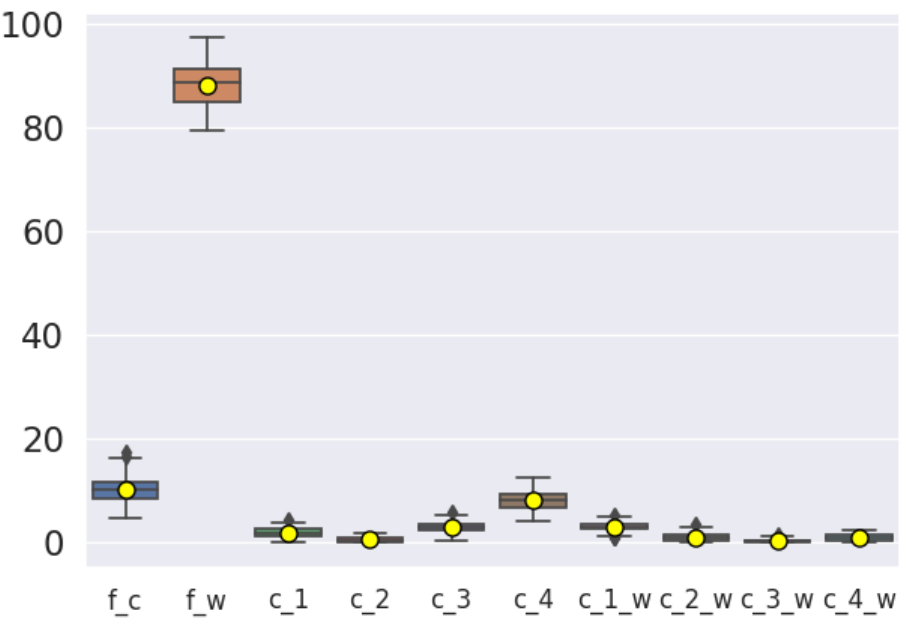}} 
\subfigure[BAG]{\label{Mob_BAG_100}\includegraphics[width=4.25cm, height=3.25cm]{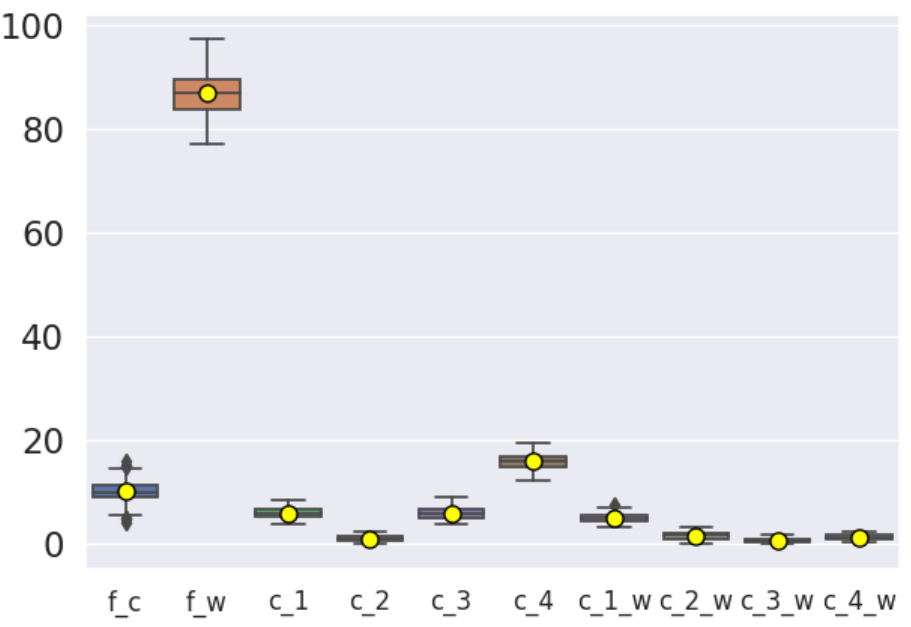}}
\vspace{-0.5em}
\subfigure[GBC]{\label{Mob_GBC_100}\includegraphics[width=4.25cm, height=3.25cm]{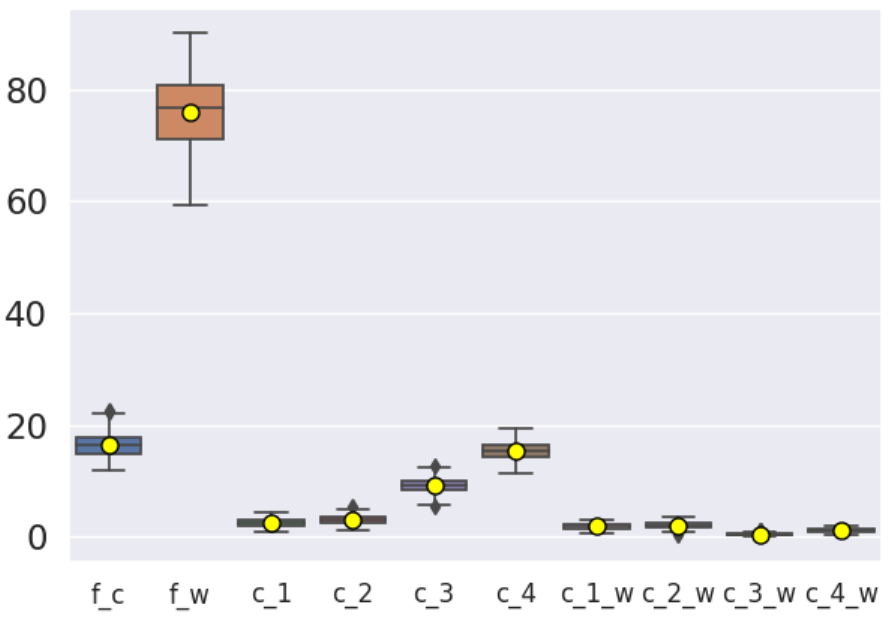}}
\caption{The distribution of $\%Reversed$, PCPD, PCPI, NCPD, and NCPI metrics, encompassing both correct and wrong predictions using PyExplainer on the cross-project mobile apps dataset. Here, $f\_c$ and $f\_w$ represent the $\%Reversed$ metric for correct and wrong predictions, respectively, while $c\_1, c\_2, c\_3, c\_4$ and $c\_1\_w, c\_2\_w, c\_3\_w, c\_4\_w$ PCPD, PCPI, NCPD, NCPI evaluation metrics for correct and wrong predictions, respectively.}
\label{Mob_100_Instance}
\vspace{-1.25em}
\end{figure}

\begin{figure}[htbp]
\centering     
\subfigure[LR]{\label{Java_LR_100}\includegraphics[width=4.25cm, height=3.25cm]{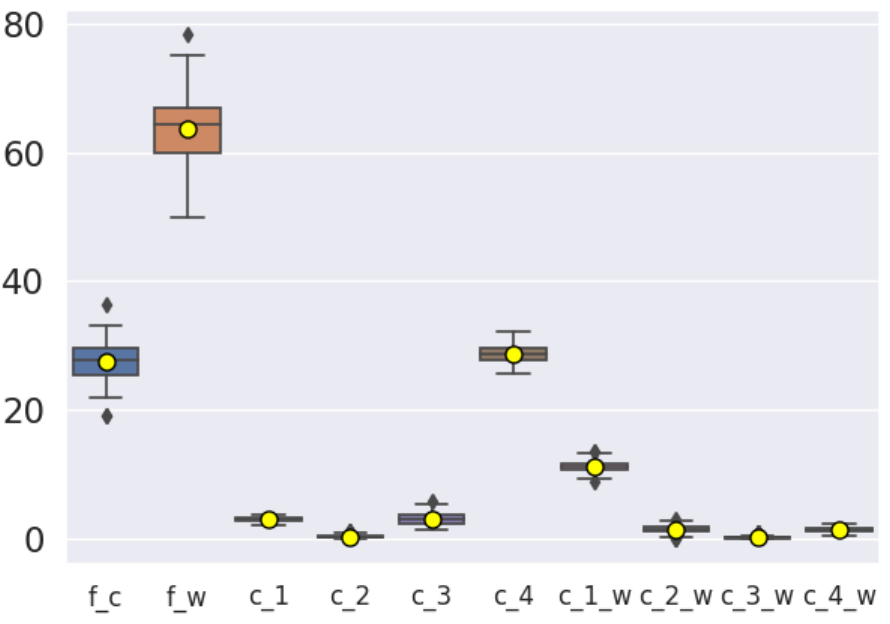}} 
\subfigure[DT]{\label{Java_DT_100}\includegraphics[width=4.25cm, height=3.25cm]{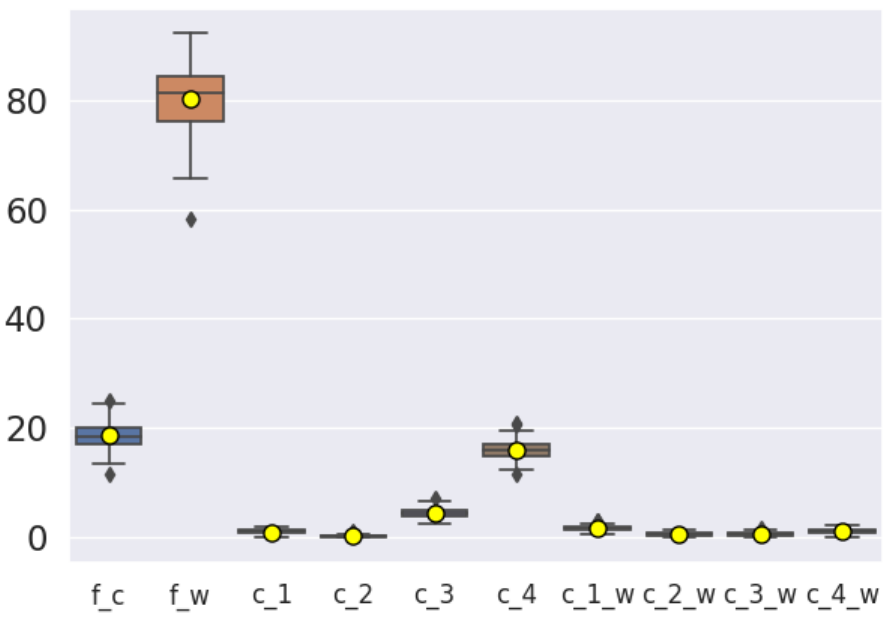}}
\subfigure[RF]{\label{Java_RF_100}\includegraphics[width=4.25cm, height=3.25cm]{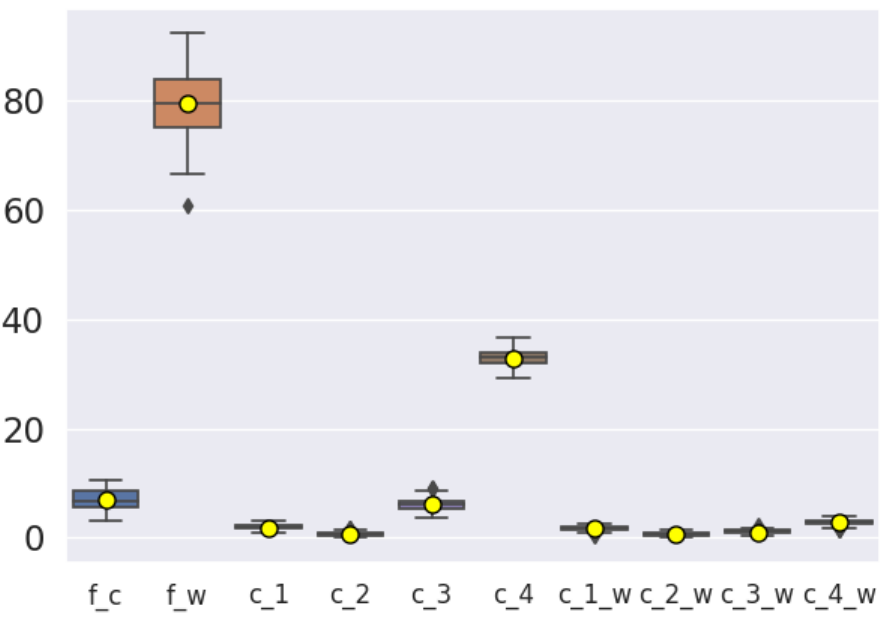}}
\subfigure[MLP]{\label{Java_MLP_100}\includegraphics[width=4.25cm, height=3.25cm]{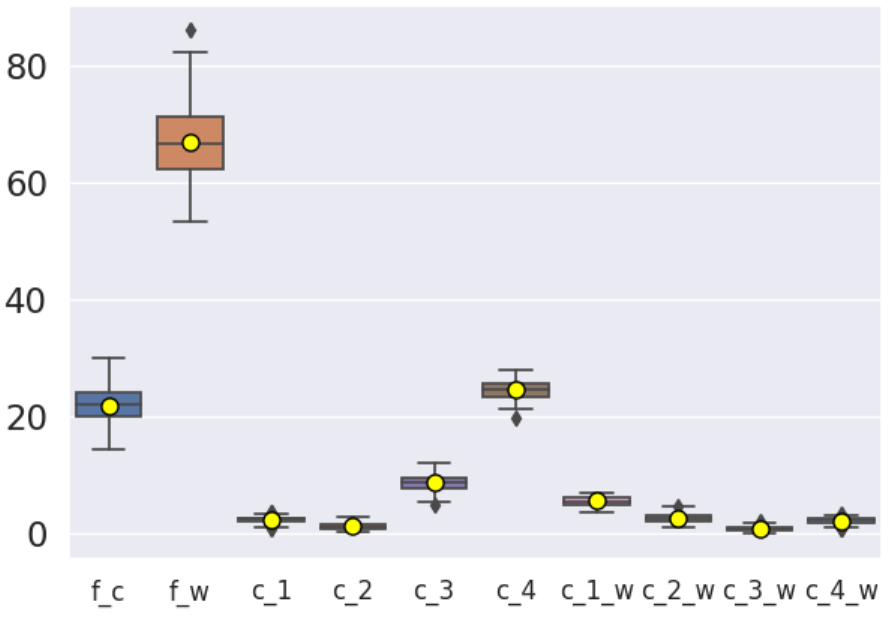}}
\subfigure[ADA]{\label{Java_ADA_100}\includegraphics[width=4.25cm, height=3.25cm]{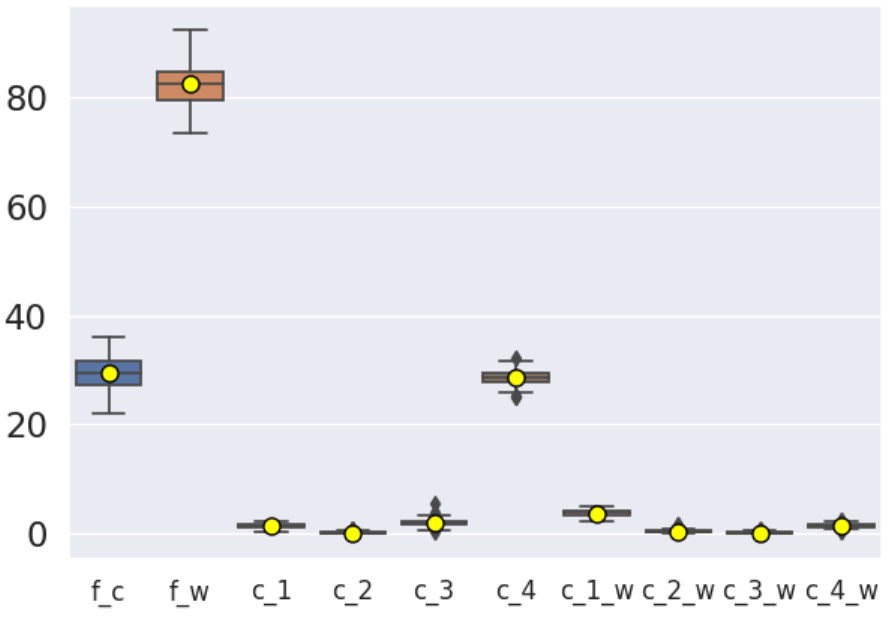}} 
\subfigure[BAG]{\label{Java_BAG_100}\includegraphics[width=4.25cm, height=3.25cm]{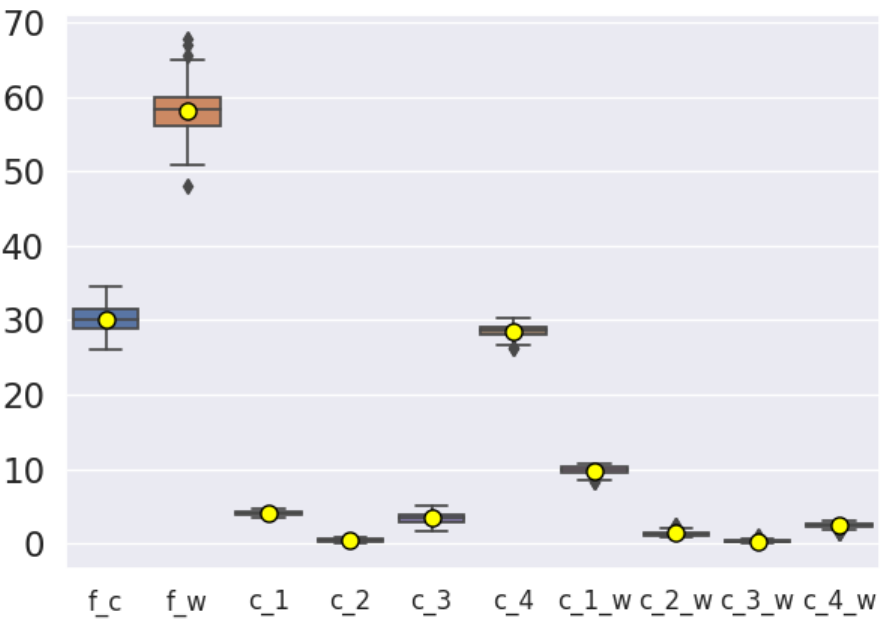}}
\subfigure[GBC]{\label{Java_GBC_100}\includegraphics[width=4.25cm, height=3.25cm]{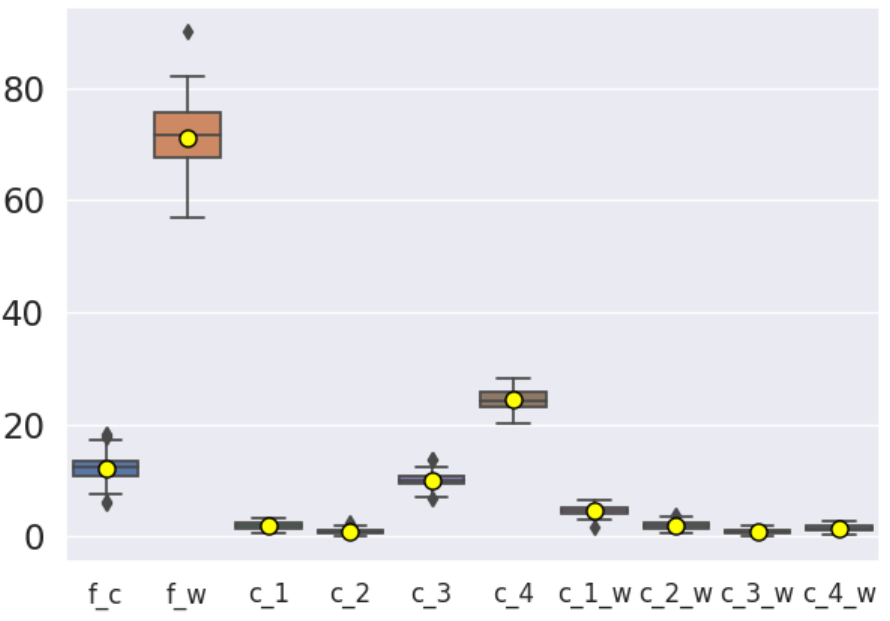}}
\caption{The distribution of $\%Reversed$, PCPD, PCPI, NCPD, and NCPI metrics, encompassing both correct and wrong predictions using PyExplainer on the Java project dataset. Here, $f\_c$ and $f\_w$ represent the $\%Reversed$ metric for correct and wrong predictions, respectively, while $c\_1, c\_2, c\_3, c\_4$ and $c\_1\_w, c\_2\_w, c\_3\_w, c\_4\_w$ PCPD, PCPI, NCPD, NCPI evaluation metrics for correct and wrong predictions, respectively.}
\label{Java_100_Instance}
\vspace{-1.25em}
\end{figure}

\begin{figure}[htbp]
\centering     
\subfigure[LR]{\label{LIME_500_mob_LR}\includegraphics[width=4.25cm, height=3.25cm]{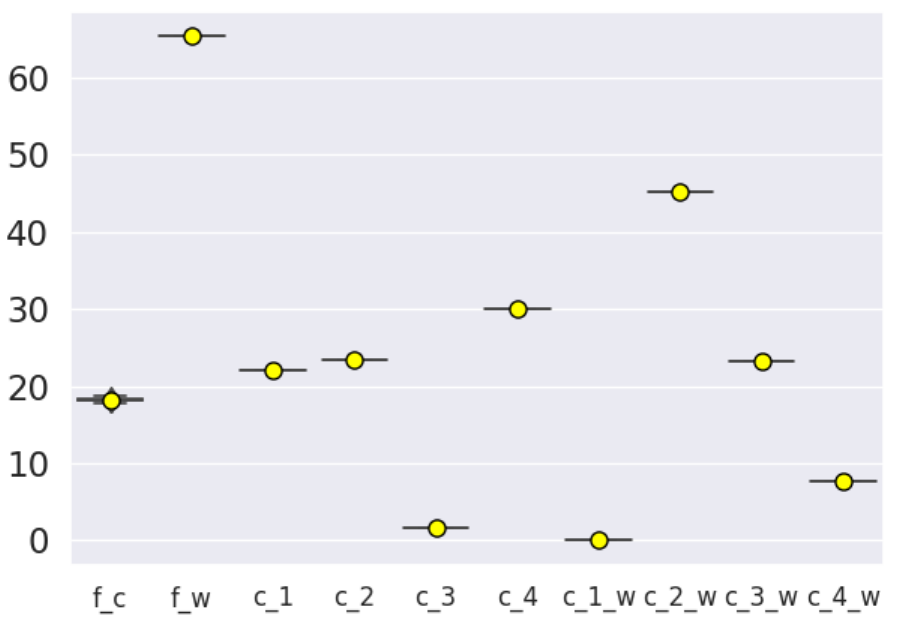}} 
\subfigure[DT]{\label{LIME_500_mob_DT}\includegraphics[width=4.25cm, height=3.25cm]{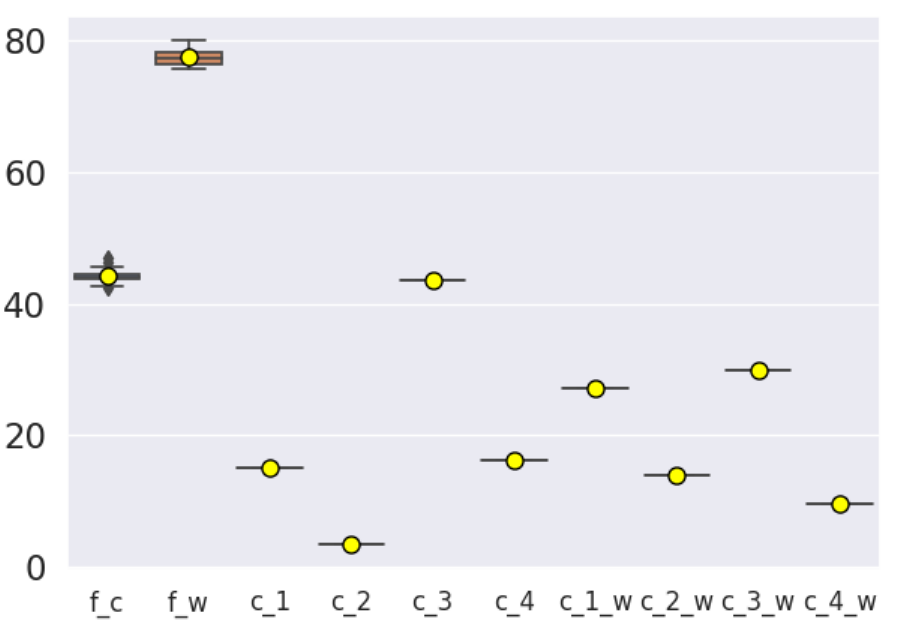}}
\subfigure[RF]{\label{LIME_500_mob_RF}\includegraphics[width=4.25cm, height=3.25cm]{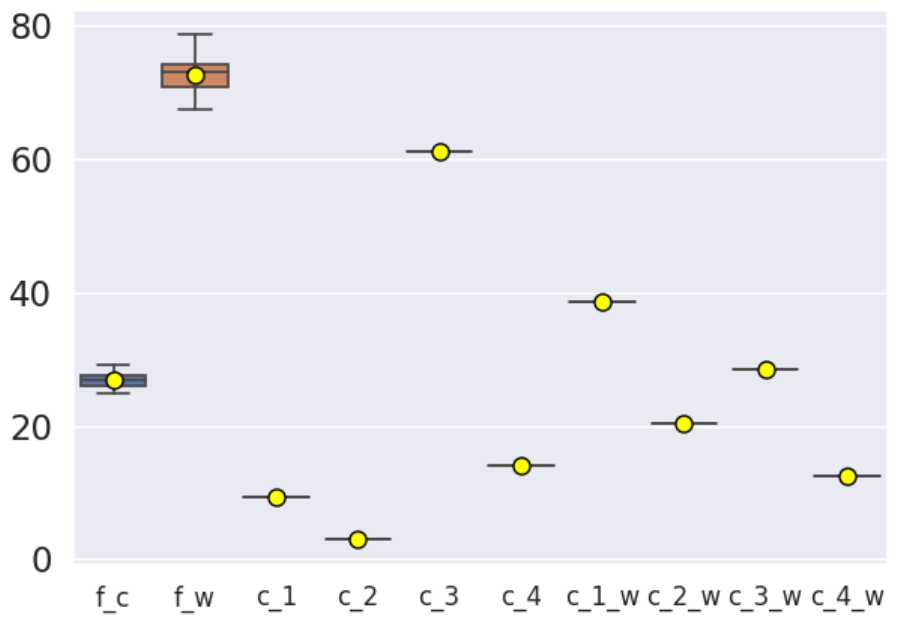}}
\subfigure[MLP]{\label{LIME_500_mob_MLP}\includegraphics[width=4.25cm, height=3.25cm]{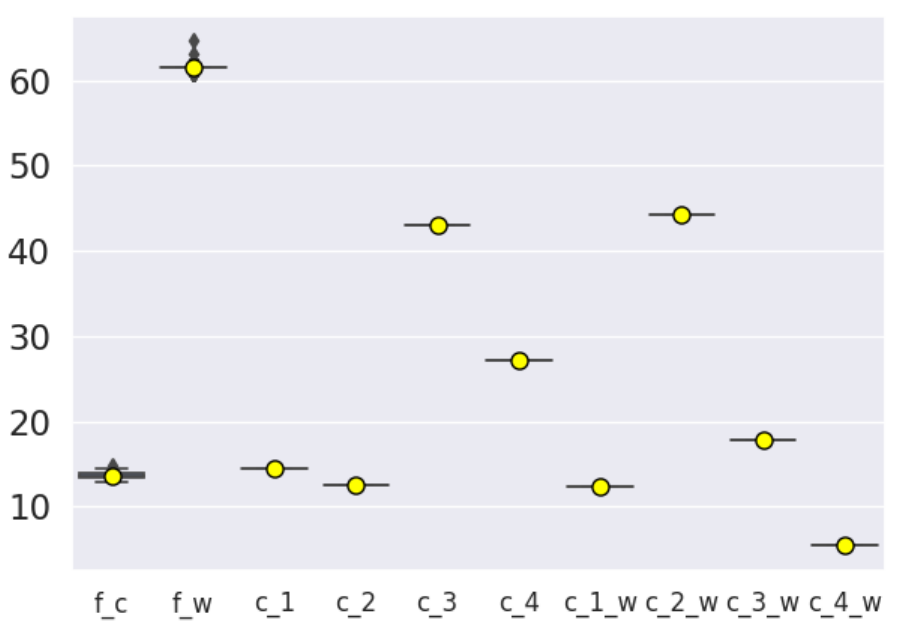}}
\subfigure[ADA]{\label{LIME_500_mob_ADA}\includegraphics[width=4.25cm, height=3.25cm]{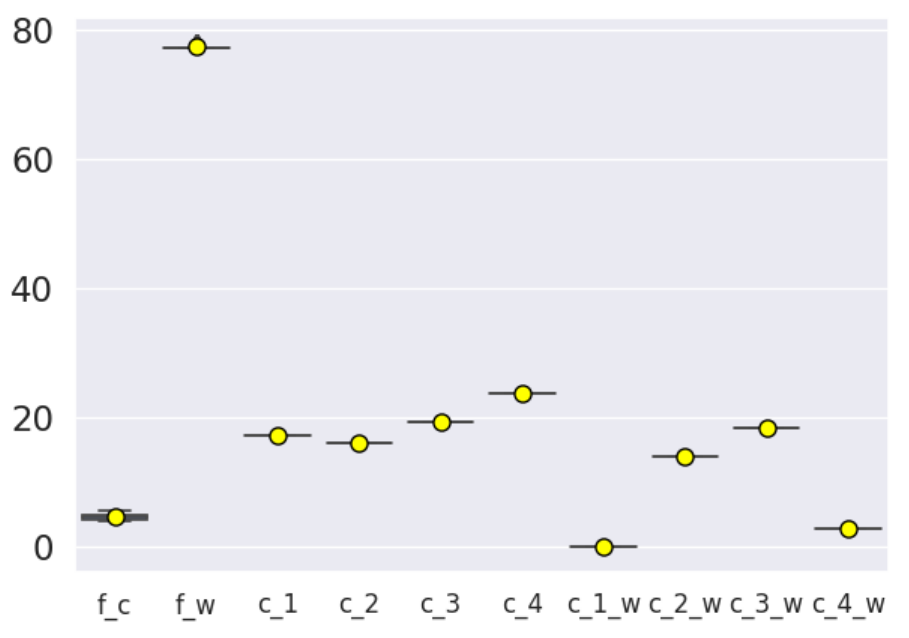}} 
\subfigure[BAG]{\label{LIME_500_mob_BAG}\includegraphics[width=4.25cm, height=3.25cm]{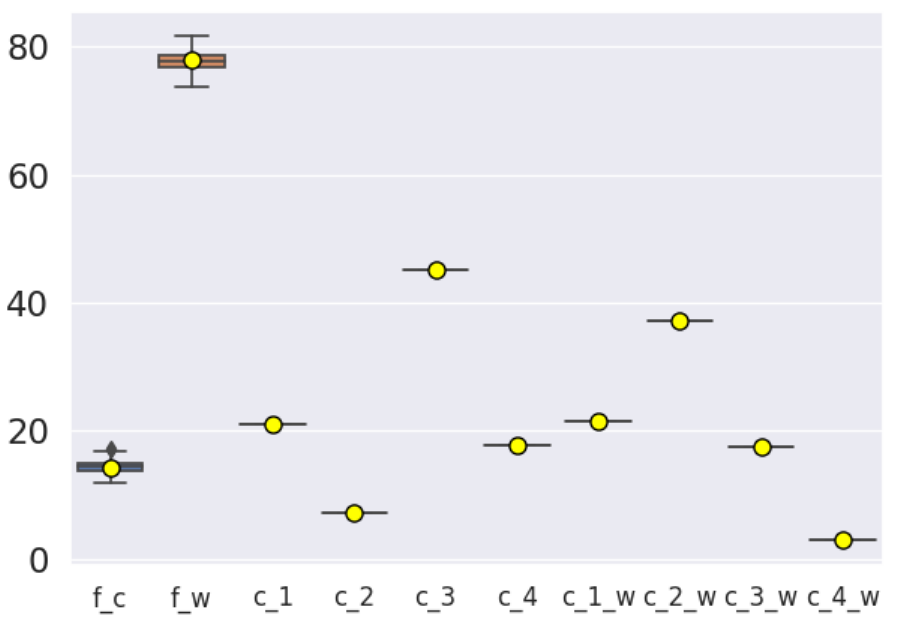}}
\subfigure[GBC]{\label{LIME_500_mob_GBC}\includegraphics[width=4.25cm, height=3.25cm]{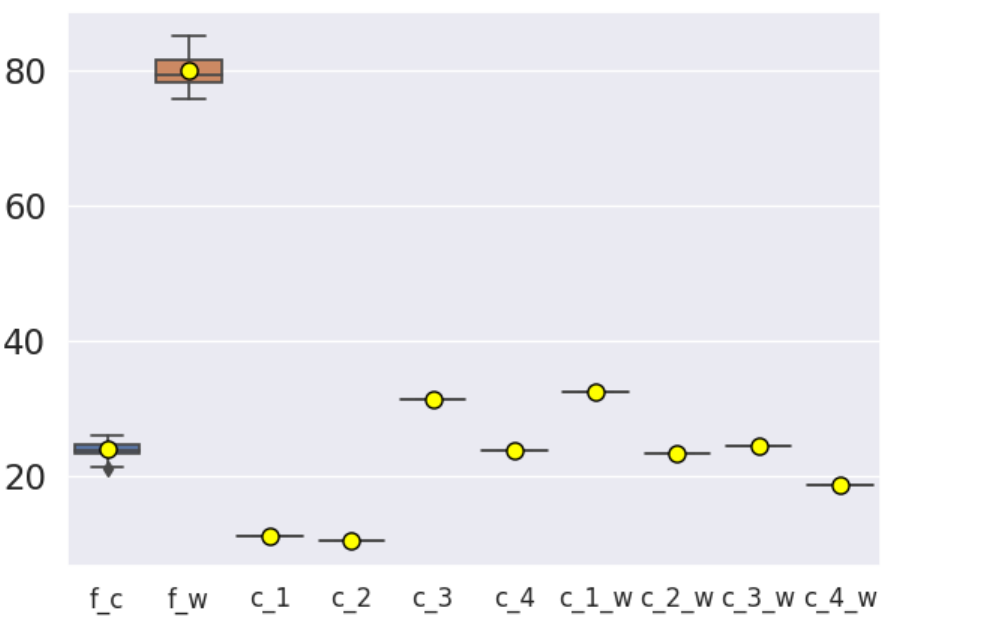}}
\caption{The distribution of $\%Reversed$, PCPD, PCPI, NCPD, and NCPI metrics, encompassing both correct and wrong predictions using LIME on the cross-project mobile apps dataset. Here, $f\_c$ and $f\_w$ represent the $\%Reversed$ metric for correct and wrong predictions, respectively, while $c\_1, c\_2, c\_3, c\_4$ and $c\_1\_w, c\_2\_w, c\_3\_w, c\_4\_w$ PCPD, PCPI, NCPD, NCPI evaluation metrics for correct and wrong predictions, respectively.}
\label{mob_lime_500}
\vspace{-1.25em}
\end{figure}

\begin{figure}[htbp]
\centering     
\subfigure[LR]{\label{LIME_500_java_LR}\includegraphics[width=4.25cm, height=3.25cm]{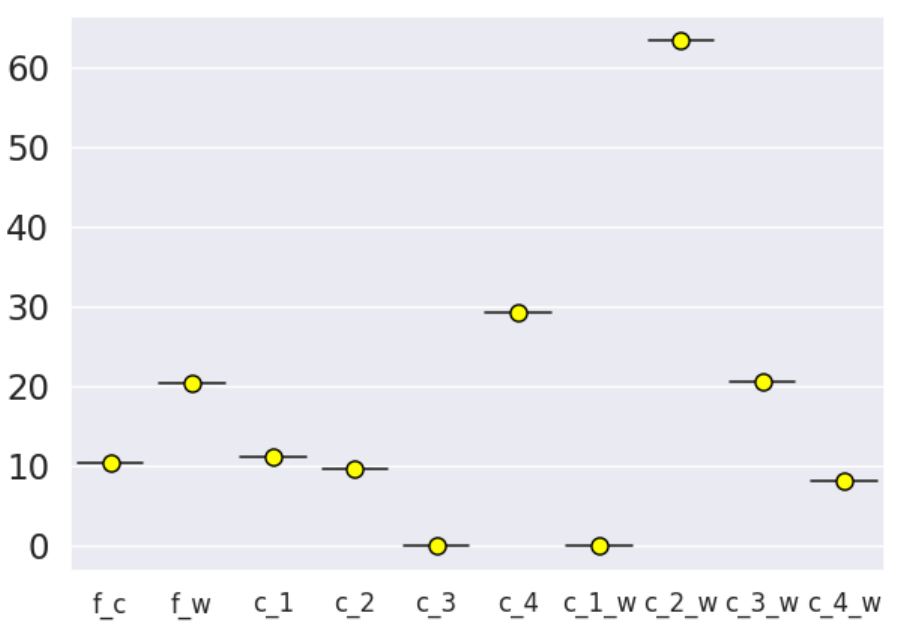}} 
\subfigure[DT]{\label{LIME_500_java_DT}\includegraphics[width=4.25cm, height=3.25cm]{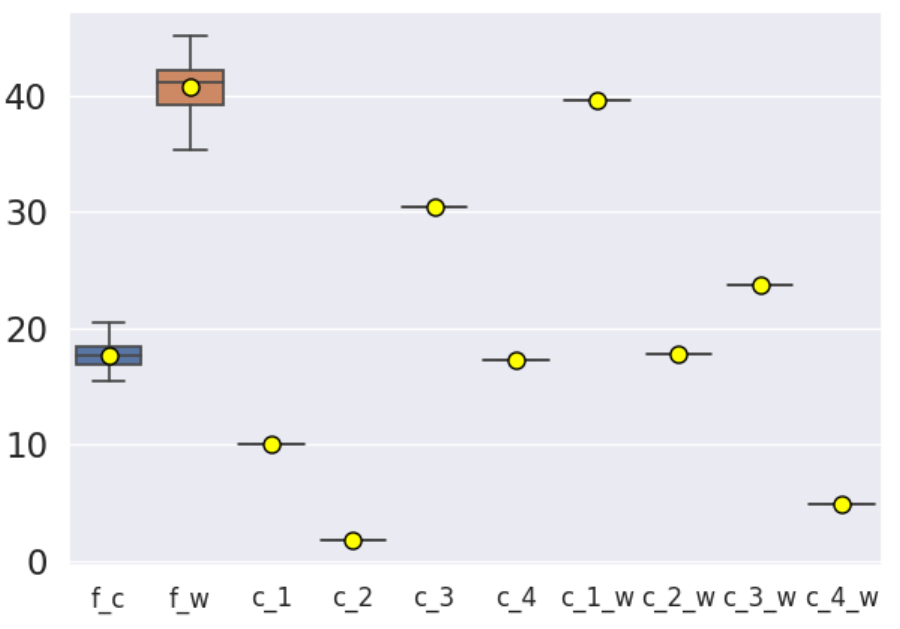}}
\subfigure[RF]{\label{LIME_500_java_RF}\includegraphics[width=4.25cm, height=3.25cm]{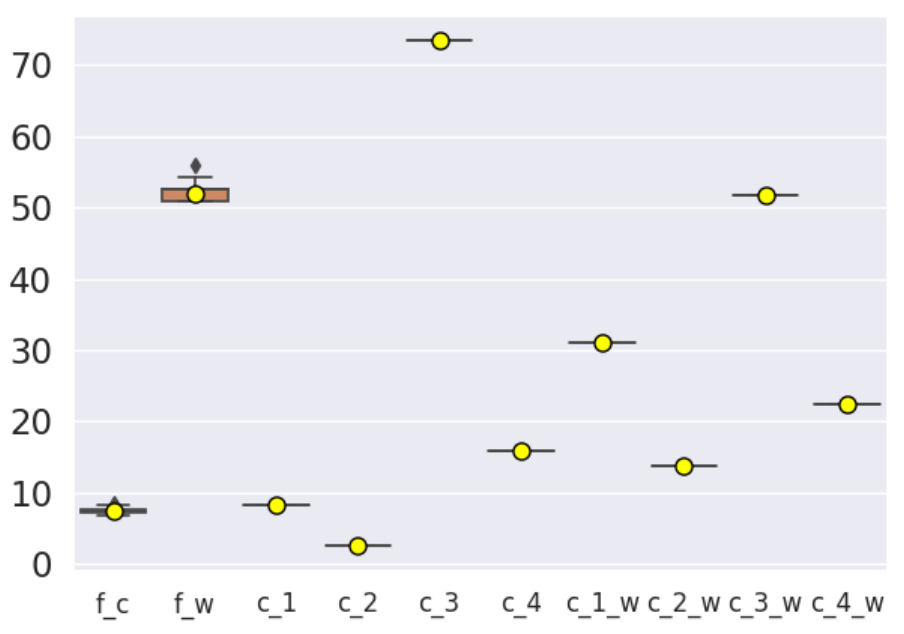}}
\subfigure[MLP]{\label{LIME_500_java_MLP}\includegraphics[width=4.25cm, height=3.25cm]{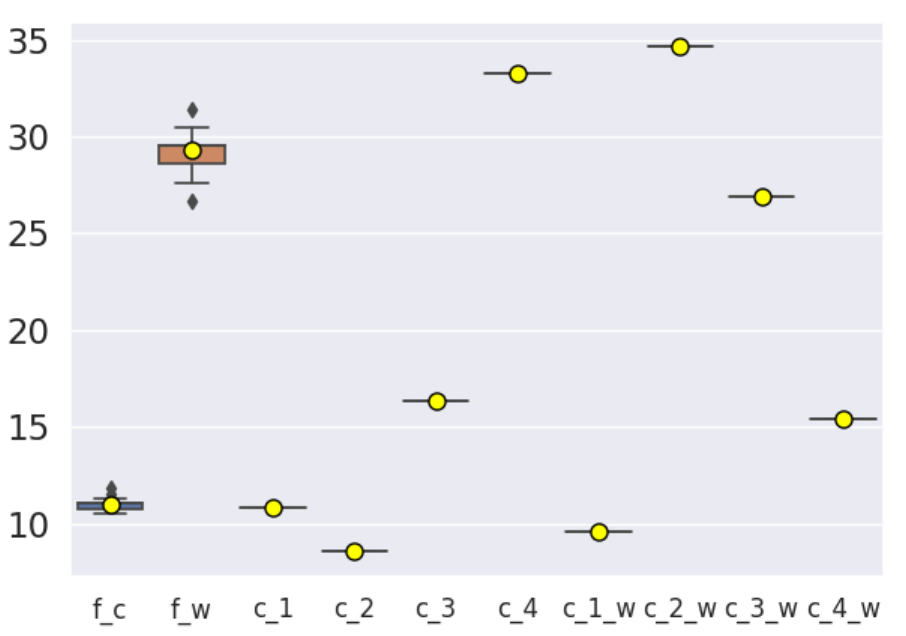}}
\subfigure[ADA]{\label{LIME_500_java_ADA}\includegraphics[width=4.25cm, height=3.25cm]{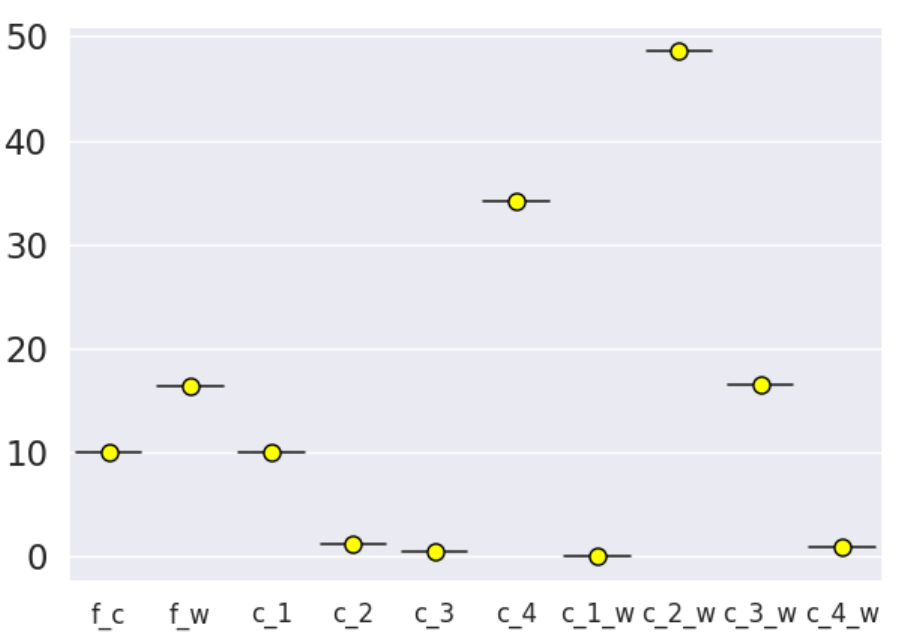}} 
\subfigure[BAG]{\label{LIME_500_java_BAG}\includegraphics[width=4.25cm, height=3.25cm]{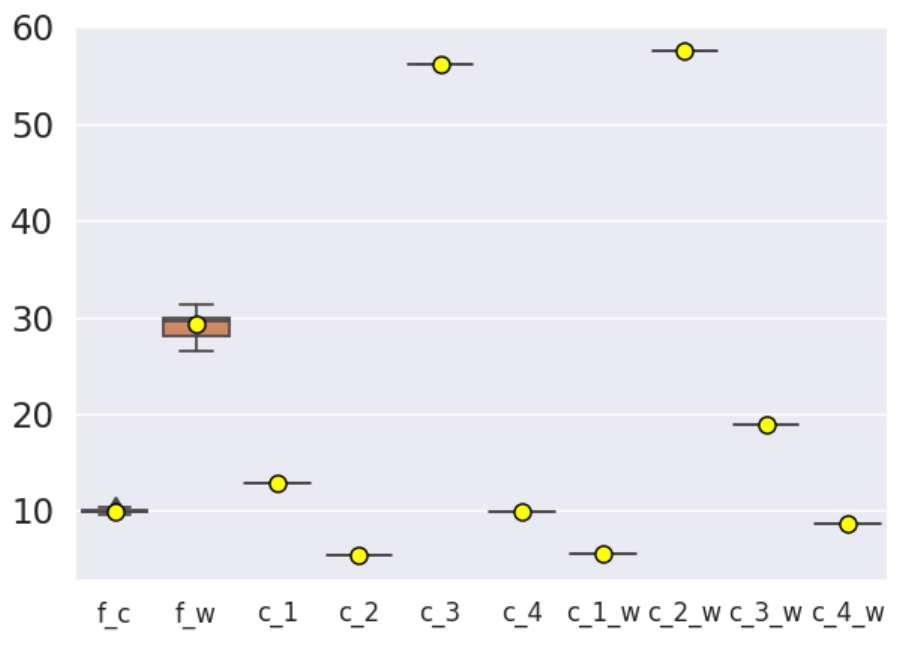}}
\subfigure[GBC]{\label{LIME_500_java_GBC}\includegraphics[width=4.25cm, height=3.25cm]{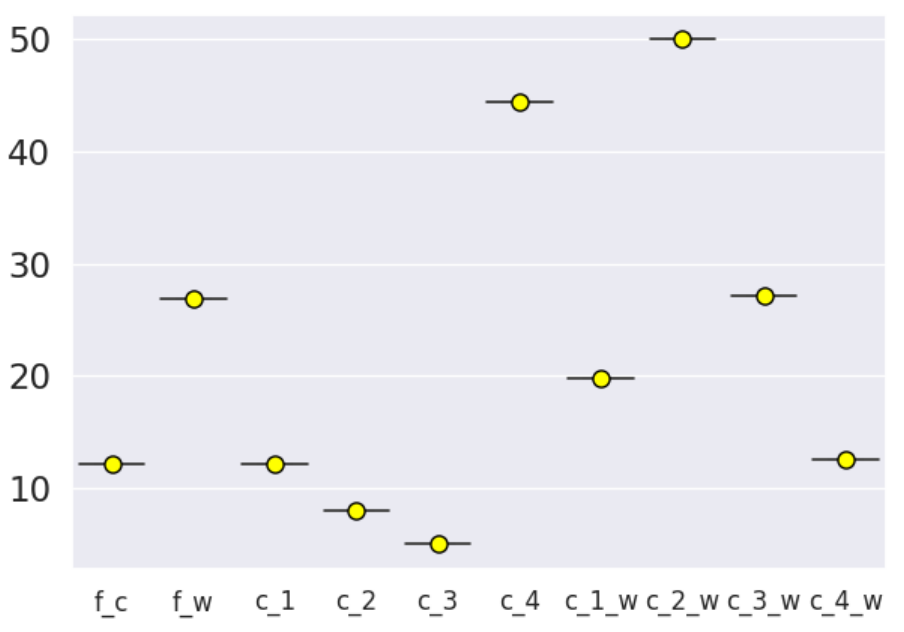}}
\caption{The distribution of $\%Reversed$, PCPD, PCPI, NCPD, and NCPI metrics, encompassing both correct and wrong predictions using LIME on the Java project dataset. Here, $f\_c$ and $f\_w$ represent the $\%Reversed$ metric for correct and wrong predictions, respectively, while $c\_1, c\_2, c\_3, c\_4$ and $c\_1\_w, c\_2\_w, c\_3\_w, c\_4\_w$ PCPD, PCPI, NCPD, NCPI evaluation metrics for correct and wrong predictions, respectively.}
\label{java_lime_500}
\vspace{-1.25em}
\end{figure}

Jiarpakdee et al. \cite{jiarpakdee2020empirical} regenerated instance explanations 100 times by randomly selecting only one instance to observe the consistency of the explanations generated by \textit{LIME}, \textit{LIME-HPO}, and \textit{BreakDown} for JIT defect prediction datasets regarding rank difference metric. In our study, we adopt a similar approach but consider 500 test instances instead of just one to evaluate the consistency of PyExplainer and LIME in explanation generation using \textit{EvaluateXAI}. We calculate the $\%Reversed$ metric and values for four granular-level (e.g., PCPD, PCPI, NCPD, NCPI) metrics for 100 iterations to plot the distribution as shown in Figures \ref{Mob_100_Instance}, \ref{Java_100_Instance}, \ref{mob_lime_500}, and \ref{java_lime_500} and observe any differences. Additionally, we consider three different executions with varying $\alpha$ values: $\alpha = 1$, $\alpha = 2$, and $\alpha = 3$ to calculate the $\%Prob\_diff$ metric for the CLCDSA and code review datasets. Keeping similarity with Jiarpakdee et al. \cite{jiarpakdee2020empirical}, we consider two defect prediction datasets, such as cross-project mobile apps and Java project, to assess the consistency of PyExplainer and LIME in explanation generation regarding the evaluation metrics except $\%Prob\_diff$.

First, we analyze the variation in $\%Reversed$ metric values across different executions for correct and wrong predictions. The distribution of $\%Reversed$ metric values in Figures \ref{Mob_100_Instance} and \ref{Java_100_Instance} reveals notable differences, irrespective of the studied datasets and ML models, when PyExplainer is employed for explanation generation. Similar patterns are observed with LIME, as depicted in Figures \ref{mob_lime_500} and \ref{java_lime_500}, with a few exceptions. For example, LR and ADA models trained on the cross-project mobile apps dataset show no difference in $\%Reversed$ metric values for wrong predictions. Similarly, LR, ADA, and GBC models trained on the Java project datasets exhibit no variation in $\%Reversed$ metric values for both correct and wrong predictions. Regarding the $\%Reversed$ metric, PyExplainer does not generate consistent explanations, while LIME demonstrates relatively more consistency, although it is not entirely consistent in explanation generation.

Secondly, we examine whether the $\%Prob\_diff$ metric values differ across multiple executions for correct predictions. By visually inspecting Figures \ref{LR_fig_cort_CLCDSA} and \ref{LR_fig_cort_CR}, we observe differences among the executions for the ML models trained on the CLCDSA and code review datasets when we apply PyExplainer to the simulated instances. With LIME, we observe differences only for the LR model trained on the code review dataset. Although it is difficult to discern the differences among the executions for the ML models by visual inspection, we quantitatively measure the differences using the Wilcoxon signed-rank test and the magnitude of the difference with Cliff's $|\delta|$ effect size. We interpret the Cliff's $|\delta|$ effect size as shown in Table \ref{cliff_delta_table}. Table \ref{p_val} presents the corresponding \textit{p}-values and Cliff's $|\delta|$ effect size for assessing the differences.

\begin{table}[htbp]
\centering
\small
\caption{Interpretation of Cliff's $|\delta|$ effect size.}
\begin{tabular}{|l|l|}
\hline
\multicolumn{1}{|c|}{\textbf{Interval}} & \multicolumn{1}{c|}{\textbf{Interpret Effect}} \\ \hline
$\delta < 0.147$                        & Negligible effect                                 \\ \hline
$0.147 \leq  0.33$                      & Small effect                                   \\ \hline
$0.33 \leq  0.474$                      & Medium effect                                  \\ \hline
$\delta \geq 0.474$                     & Large effect                                   \\ \hline
\end{tabular}
\label{cliff_delta_table}
\vspace{-1.0em}
\end{table}

\begin{table*}[t]
\caption{$p$\_values and Cliff's $|\delta|$ effect size between different executions on the same instance across different datasets under the same $\alpha$ value. Here, $R\_i\_R\_j$ denotes the difference between $i^{th}$ and $j^{th}$ execution.}
\centering
\resizebox{\linewidth}{!}{
\begin{tabular}{l|l|l|llllllllllll}
\hline
                                                                                    &                                                                               &                                                                             & \multicolumn{12}{c}{\textbf{Dataset}}                                                                                                                                                                                                                                                                                                                                                                                                                                                                                                                                                                                                                                                              \\ \cline{4-15} 
                                                                                    &                                                                               &                                                                             & \multicolumn{6}{c|}{\textbf{CLCDSA}}                                                                                                                                                                                                                                                                                             & \multicolumn{6}{c}{\textbf{Code   review}}                                                                                                                                                                                                                                                                                                                      \\ \cline{4-15} 
                                                                                    &                                                                               &                                                                             & \multicolumn{2}{c|}{\textbf{R\_1\_R\_2}}                                                                                  & \multicolumn{2}{c|}{\textbf{R\_1\_R\_3}}                                                                                 & \multicolumn{2}{c|}{\textbf{R\_2\_R\_3}}                                  & \multicolumn{2}{c|}{\textbf{R\_1\_R\_2}}                                                                                  & \multicolumn{2}{c|}{\textbf{R\_1\_R\_3}}                                                                                  & \multicolumn{2}{c}{\textbf{R\_2\_R\_3}}                                                                 \\ \cline{4-15} 
\multirow{-4}{*}{\textbf{\begin{tabular}[c]{@{}l@{}}XAI \\ Technique\end{tabular}}} & \multirow{-4}{*}{\textbf{\begin{tabular}[c]{@{}l@{}}ML\\ Model\end{tabular}}} & \multirow{-4}{*}{\begin{tabular}[c]{@{}l@{}}$\alpha$\\ Values\end{tabular}} & \multicolumn{1}{c|}{p\_val}                                  & \multicolumn{1}{c|}{Cliff’s   $|\delta|$}                  & \multicolumn{1}{c|}{p\_val}                                 & \multicolumn{1}{c|}{Cliff’s   $|\delta|$}                  & \multicolumn{1}{c|}{P\_val}   & \multicolumn{1}{c|}{Cliff’s   $|\delta|$} & \multicolumn{1}{c|}{p\_val}                                  & \multicolumn{1}{c|}{Cliff’s   $|\delta|$}                  & \multicolumn{1}{c|}{p\_val}                                  & \multicolumn{1}{c|}{Cliff’s   $|\delta|$}                  & \multicolumn{1}{c|}{p\_val}                                  & \multicolumn{1}{c}{Cliff’s   $|\delta|$} \\ \hline
                                                                                    &                                                                               & $\alpha = 1$                                                                  & \multicolumn{1}{l|}{0.00058}                                 & \multicolumn{1}{l|}{0.22}                                  & \multicolumn{1}{l|}{0.00099}                                & \multicolumn{1}{l|}{0.31}                                  & \multicolumn{1}{l|}{0.00025}  & \multicolumn{1}{l|}{0.11}        & \multicolumn{1}{l|}{0.00043}                                 & \multicolumn{1}{l|}{0.23}                                  & \multicolumn{1}{l|}{0.00033}                                 & \multicolumn{1}{l|}{0.29}                                  & \multicolumn{1}{l|}{0.00054}                                 & 0.11                                      \\ \cline{3-15} 
                                                                                    &                                                                               & $\alpha = 2$                                                                  & \multicolumn{1}{l|}{0.0002}                                  & \multicolumn{1}{l|}{0.23}                                  & \multicolumn{1}{l|}{0.00015}                                & \multicolumn{1}{l|}{0.33}                                  & \multicolumn{1}{l|}{0.000022} & \multicolumn{1}{l|}{0.11}        & \multicolumn{1}{l|}{0.00011}                                 & \multicolumn{1}{l|}{0.15}                                  & \multicolumn{1}{l|}{0.00045}                                 & \multicolumn{1}{l|}{0.26}                                  & \multicolumn{1}{l|}{0.00011}                                 & 0.15                                      \\ \cline{3-15} 
                                                                                    & \multirow{-3}{*}{\textbf{LR}}                                                 & $\alpha = 3$                                                                  & \multicolumn{1}{l|}{0.00001}                                 & \multicolumn{1}{l|}{0.25}                                  & \multicolumn{1}{l|}{0.000017}                               & \multicolumn{1}{l|}{0.34}                                  & \multicolumn{1}{l|}{0.000033} & \multicolumn{1}{l|}{0.12}        & \multicolumn{1}{l|}{0.00051}                                 & \multicolumn{1}{l|}{0.18}                                  & \multicolumn{1}{l|}{0.00014}                                 & \multicolumn{1}{l|}{0.29}                                  & \multicolumn{1}{l|}{0.00026}                                 & 0.14                                      \\ \cline{2-15} 
                                                                                    &                                                                               & $\alpha = 1$                                                                  & \multicolumn{1}{l|}{0.00039}                                 & \multicolumn{1}{l|}{0.02}                                  & \multicolumn{1}{l|}{0.00069}                                & \multicolumn{1}{l|}{0.03}                                  & \multicolumn{1}{l|}{0.00018}  & \multicolumn{1}{l|}{0.01}                 & \multicolumn{1}{l|}{\cellcolor[HTML]{D3D3D3}\textbf{0.7464}} & \multicolumn{1}{l|}{\cellcolor[HTML]{D3D3D3}\textbf{0.01}} & \multicolumn{1}{l|}{\cellcolor[HTML]{D3D3D3}\textbf{0.2012}} & \multicolumn{1}{l|}{\cellcolor[HTML]{D3D3D3}\textbf{0.04}} & \multicolumn{1}{l|}{\cellcolor[HTML]{D3D3D3}\textbf{0.0665}} & \cellcolor[HTML]{D3D3D3}\textbf{0.03}     \\ \cline{3-15} 
                                                                                    &                                                                               & $\alpha = 2$                                                                  & \multicolumn{1}{l|}{0.00011}                                 & \multicolumn{1}{l|}{0.02}                                  & \multicolumn{1}{l|}{0.00068}                                & \multicolumn{1}{l|}{0.03}                                  & \multicolumn{1}{l|}{0.00026}  & \multicolumn{1}{l|}{0.01}                 & \multicolumn{1}{l|}{\cellcolor[HTML]{D3D3D3}\textbf{0.0643}} & \multicolumn{1}{l|}{\cellcolor[HTML]{D3D3D3}\textbf{0.02}} & \multicolumn{1}{l|}{0.017}                                   & \multicolumn{1}{l|}{0.04}                                  & \multicolumn{1}{l|}{\cellcolor[HTML]{D3D3D3}\textbf{0.0972}} & \cellcolor[HTML]{D3D3D3}\textbf{0.024}    \\ \cline{3-15} 
\multirow{-6}{*}{\textbf{PyExplainer}}                                              & \multirow{-3}{*}{\textbf{RF}}                                                 & $\alpha = 3$                                                                  & \multicolumn{1}{l|}{0.00033}                                 & \multicolumn{1}{l|}{0.01}                                  & \multicolumn{1}{l|}{0.00052}                                & \multicolumn{1}{l|}{0.02}                                  & \multicolumn{1}{l|}{0.00073}  & \multicolumn{1}{l|}{0.08}                 & \multicolumn{1}{l|}{0.029}                                   & \multicolumn{1}{l|}{0.036}                                 & \multicolumn{1}{l|}{0.0037}                                  & \multicolumn{1}{l|}{0.068}                                 & \multicolumn{1}{l|}{0.0036}                                  & 0.03                                      \\ \hline
                                                                                    &                                                                               & $\alpha = 1$                                                                  & \multicolumn{1}{l|}{0.00016}                                 & \multicolumn{1}{l|}{0.04}                                  & \multicolumn{1}{l|}{0.00272}                                & \multicolumn{1}{l|}{0.05}                                  & \multicolumn{1}{l|}{0.00061}  & \multicolumn{1}{l|}{0.01}                 & \multicolumn{1}{l|}{\cellcolor[HTML]{D3D3D3}\textbf{0.0536}} & \multicolumn{1}{l|}{\cellcolor[HTML]{D3D3D3}\textbf{0.01}} & \multicolumn{1}{l|}{0.0038}                                  & \multicolumn{1}{l|}{0.001}                                 & \multicolumn{1}{l|}{0.0003}                                  & 0.02                                      \\ \cline{3-15} 
                                                                                    &                                                                               & $\alpha = 2$                                                                  & \multicolumn{1}{l|}{0.00045}                                 & \multicolumn{1}{l|}{0.03}                                  & \multicolumn{1}{l|}{0.00065}                                & \multicolumn{1}{l|}{0.04}                                  & \multicolumn{1}{l|}{0.00037}  & \multicolumn{1}{l|}{0.01}                 & \multicolumn{1}{l|}{0.035}                                   & \multicolumn{1}{l|}{0.006}                                 & \multicolumn{1}{l|}{0.0062}                                  & \multicolumn{1}{l|}{0.006}                                 & \multicolumn{1}{l|}{0.0014}                                  & 0.013                                     \\ \cline{3-15} 
                                                                                    & \multirow{-3}{*}{\textbf{LR}}                                                 & $\alpha = 3$                                                                  & \multicolumn{1}{l|}{0.00032}                                 & \multicolumn{1}{l|}{0.03}                                  & \multicolumn{1}{l|}{0.00041}                                & \multicolumn{1}{l|}{0.04}                                  & \multicolumn{1}{l|}{0.00048}  & \multicolumn{1}{l|}{0.01}                 & \multicolumn{1}{l|}{0.0250}                                  & \multicolumn{1}{l|}{0.003}                                 & \multicolumn{1}{l|}{0.0051}                                  & \multicolumn{1}{l|}{0.002}                                 & \multicolumn{1}{l|}{0.0016}                                  & 0.02                                      \\ \cline{2-15} 
                                                                                    &                                                                               & $\alpha = 1$                                                                  & \multicolumn{1}{l|}{\cellcolor[HTML]{D3D3D3}\textbf{0.237}}  & \multicolumn{1}{l|}{\cellcolor[HTML]{D3D3D3}\textbf{0.01}} & \multicolumn{1}{l|}{\cellcolor[HTML]{D3D3D3}\textbf{0.621}} & \multicolumn{1}{l|}{\cellcolor[HTML]{D3D3D3}\textbf{0.01}} & \multicolumn{1}{l|}{0.0169}   & \multicolumn{1}{l|}{0.001}                & \multicolumn{1}{l|}{0.015}                                   & \multicolumn{1}{l|}{0.05}                                  & \multicolumn{1}{l|}{0.00046}                                 & \multicolumn{1}{l|}{0.07}                                  & \multicolumn{1}{l|}{0.0315}                                  & 0.02                                      \\ \cline{3-15} 
                                                                                    &                                                                               & $\alpha = 2$                                                                  & \multicolumn{1}{l|}{\cellcolor[HTML]{D3D3D3}\textbf{0.8455}} & \multicolumn{1}{l|}{\cellcolor[HTML]{D3D3D3}\textbf{0.01}} & \multicolumn{1}{l|}{\cellcolor[HTML]{D3D3D3}\textbf{0.378}} & \multicolumn{1}{l|}{\cellcolor[HTML]{D3D3D3}\textbf{0.01}} & \multicolumn{1}{l|}{0.0174}   & \multicolumn{1}{l|}{0.01}                 & \multicolumn{1}{l|}{0.0244}                                  & \multicolumn{1}{l|}{0.046}                                 & \multicolumn{1}{l|}{0.00039}                                 & \multicolumn{1}{l|}{0.067}                                 & \multicolumn{1}{l|}{0.0185}                                  & 0.02                                      \\ \cline{3-15} 
\multirow{-6}{*}{\textbf{LIME}}                                                     & \multirow{-3}{*}{\textbf{RF}}                                                 & $\alpha = 3$                                                                  & \multicolumn{1}{l|}{\cellcolor[HTML]{D3D3D3}\textbf{0.419}}  & \multicolumn{1}{l|}{\cellcolor[HTML]{D3D3D3}\textbf{0.01}} & \multicolumn{1}{l|}{\cellcolor[HTML]{D3D3D3}\textbf{0.524}} & \multicolumn{1}{l|}{\cellcolor[HTML]{D3D3D3}\textbf{0.01}} & \multicolumn{1}{l|}{0.00045}  & \multicolumn{1}{l|}{0.01}                 & \multicolumn{1}{l|}{\cellcolor[HTML]{D3D3D3}\textbf{0.072}}  & \multicolumn{1}{l|}{\cellcolor[HTML]{D3D3D3}\textbf{0.03}} & \multicolumn{1}{l|}{0.00082}                                 & \multicolumn{1}{l|}{0.058}                                 & \multicolumn{1}{l|}{0.0036}                                  & 0.026                                    \\ \hline
\end{tabular}}
\label{p_val}
\end{table*}

To evaluate the consistency of the explanations generated by rule-based XAI techniques, we start with PyExplainer and subsequently present the experimental results of LIME. From Table \ref{p_val}, we see that in the case of the RF model trained on the code review dataset, there is no significant $\%Prob\_diff$ metric difference when the value of $\alpha$ is 1 for multiple executions. For the same model, we observe a difference when the value of $\alpha$ is $1$ between the first and third executions, although the effect size is small. In contrast, for all other cases, the differences between multiple executions are clearly visible despite the small effect size values. Moving to LIME, we find no significant $\%Prob\_diff$ metric difference between the first and second executions for the RF model trained on the CLCDSA dataset. Furthermore, for both models trained on the code review dataset, there is no significant difference between the first and second executions. On the other hand, in all other cases, differences between multiple executions are noticeable despite having negligible and small effect size values. Lundberg et al. \cite{lundberg2017unified} argued that ``instance explanation generation must remain consistent upon regeneration for the same instance," emphasizing that even negligible or small effect sizes can impact the consistency of explanations generated by XAI techniques. However, when considering the $\%Prob\_diff$ metric, the experimental results mentioned above appear to contradict this statement.

Finally, we expect PyExplainer and LIME to produce the same output for other granular-level (e.g., PCPD, PCPI, NCPD, NCPI) metrics for multiple executions, considering correct and wrong predictions. However, Figures \ref{Mob_100_Instance} and \ref{Java_100_Instance} demonstrate that regardless of the studied datasets and ML models, PyExplainer generates inconsistent instance explanations. Upon visual inspection of these figures, it becomes evident that there are noticeable differences in the four granular-level evaluation metrics (e.g., PCPD, PCPI, NCPD, NCPI), both for correct and wrong predictions. In contrast, as depicted in Figures \ref{mob_lime_500} and \ref{java_lime_500}, we observe consistent values for PCPD, PCPI, NCPD, and NCPI metrics, indicating that LIME exhibits greater consistency in generating explanations for ML models in software analytics. Table \ref{p_val} quantitatively demonstrates that PyExplainer provides $(31/36)*100=86.11\%$ inconsistent explanations for all the test instances of CLCDSA and code review datasets. In contrast, LIME provides $(28/36)*100=77.78\%$ inconsistent explanations.

\begin{figure*}[t]
\centering     
\subfigure[LR]{\label{1_LR_JavaPrject_Balanced_Granular}\includegraphics[width=4.25cm, height=3.25cm]{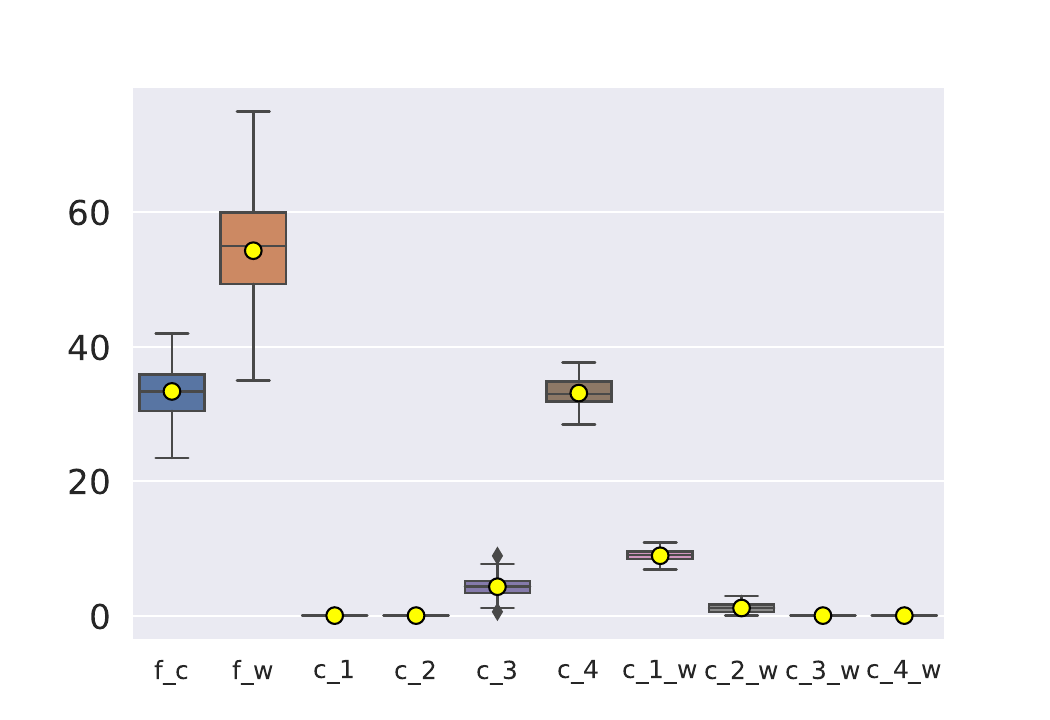}} 
\subfigure[DT]{\label{2_DT_JavaPrject_Balanced_Granular}\includegraphics[width=4.25cm, height=3.25cm]{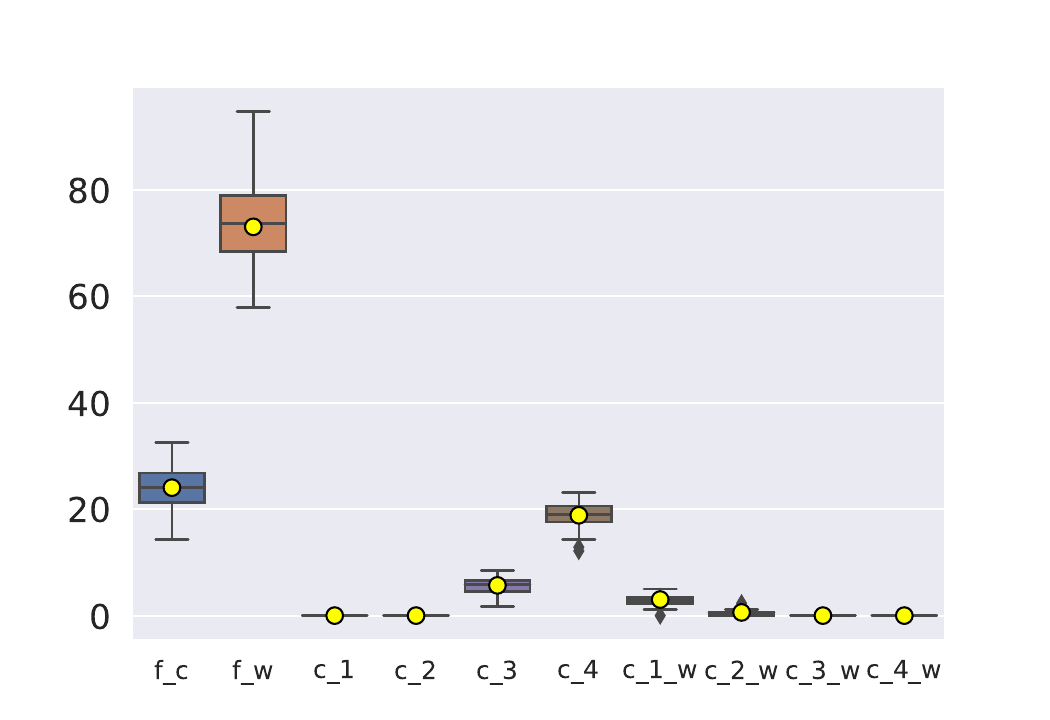}}
\subfigure[RF]{\label{3_RF_JavaPrject_Balanced_Granular}\includegraphics[width=4.25cm, height=3.25cm]{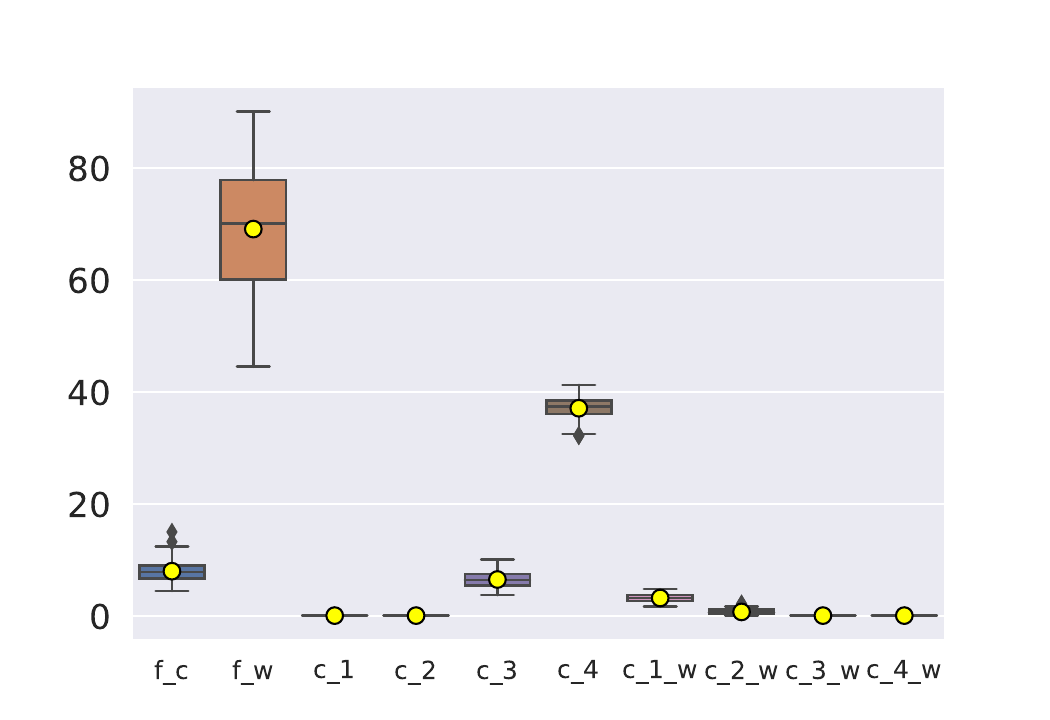}}
\vspace{-0.5em}
\subfigure[MLP]{\label{4_MLP_JavaPrject_Balanced_Granular}\includegraphics[width=4.25cm, height=3.25cm]{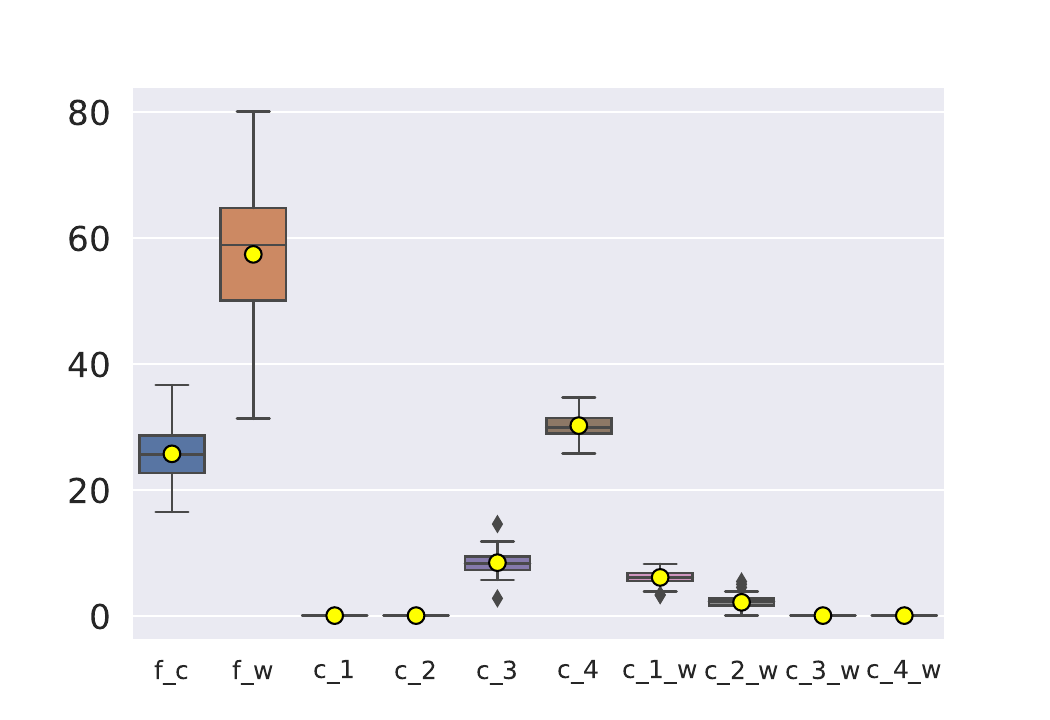}}
\subfigure[ADA]{\label{5_ADA_JavaPrject_Balanced_Granular}\includegraphics[width=4.25cm, height=3.25cm]{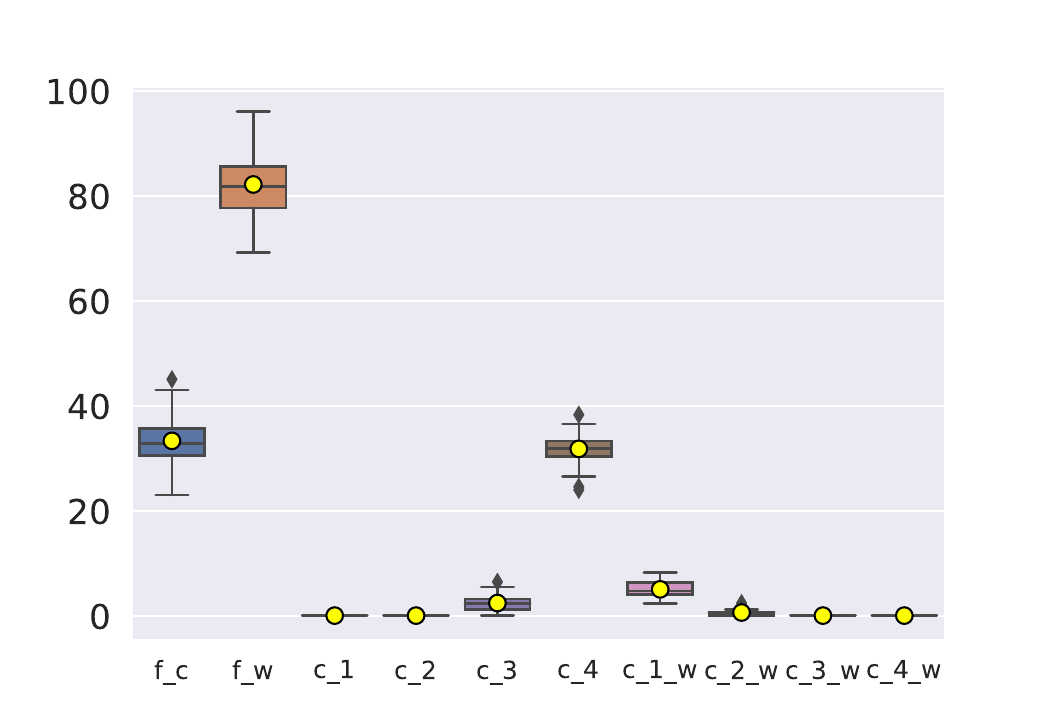}} 
\subfigure[BAG]{\label{6_BAG_JavaPrject_Balanced_Granular}\includegraphics[width=4.25cm, height=3.25cm]{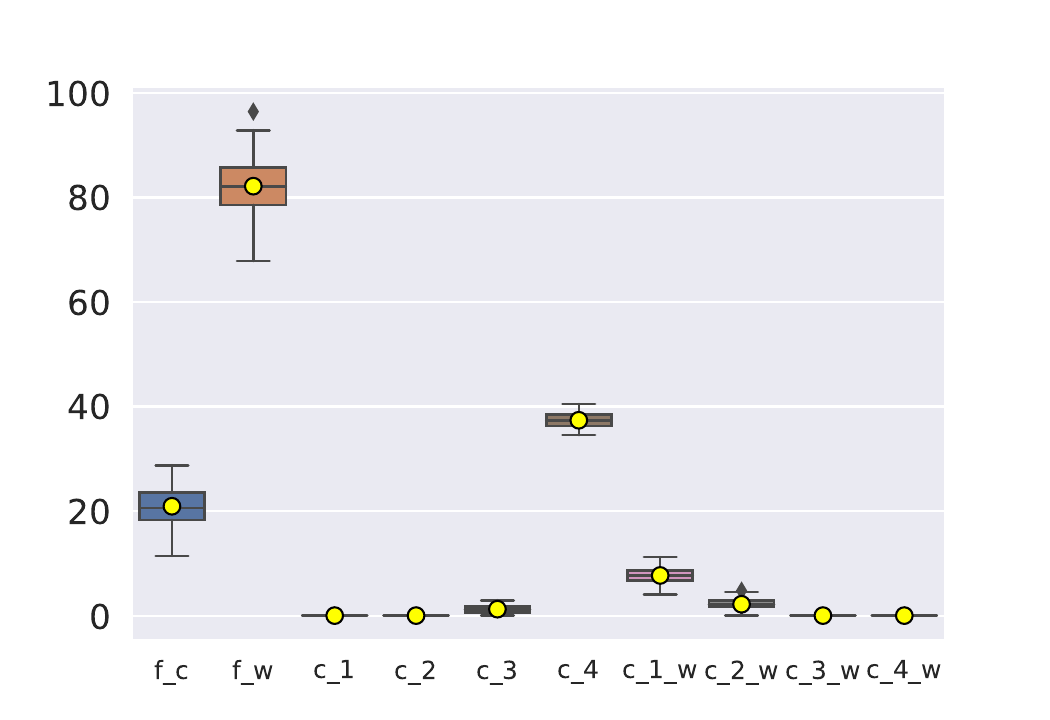}}
\vspace{-0.5em}
\subfigure[GBC]{\label{7_GBC_JavaPrject_Balanced_Granular}\includegraphics[width=4.25cm, height=3.25cm]{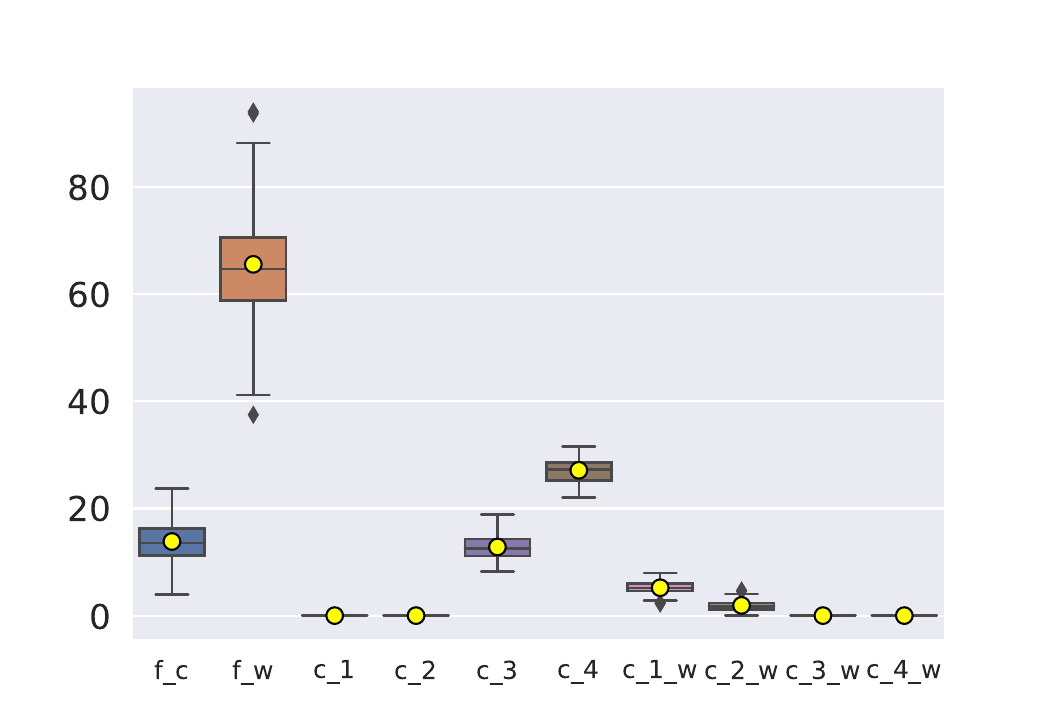}}
\caption{The distribution of $\%Reversed$, PCPD, PCPI, NCPD, and NCPI metrics, encompassing both correct and wrong predictions using PyExplainer on the Java project dataset. Here, $f\_c$ and $f\_w$ represent the $\%Reversed$ metric for correct and wrong predictions, respectively, while $c\_1, c\_2, c\_3, c\_4$ and $c\_1\_w, c\_2\_w, c\_3\_w, c\_4\_w$ PCPD, PCPI, NCPD, NCPI evaluation metrics for correct and wrong predictions, respectively.}
\label{balance_consistency_all_metric}
\end{figure*}

Similar to evaluating the reliability of the explanations generated by PyExplainer using the balanced test set of the Java project dataset for addressing RQ1, we conduct the same experiments originally performed on the imbalanced test dataset to assess consistency using the balanced test data. Figure \ref{balance_consistency_all_metric} shows the distributions of $\%Reversed$, PCPD, PCPI, NCPD, and NCPI metrics. From this figure, we observe notable differences across multiple executions, considering different metrics, irrespective of the studied ML models when PyExplainer is employed for explanation generation. We also observe similar results for the $\%Prob\_diff$ metric with the balanced test set of the Java project dataset. The detailed results can be found in our replication package, following the instructions in the README.md file. Therefore, despite using the balanced test set, PyExplainer still produces inconsistent explanations.

Therefore, using \textit{EvaluateXAI} and considering the \textit{\%Reversed}, $\%Prob\_diff$ and four granular-level evaluation metrics (e.g., PCPD, PCPI, NCPD, NCPI), we can conclude that PyExplainer does not provide consistent explanations for ML models in software analytics tasks. In addition, although LIME offers a bit more consistent explanation regarding granular-level evaluation metrics, its inconsistency can be assessed using \textit{EvaluateXAI} by considering the $\%Reversed$ and $\%Prob\_diff$ metrics.


\vspace{1em}
\noindent
\fbox{\begin{minipage}{24em}
\textbf{Result RQ2}: In the case of 36 experimental combinations involving the CLCDSA and code review datasets, PyExplainer and LIME provide inconsistent explanations for 31 and 28 cases, representing 86.11\% and 77.78\%, respectively. These findings corroborate the inconsistent explanations produced by PyExplainer and LIME for ML models in software analytics tasks.
\end{minipage}}
\vspace{1em}

\section{Key Findings \& Guidelines}
\label{disc}
In this study, we design a novel framework called \textit{EvaluateXAI}, along with six evaluation metrics, to evaluate the effectiveness of rule-based XAI techniques in generating reliable and consistent explanations for ML models trained on tabular data. We summarize the key findings from our study as follows:

\begin{enumerate}
    \item PyExplainer and LIME generate completely different explanations, as shown in Figures \ref{pyexp explanation} and \ref{lime_anomaly} for the same instance regarding multiple executions. These inconsistent explanations may confuse its practitioners and discourage them from using PyExplainer and LIME to explain the outcome of JIT defect prediction models. Therefore, further research is necessary to generate consistent explanations across multiple executions of the same instance.

    \item While manually investigating the consistency of the explanations generated by PyExplainer and LIME, we found that LIME generates more consistent rules in each explanation than PyExplainer. For instance, in Figures \ref{lime_a} and \ref{lime_b}, we observe two different explanations from LIME for the same instance. However, the generated rules are the same, with only a change in orientation. In contrast, Figure \ref{pyexp explanation} shows that PyExplainer produces four different explanations with distinct rules for the same instance. Additionally, we observe the rule values are different, as shown in Figures \ref{motivation_a} and \ref{motivation_b}. For example, in Figure \ref{motivation_a}, PyExplainer generates the explanation where the feature value \textit{LOC} is less than $11.0$. On the other hand, in Figure \ref{motivation_b}, PyExplainer generates the explanation where the feature value \textit{LOC} is more than $11.0$. Our experimental results also indicate that LIME provides more consistent rules and explanations than PyExplainer across four granular-level evaluation metrics: PCPD, PCPI, NCPD, and NCPI. Therefore, generating consistent rules is crucial for ensuring the effectiveness of XAI techniques in practical applications.
    
    \item Our study only considers the rule-based XAI techniques PyExplainer and LIME. We adopt two metrics from the existing study by Pornprasit et al. \cite{pornprasit2021pyexplainer} and design four other granular-level evaluation metrics to evaluate the reliability and consistency of generated explanations by PyExplainer and LIME. However, the in-depth experimental results possibly threaten the effectiveness of the PyExplainer and LIME in providing reliable and consistent explanations for the JIT defect prediction models. In addition, the empirical findings also demonstrate that PyExplainer does not always perform well in generating explanations for the ML models trained on the CLCDSA and code review datasets, meaning it is less generalizable. Therefore, further actions are warranted to advance the XAI research in software analytics. 
    
    \item Pornprasit et al. \cite{pornprasit2021pyexplainer} changed the original feature value by one STD (a standard deviation of that particular feature found from the training dataset) from the rule threshold. We also apply the same technique to change the feature values by one STD while generating simulated instances. However, there is a high chance of generating invalid feature values. For example, for the given instance, as shown in Table \ref{instance}, the value of \textit{LOC} feature is $11.0$, and the approximate threshold value is $16$ found from the explanation as shown in Figure \ref{motivation_b}. Now, consider one STD value of this feature is $50$, and for ``\textit{more than}" rule, we have to subtract one STD from the rule threshold value to generate a simulated instance. Any further subtraction operation will yield a negative feature value (e.g., $16-50=-34$ ), which is not accepted as the \textit{Line of Code (LOC)} cannot be negative in a file. Therefore, further research is essential for the effective generation of simulated instances.
    
    \item We design \textit{EvaluateXAI} and the granular-level evaluation metrics based on the PyExplainer. However, our simulated instance generation process for LIME, along with comprehensive experimental results, demonstrates that \textit{EvaluateXAI} is adaptable to any rule-based XAI technique. This adaptability allows us to evaluate their effectiveness in generating reliable and consistent explanations for ML models trained on tabular data in software analytics tasks.
\end{enumerate}

Finally, based on the in-depth experiments conducted to evaluate the effectiveness of PyExplainer and LIME in generating reliable and consistent explanations using \textit{EvaluateXAI} and evaluation metrics, we offer the following recommendations to XAI researchers and users:

\begin{enumerate}
    \item Ensure that the generated explanations remain consistent even with multiple executions on the same instance.
    \item Ensure that the generated rules in each explanation remain consistent even with multiple executions on the same instance.
    \item The rule-based XAI techniques must satisfy the granular-level evaluation metrics (e.g., PCPD, PCPI, NCPD, NCPI) outlined in our study to ensure the reliability of the generated explanations.
    \item If practitioners use JIT defect prediction models to prioritize high-risk commits, we suggest exercising caution when allocating SQA resources based on the explanations provided by rule-based XAI techniques (e.g., PyExplainer or LIME).
\end{enumerate}

\section{Threats to Validity}
\label{threat}
This section briefly explains the internal and external threats associated with our research and the methods we employed to address them.

\subsection{Internal Threat}
\label{internal}
The first potential internal threat is the accuracy of the selected ML models. However, we addressed this issue by adopting the parameter and hyperparameter configurations outlined in previous studies \cite{yatish2019mining, pornprasit2021pyexplainer, jiarpakdee2020empirical, roy2022don}. To handle imbalanced datasets and multicollinearity issues in training ML models, we also utilized the SMOTE and Autospearman parameter settings defined in \cite{tantithamthavorn2018impact}. Furthermore, we applied random search, grid search, and Bayesian optimization to select the best hyper-parameter combinations, resulting in highly accurate and non-overfitted ML models. Finally, we tested our trained models based on AUC values, confirming their high accuracy. Thus, we successfully mitigated this internal threat.

Another internal threat could be generating simulated instances to flip predictions, find the prob-diff, and calculate different granular-level evaluation metrics (e.g., PCPD, PCPI, NCPD, NCPI) values. Nevertheless, we addressed this threat by changing the feature values by one standard deviation from the threshold values found by PyExplainer, as described in \cite{pornprasit2021pyexplainer}.

\subsection{External Threat}
\label{external}
The generalizability of our experimental results can be considered a potential threat to the external validity of \textit{EvaluateXAI}. However, we addressed this issue by selecting seven state-of-the-art ML models and five datasets previously used in other studies \cite{catolino2019cross, pornprasit2021pyexplainer, roy2022don, nguyen2023fix}. Additionally, our extensive experiments considered $7 \times 5 = 35$ combinations, indicating that our findings are possibly sufficiently robust to generalize the performance of \textit{EvaluateXAI}. The generalizability of the granular-level evaluation metrics could present an external threat to our study. Nevertheless, we mitigated this threat by adopting two metrics from Pornprasit et al. \cite{pornprasit2021pyexplainer} and designing four additional granular-level metrics based on the inconsistencies and anomalies found in PyExplainer. Furthermore, we demonstrated the adaptability of these metrics in the case of another XAI technique called LIME to assess the reliability and consistency of the explanations generated by LIME. Therefore, we believe that these granular-level evaluation metrics are applicable to any rule-based XAI techniques as long as they adhere to the conventions of the generated rules \cite{almutairi2021reg}.

\section{Related Work}
\label{RW}
There has been a growing interest in Explainable AI (XAI) in recent years, specifically focusing on making ML models interpretable and transparent to practitioners. At the same time, more research is being conducted to propose methods for formally assessing and contrasting various XAI methods. Therefore, obtaining practical knowledge to evaluate interpretability across multiple tools and establish a collaborative agreement among the community for future progress is crucial. This section briefly discusses existing works that aim to evaluate XAI methods.

In computer vision and image processing, many XAI methods such as saliency map \cite{simonyan2013deep}, occlusion sensitivity \cite{zeiler2014visualizing}, integrated gradient \cite{sundararajan2017axiomatic}, Grad-CAM \cite{selvaraju2017grad}, and SmoothGrad \cite{smilkov2017smoothgrad} have been proposed to explain the outcome of ML models. However, Ghorbani et al. \cite{ghorbani2019interpretation} and Kindermans et al. \cite{kindermans2019reliability} demonstrated that it is possible to attack the saliency map methods by manipulating the derived explanations. Samek et al. \cite{samek2016evaluating} and Montavon et al. \cite{montavon2018methods} both suggested a method for evaluating the effectiveness of saliency map methods by introducing perturbations to the input. Adebayo et al. \cite{adebayo2018sanity} proposed a practical framework for assessing the types of explanations the saliency map method is capable and incapable of providing. By conducting thorough experiments, the authors demonstrated that certain saliency map techniques are not dependent on either the model or the process of generating data. The above research evaluates the effectiveness of different saliency map methods in generating explanations in computer vision targeting image data. Additionally, those studies did not consider classical ML models trained on tabular data when assessing the effectiveness of rule-based XAI techniques. However, this study aims to evaluate the reliability and consistency of rule-based XAI techniques in generating explanations for classical ML models trained on tabular data for software analytics tasks.

Doshi-Velez et al. \cite{doshi2017towards} suggested three primary methods for evaluating interpretability: application-based, human-based, and function-based. These methods vary from having human participation to no human involvement at all. Robnik et al. \cite{robnik2018perturbation} proposed several properties (e.g., Accuracy, Fidelity, Consistency, Stability, Comprehensibility, Certainty, Importance, Novelty, and Representatives) to assess the generated explanation in a functional context. However, there is uncertainty about evaluating these properties, and how useful they are for XAI needs to be clarified. Miller et al. \cite{miller2019explanation} conducted an extensive investigation to assess the explanation regarding constructiveness, selectivity, social aspects, emphasis on abnormality, truthfulness, consistency, and generality. Kacper et al. \cite{sokol2020explainability} systematically assessed XAI techniques across five significant dimensions: functional, operational, usability, safety, and validation. Their study involved comparing XAI methods to understand better how they differ. The above studies evaluate different model-agnostic XAI methods considering tabular data and ML models. Elshawi et al. \cite{elshawi2021interpretability} evaluated three model-agnostic interpretability techniques (e.g., SHAP, LIME, and Anchors) on healthcare data using different evaluation metrics: identity, stability, separability, similarity, execution time, and bias detection. Hailemariam et al. \cite{hailemariam2020empirical} conducted an empirical study on SHAP and LIME to gain insight into their operations on Artificial Neural Networks (ANNs) and Convolutional Neural Networks (CNNs). Their findings show that SHAP performed slightly better than LIME regarding Identity, Stability, and Separability. None of the studies described above consider classical ML models trained on tabular data for various software analytics tasks. Furthermore, our proposed novel framework and six evaluation metrics differ from the criteria proposed in the studies above. These criteria include accuracy, fidelity, consistency, stability, comprehensibility, certainty, importance, novelty, and representativeness to assess the generated explanation in a functional context. Finally, we demonstrate how our framework is utilized to evaluate the reliability and consistency of rule-based XAI techniques for classical ML models trained on tabular data through an extensive study of software analytics tasks.

Jiarpakdee et al. \cite{jiarpakdee2020empirical} conducted an empirical study to evaluate three model-agnostic techniques for defect prediction models. Their experimental results suggest that contrastive explanations are both essential and valuable in comprehending the predictions of defect models. Hase et al. \cite{hase2020evaluating} carried out human subject tests utilizing both tabular and text data to evaluate five XAI methods: LIME, Anchors, Decision Boundary, a Prototype Model, and a Composite approach that combines explanations from each technique. Gao et al. \cite{gao2022evaluating} evaluated the effectiveness of local explanation techniques on defect prediction models trained on source code. Their study found that local explanation techniques are effective on token frequency-based models but not on deep learning-based models. Ledel et al. \cite{ledel2022studying} evaluated the quality of the explanations generated by both SHAP and LIME on the bug and non-bug issues. Their study assessed the quality of the explanations based on human expectations. Shit et al. \cite{shin4328070empirical} conducted an empirical study on the stability of two XAI techniques, i.e., LIME-HPO and BreakDown for defect prediction models using two evaluation metrics, i.e., \textit{hit\_rate} and \textit{rank\_diff}. However, their experiment did not consider PyExplainer for stability check in explanation generation. Roy et al. \cite{roy2022don} showed how to resolve the disagreement problems among different XAI techniques for classical ML models trained on tabular data for JIT defect prediction tasks. However, they did not evaluate the reliability and consistency of the explanations generated by rule-based XAI techniques.

All the studies above evaluated different XAI methods, considering deep learning and ML models trained on image, text, and tabular data in various domains. However, to the best of our knowledge, studies have yet to consider evaluating the reliability and consistency of rule-based XAI techniques such as LIME and PyExplainer through the development of a framework and granular-level evaluation metrics. Furthermore, the evaluation criteria and proposed techniques significantly differ from our novel framework, and the inclusion of six evaluation metrics makes our work distinctive.

\section{Conclusion}
\label{end}
Machine learning (ML)-based approaches have enhanced various software analytics tasks in software maintenance and evolution. However, the lack of explainability behind the reasoning behind ML models can hinder practitioners from effectively applying ML-based approaches to improve software analytics tasks. To address this issue, novel rule-based XAI techniques, such as PyExplainer and LIME, have been utilized to elucidate the predictions of ML models. This allows practitioners to understand why a particular decision is made and take appropriate actions when needed. In this paper, we assess the ability of these techniques, particularly the state-of-the-art PyExplainer and LIME, to generate reliable and consistent explanations for ML models across various software analytics tasks. To do so, we propose a novel framework called \textit{EvaluateXAI} and six evaluation metrics to assess the reliability and consistency of the explanations generated by rule-based XAI techniques for ML models in software analytics tasks. Our extensive investigations using \textit{EvaluateXAI} conclude that none of the evaluation metrics reached 100\%, indicating the unreliability of the explanations generated by XAI techniques. Moreover, PyExplainer and LIME could not offer consistent explanations in 86.11\% and 77.78\% of the experimental combinations. Finally, we show that \textit{EvaluateXAI} could be adapted to any rule-based XAI techniques for evaluating their effectiveness in generating reliable and consistent explanations for the ML models in software analytics tasks. Our findings warrant further efforts to advance XAI research in software analytics.

\section*{Acknowledgment}
This research is supported in part by the Natural Sciences and Engineering Research Council of Canada (NSERC), and by the industry-stream NSERC CREATE in Software Analytics Research (SOAR).

\section*{Declaration of generative AI and AI-assisted technologies in the writing process}
During the preparation of this work, the author(s) used Grammarly \footnote{https://app.grammarly.com/} and ChatGPT 3.5\footnote{https://chat.openai.com/} to find grammatical mistakes and improve sentence clarity/ presentation. After using these tools/services, the author(s) reviewed and edited the content as needed and take(s) full responsibility for the content of the publication.

\bibliographystyle{elsarticle-num-names} 
\bibliography{manuscript.bib}

\appendix

\section{AUC--ROC Curve}
\label{AUC_ROC_Appendix}
\begin{figure}[htbp]
\footnotesize
\centering     
\subfigure[Cross-project mobile apps]{\label{AUC_MOB}\includegraphics[width=4.25cm, height=3.5cm]{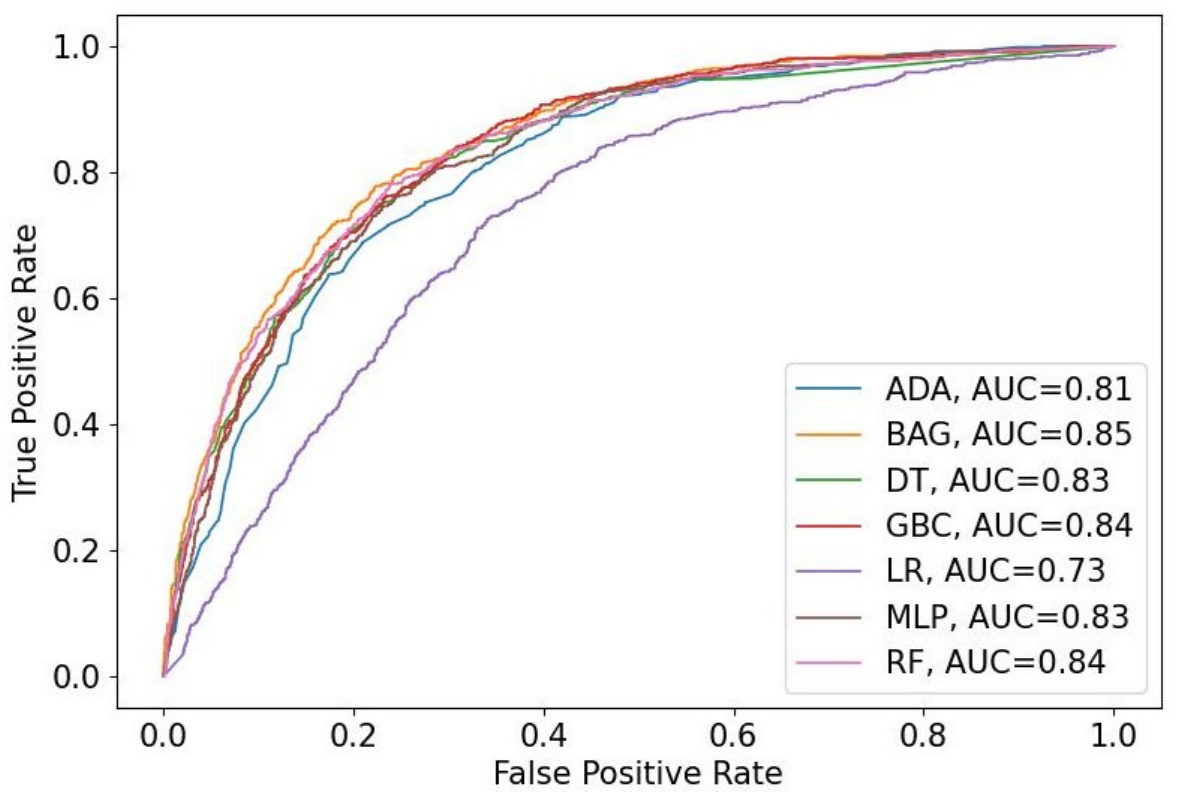}} 
\subfigure[Java]{\label{AUC_Java}\includegraphics[width=4.25cm, height=3.5cm]{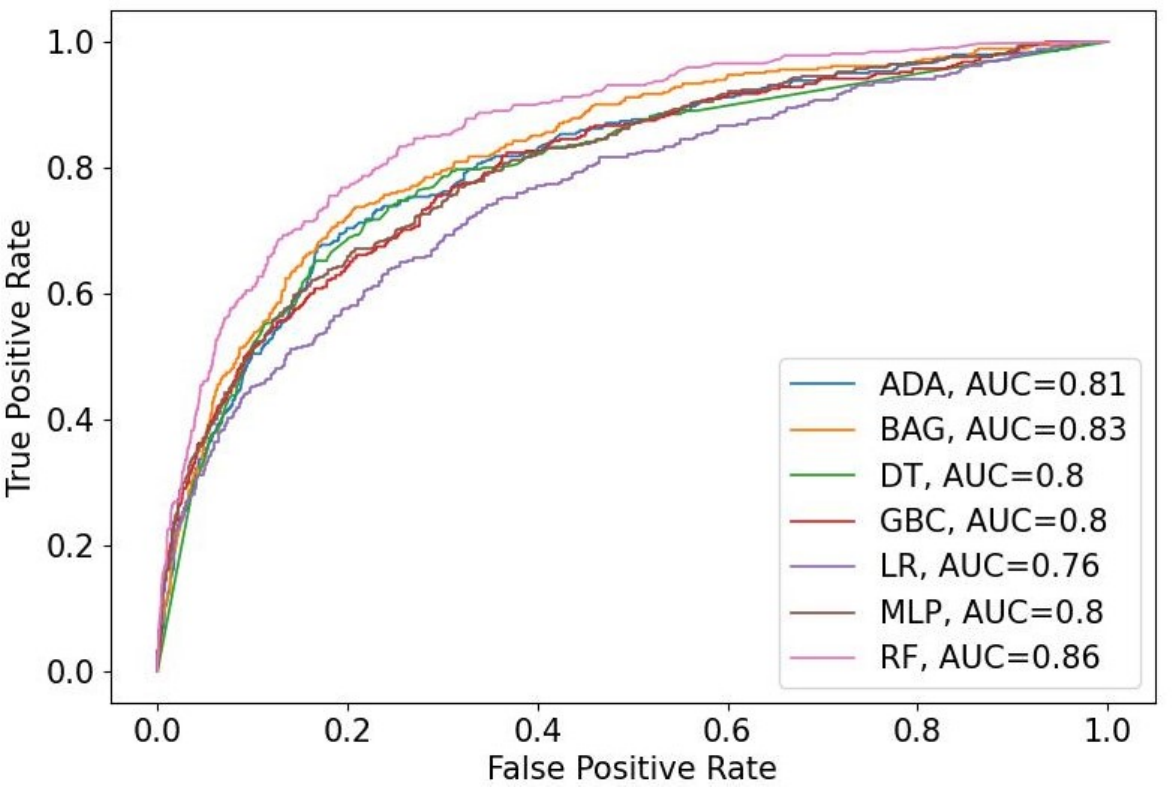}}
\subfigure[Postgres]{\label{AUC_Desk}\includegraphics[width=4.25cm, height=3.5cm]{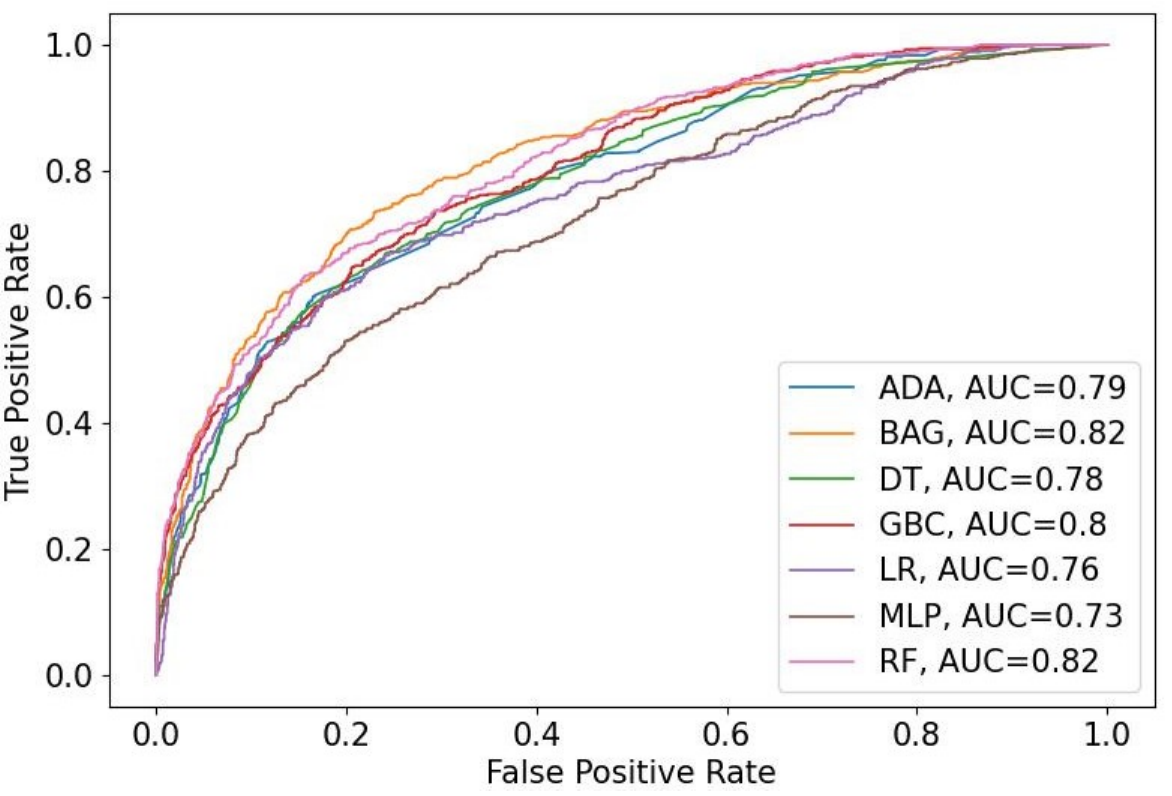}} 
\vspace{-0.5em}
\subfigure[CLCDSA]{\label{AUC_CLCDSA}\includegraphics[width=4.25cm, height=3.5cm]{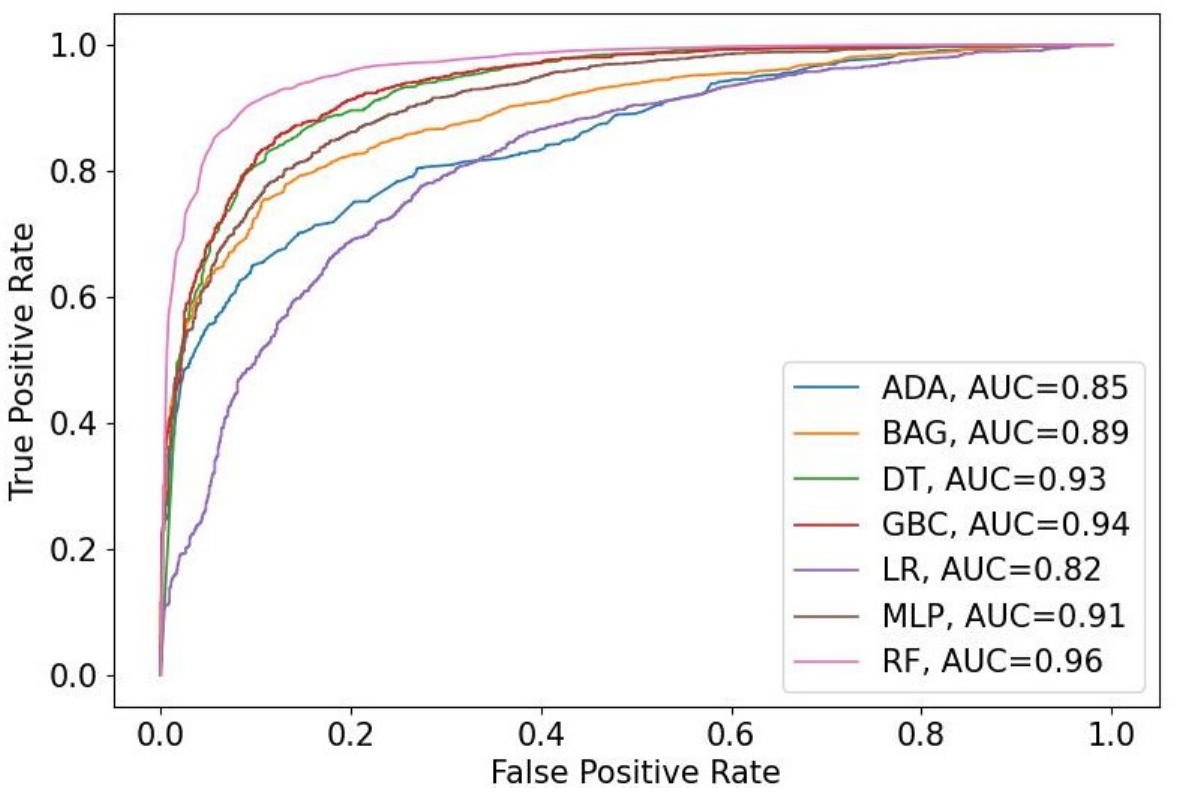}}
\subfigure[Code review]{\label{AUC_CR}\includegraphics[width=4.25cm, height=3.5cm]{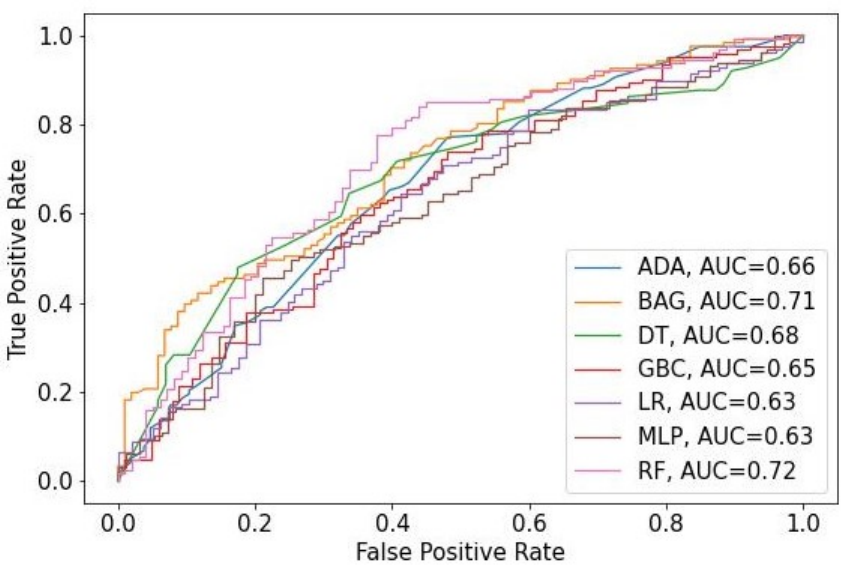}}
\caption{Performance of ML models across different datasets based on AUC-ROC curve. Figures \ref{AUCMOB}, \ref{AUC_Java}, \ref{AUC_Desk}, \ref{AUC_CLCDSA}, and \ref{AUC_CR} depict the AUC-ROC curves for ML models trained on the cross-project mobile apps, Java project, Postgres, CLCDSA, and code review datasets, respectively.}
\label{AUC_ROC}
\vspace{-1.0em}
\end{figure}

Figure \ref{AUC_ROC} shows the AUC\_ROC curves for the selected ML models trained on different datasets. From Figure \ref{AUC_MOB}, we observe that the LR model exhibits the lowest AUC value ($0.73$), whereas the BAG model shows the highest AUC value ($0.85$) when trained on the cross-project mobile apps dataset. The ML models trained on the CLCDSA dataset show the highest AUC values, ranging from $0.82$ to $0.96$. Additionally, Figure \ref{AUC_CR} demonstrates comparatively low AUC values for different ML models trained on the code review dataset. However, Figure \ref{AUC_ROC} indicates that the AUC value varies from $0.63$ to $0.96$.

\section{\textit{\%Prob\_diff Metric}}
\label{prob_diff_appendix}
\begin{figure}[htbp]
\centering     
\subfigure[LR + CLCDSA + PyExplainer]{\label{LR_fig_wrng_CLCDSA}\includegraphics[width=4.25cm, height=3.5cm]{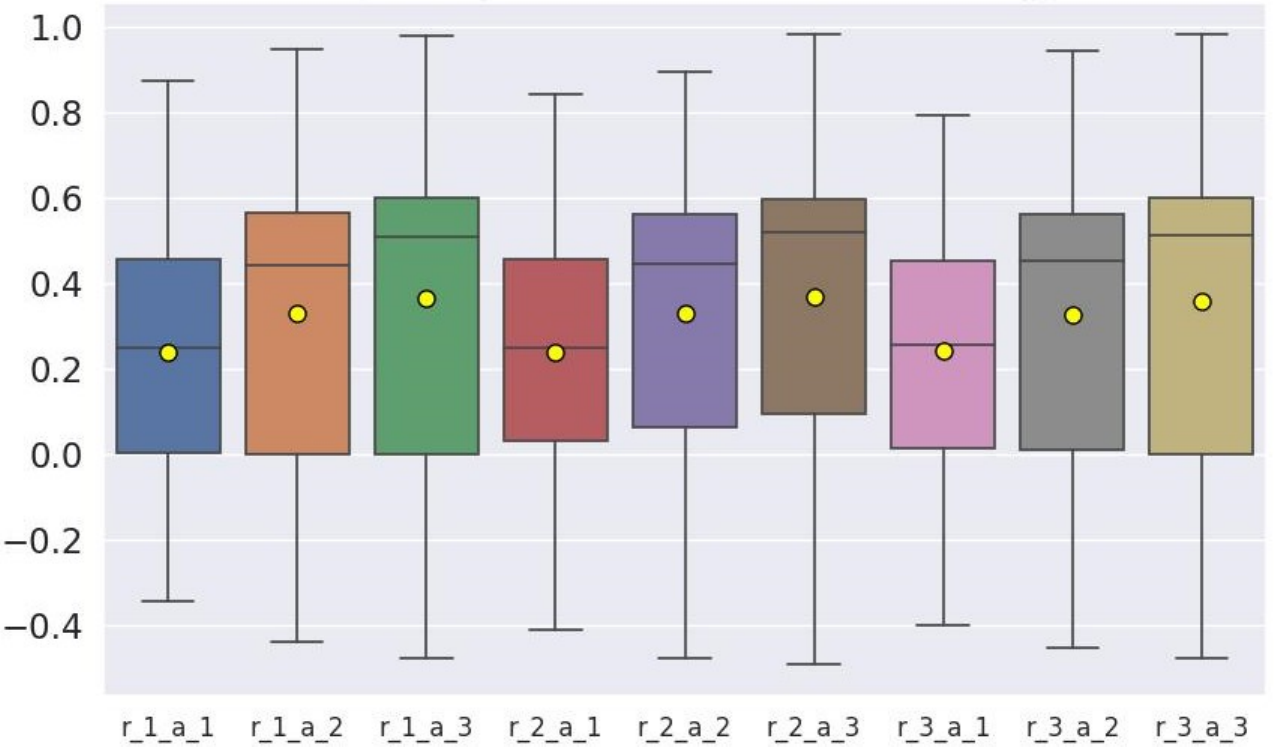}} 
\subfigure[LR + code review + PyExplainer]{\label{LR_fig_wrng_CR}\includegraphics[width=4.25cm, height=3.5cm]{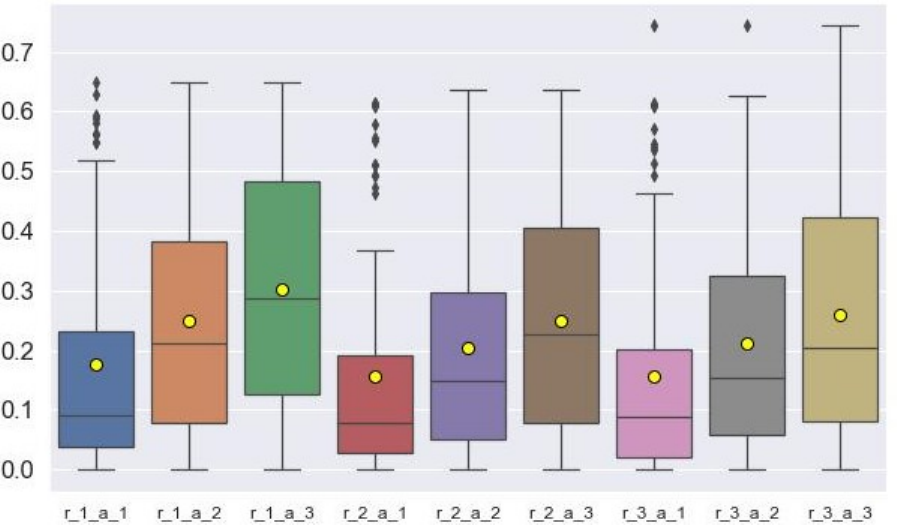}}
\vspace{-0.5em}
\subfigure[LR + CLCDSA + LIME]{\label{LR_LIME_CLCDSA_wrng}\includegraphics[width=4.25cm, height=3.5cm]{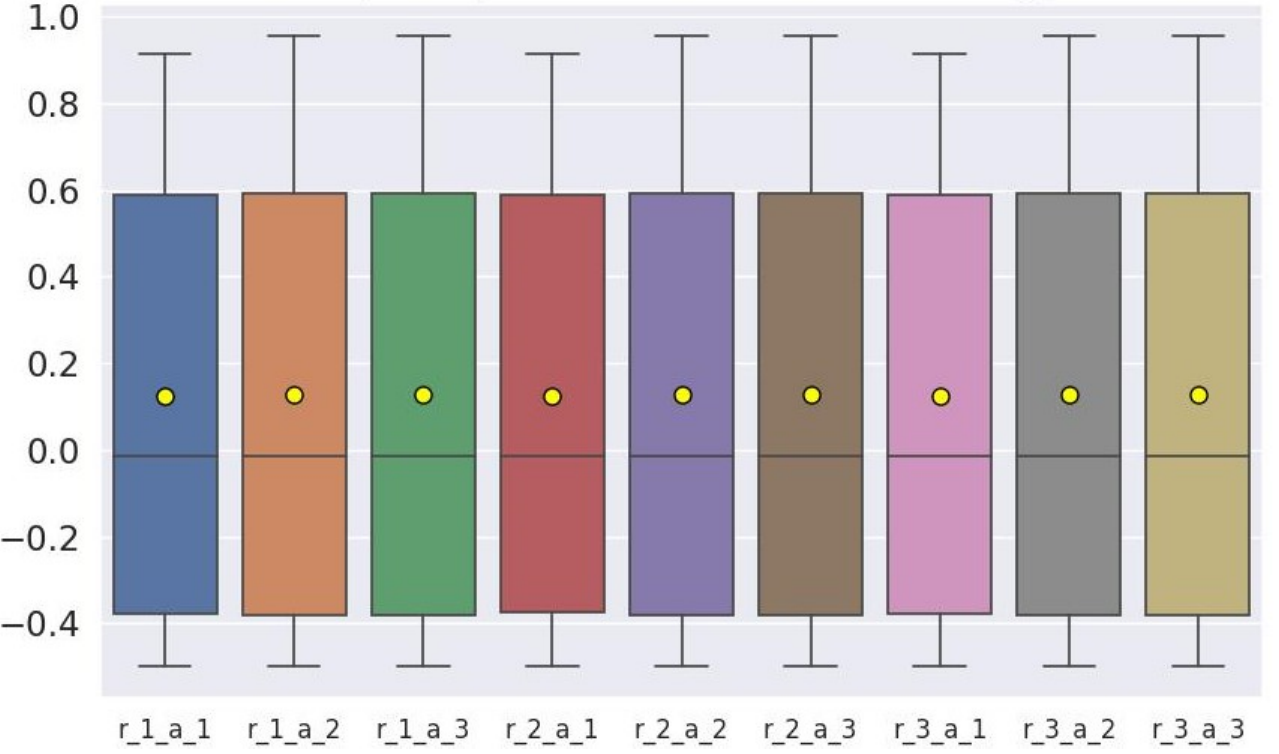}}
\subfigure[LR + code review + LIME]{\label{LR_LIME_Code_wrng}\includegraphics[width=4.25cm, height=3.5cm]{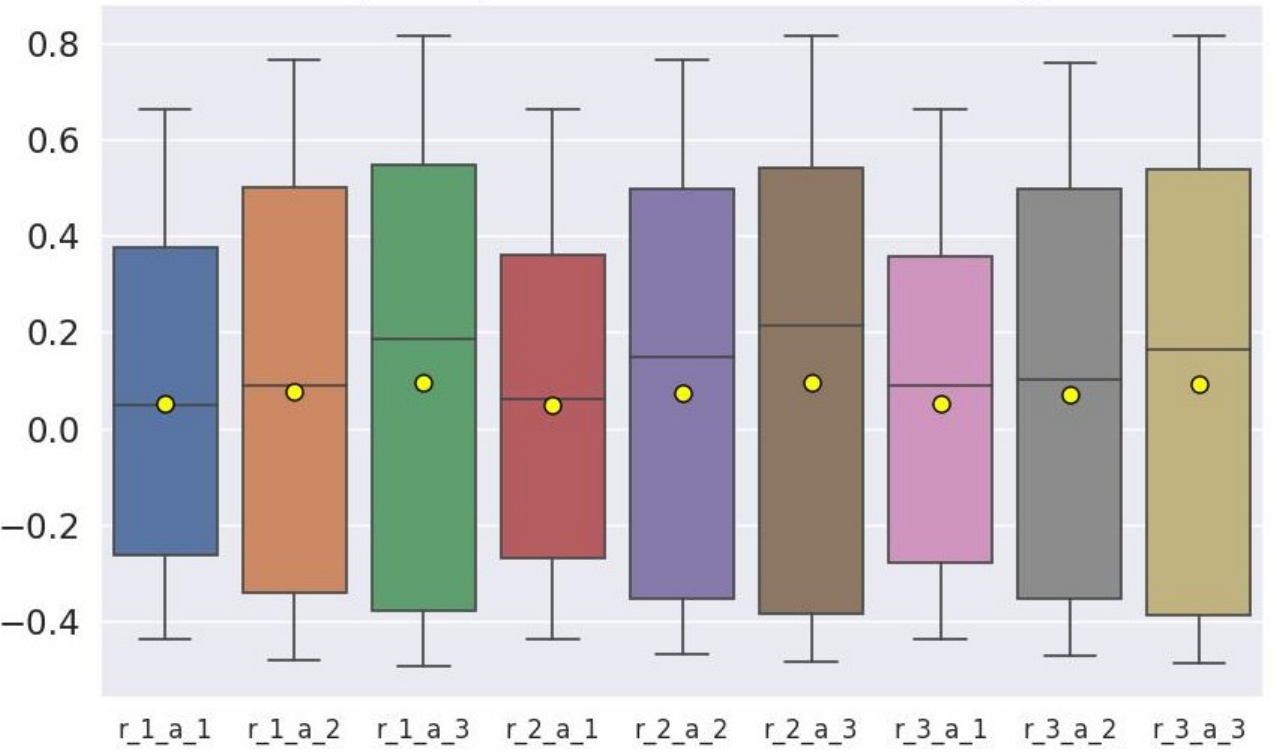}}
\subfigure[RF + CLCDSA + PyExplainer]{\label{RF_fig_wrng_CLCDSA}\includegraphics[width=4.25cm, height=3.5cm]{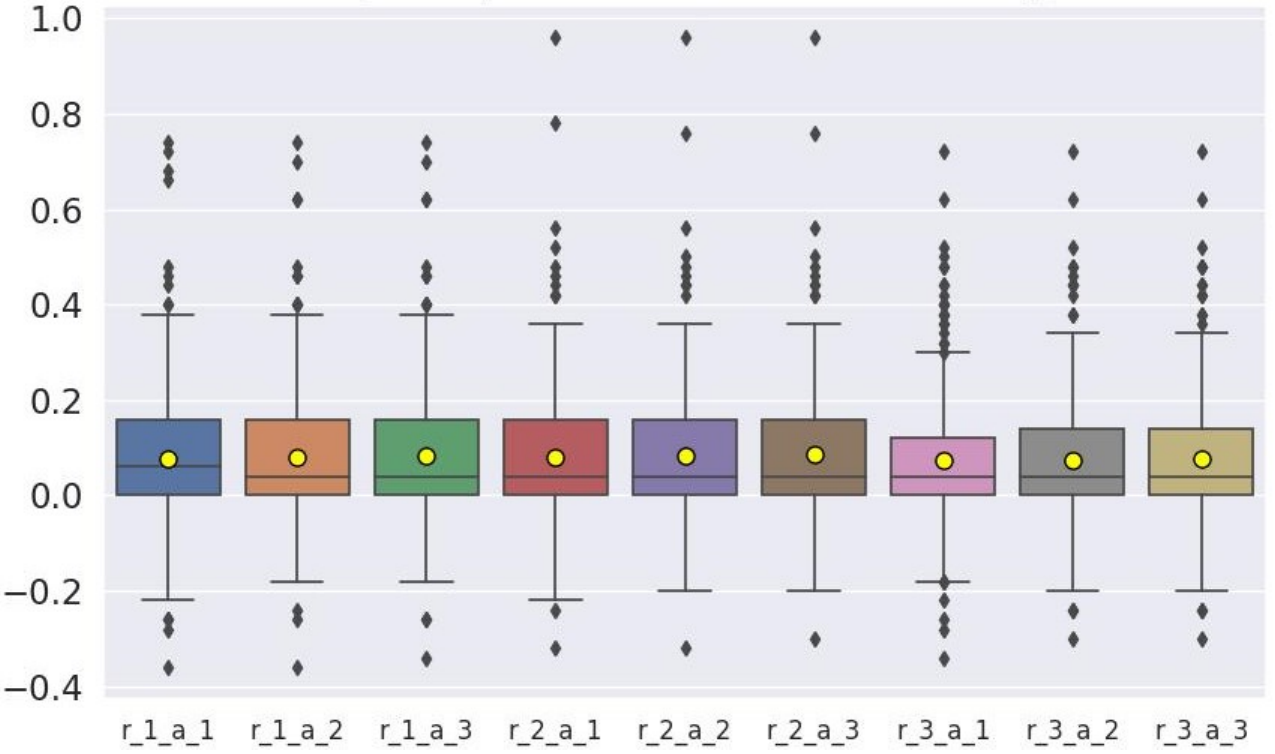}} 
\vspace{-0.5em}
\subfigure[RF + code review + PyExplainer]{\label{RF_fig_wrng_CR}\includegraphics[width=4.25cm, height=3.5cm]{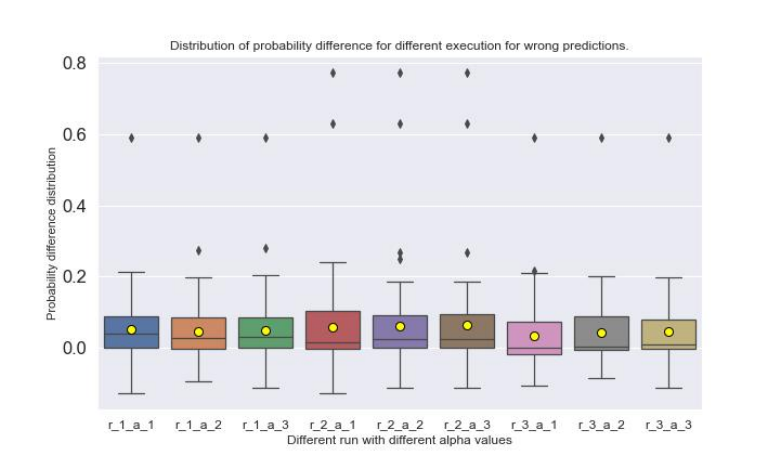}}
\vspace{-0.5em}
\subfigure[RF + CLCDSA + LIME]{\label{RF_LIME_CLCDSA_wrng}\includegraphics[width=4.25cm, height=3.5cm]{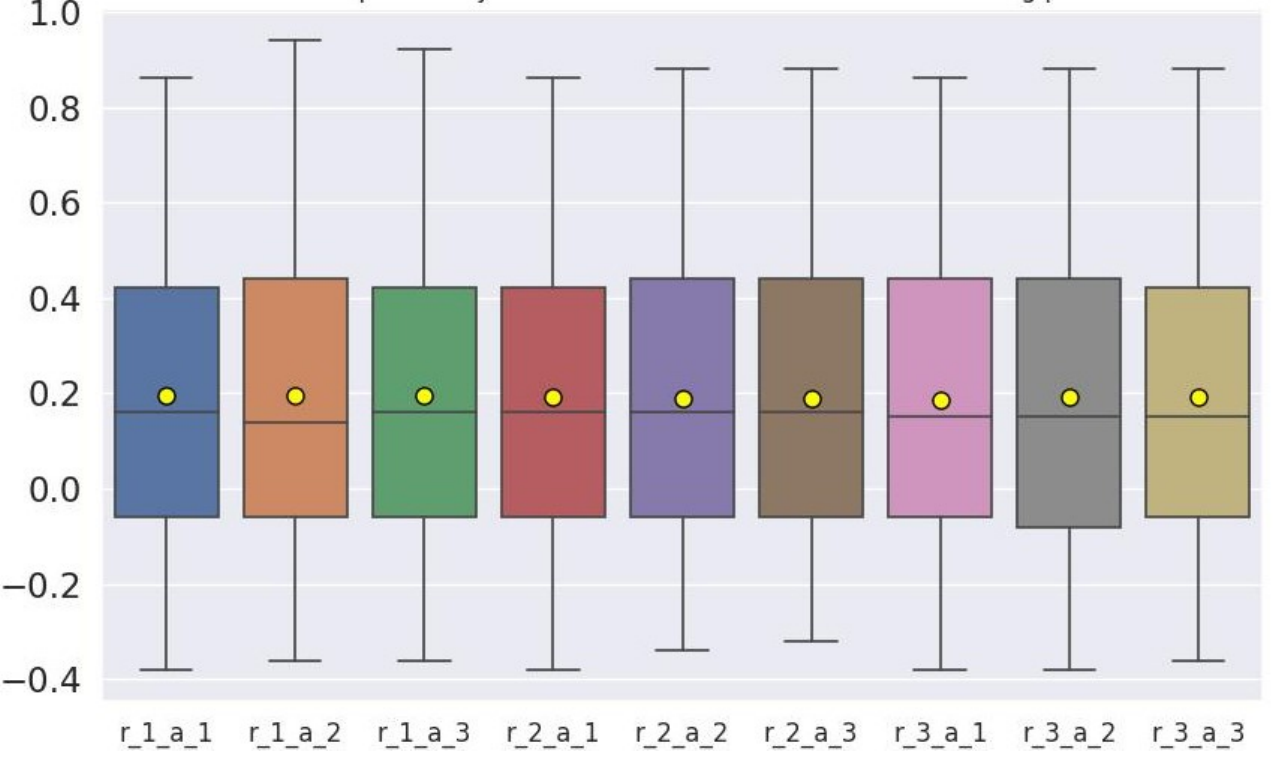}}
\subfigure[RF + code review + LIME]{\label{RF _LIME_Code_wrng}\includegraphics[width=4.25cm, height=3.5cm]{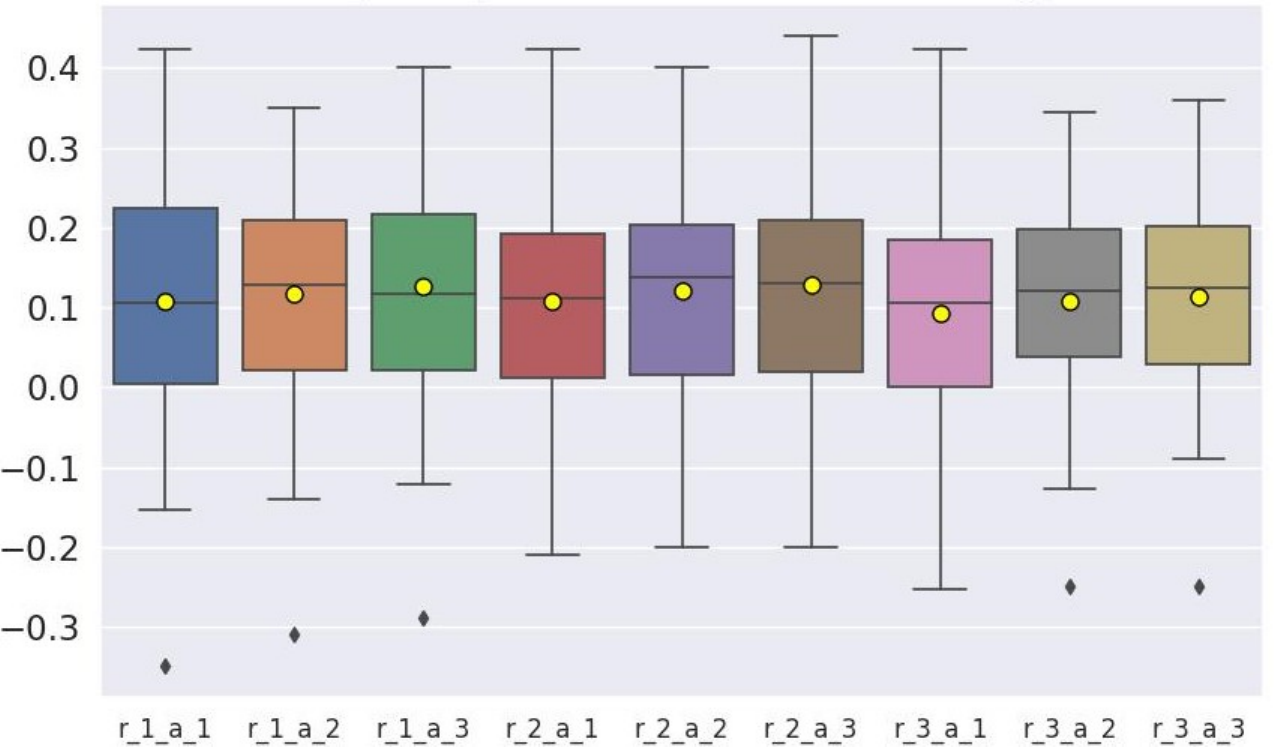}}
\caption{Distribution of the probability difference between the original and the simulated instances. Figures \ref{LR_fig_wrng_CLCDSA} and \ref{LR_fig_wrng_CR} show the probability difference distribution for the LR model trained on CLCDSA and code review datasets when we use PyExplainer for explanation generation. Figures \ref{LR_LIME_CLCDSA_wrng} and \ref{LR_LIME_Code_wrng} show the probability difference distribution for the LR model trained on the CLCDSA and code review datasets when we use LIME for explanation generation. Figures \ref{RF_fig_wrng_CLCDSA} and \ref{RF_fig_wrng_CR} show the probability difference distribution for the RF model trained on CLCDSA and code review datasets when we use PyExplainer for explanation generation. Figures \ref{RF_LIME_CLCDSA_wrng} and \ref{RF_LIME_CLCDSA_wrng} show the probability difference distribution for the RF model trained on the CLCDSA and code review datasets when we use LIME for explanation generation. Here, r\_i\_a\_j denotes the $i^{th}$ execution with $\alpha=j$.}
\label{prob_diff_wrng_appendix}
\vspace{-1.5em}
\end{figure}

Figure \ref{prob_diff_wrng_appendix} shows the distribution of the probability difference between the original instances and the simulated instances for the ML models making wrong predictions. Similar to the results of the $\%Prob_diff$ metric for correct predictions, as described in Sub-section \ref{present_exp_resutls}, we observe similar patterns for the ML models making wrong predictions.

\section{Granular-level Evaluation Metrics}
\label{granular_level_appendix}

Table \ref{granular_PyExp_appendix} presents the experimental results of four granular-level evaluation metrics (i.e., PCPD, PCPI, NCPD, and NCPI) for PyExplainer applied to ML models, both for correct and wrong predictions. In an ideal scenario, we expect an increase in each evaluation metric for each simulated instance. However, from Table \ref{granular_PyExp_appendix}, it is evident that we observe very low values for PCPD, PCPI, NCPD, and NCPI metrics in the case of correct predictions, with only a few exceptions. For example, when considering the LR model trained on the CLCDSA dataset, we find a PCPD value of 24.59 and an NCPI value of 34.68. In contrast, we consistently observe very low PCPI and NCPD values for correct predictions.

For wrong predictions, we observe relatively high \textit{PCPD} values for all the ML models. However, this is not the case for the \textit{PCPI} metric, with only a few exceptions. For example, the \textit{PCPI} values for the BAG and GBC models are $32.47$ and $23.08$, respectively. Similarly, the \textit{NCPD} values are low for the simulated instances, with only a few exceptions. Finally, we observe comparatively high \textit{NCPI} values for all the models trained on the CLCDSA dataset. In contrast, for the cross-project mobile apps dataset, the \textit{NCPI} values are low. In summary, when considering both correct and wrong predictions, the four granular-level evaluation metrics consistently yield low values, raising concerns about the reliability of the explanations generated by PyExplainer.

We observe similar results for LIME, as shown in Table \ref{granular_LIME_appendix}. However, the granular-level evaluation metric values do not consistently increase with the rise of $\alpha$ values for LIME.

\begin{table*}[t]
\centering
\caption{Granular-level evaluation metric values for PyExplainer executed on the same instance across the entire Java project, Postgres, CLCDSA, and code review datasets, with varying $\alpha$ values.}
\resizebox{\linewidth}{!}{
\begin{tabular}{|lllllllllllllllllllllllll|}
\hline
\multicolumn{1}{|l|}{\multirow{4}{*}{\textbf{\begin{tabular}[c]{@{}l@{}}ML\\ \\ Model\end{tabular}}}} & \multicolumn{24}{c|}{\textbf{Java Project}}                                                                                                                                                                                                                                                                                                                                                                                                                                                                                                                                                                                                                                                              \\ \cline{2-25} 
\multicolumn{1}{|l|}{}                                                                                & \multicolumn{6}{c|}{\textbf{PCPD}}                                                                                                                                          & \multicolumn{6}{c|}{\textbf{PCPI}}                                                                                                                                          & \multicolumn{6}{c|}{\textbf{NCPD}}                                                                                                                                          & \multicolumn{6}{c|}{\textbf{NCPI}}                                                                                                                     \\ \cline{2-25} 
\multicolumn{1}{|l|}{}                                                                                & \multicolumn{2}{c|}{$\alpha=1$}                         & \multicolumn{2}{c|}{$\alpha=2$}                         & \multicolumn{2}{c|}{$\alpha=3$}                         & \multicolumn{2}{c|}{$\alpha=1$}                         & \multicolumn{2}{c|}{$\alpha=2$}                         & \multicolumn{2}{c|}{$\alpha=3$}                         & \multicolumn{2}{c|}{$\alpha=1$}                         & \multicolumn{2}{c|}{$\alpha=2$}                         & \multicolumn{2}{c|}{$\alpha=3$}                         & \multicolumn{2}{c|}{$\alpha=1$}                         & \multicolumn{2}{c|}{$\alpha=2$}                         & \multicolumn{2}{c|}{$\alpha=3$}    \\ \cline{2-25} 
\multicolumn{1}{|l|}{}                                                                                & \multicolumn{1}{l|}{Cort.} & \multicolumn{1}{l|}{Wrng.} & \multicolumn{1}{l|}{Cort.} & \multicolumn{1}{l|}{Wrng.} & \multicolumn{1}{l|}{Cort.} & \multicolumn{1}{l|}{Wrng.} & \multicolumn{1}{l|}{Cort.} & \multicolumn{1}{l|}{Wrng.} & \multicolumn{1}{l|}{Cort.} & \multicolumn{1}{l|}{Wrng.} & \multicolumn{1}{l|}{Cort.} & \multicolumn{1}{l|}{Wrng.} & \multicolumn{1}{l|}{Cort.} & \multicolumn{1}{l|}{Wrng.} & \multicolumn{1}{l|}{Cort.} & \multicolumn{1}{l|}{Wrng.} & \multicolumn{1}{l|}{Cort.} & \multicolumn{1}{l|}{Wrng.} & \multicolumn{1}{l|}{Cort.} & \multicolumn{1}{l|}{Wrng.} & \multicolumn{1}{l|}{Cort.} & \multicolumn{1}{l|}{Wrng.} & \multicolumn{1}{l|}{Cort.} & Wrng. \\ \hline
\multicolumn{1}{|l|}{LR}                                                                              & \multicolumn{1}{l|}{3.97}  & \multicolumn{1}{l|}{61.69} & \multicolumn{1}{l|}{3.73}  & \multicolumn{1}{l|}{55.72} & \multicolumn{1}{l|}{3.51}  & \multicolumn{1}{l|}{53.9}  & \multicolumn{1}{l|}{0.45}  & \multicolumn{1}{l|}{10.78} & \multicolumn{1}{l|}{0.5}   & \multicolumn{1}{l|}{8.96}  & \multicolumn{1}{l|}{0.53}  & \multicolumn{1}{l|}{8.46}  & \multicolumn{1}{l|}{3.11}  & \multicolumn{1}{l|}{2.32}  & \multicolumn{1}{l|}{3.54}  & \multicolumn{1}{l|}{2.32}  & \multicolumn{1}{l|}{3.63}  & \multicolumn{1}{l|}{2.49}  & \multicolumn{1}{l|}{29.44} & \multicolumn{1}{l|}{14.59} & \multicolumn{1}{l|}{27.34} & \multicolumn{1}{l|}{12.94} & \multicolumn{1}{l|}{27.03} & 12.77 \\ \hline
\multicolumn{1}{|l|}{DT}                                                                              & \multicolumn{1}{l|}{0.97}  & \multicolumn{1}{l|}{23.91} & \multicolumn{1}{l|}{0.94}  & \multicolumn{1}{l|}{21.83} & \multicolumn{1}{l|}{0.85}  & \multicolumn{1}{l|}{21.41} & \multicolumn{1}{l|}{0.51}  & \multicolumn{1}{l|}{7.69}  & \multicolumn{1}{l|}{0.58}  & \multicolumn{1}{l|}{8.32}  & \multicolumn{1}{l|}{0.55}  & \multicolumn{1}{l|}{8.73}  & \multicolumn{1}{l|}{5.23}  & \multicolumn{1}{l|}{3.53}  & \multicolumn{1}{l|}{4.9}   & \multicolumn{1}{l|}{2.91}  & \multicolumn{1}{l|}{4.81}  & \multicolumn{1}{l|}{2.7}   & \multicolumn{1}{l|}{15.27} & \multicolumn{1}{l|}{7.48}  & \multicolumn{1}{l|}{14.78} & \multicolumn{1}{l|}{7.48}  & \multicolumn{1}{l|}{14.6}  & 7.69  \\ \hline
\multicolumn{1}{|l|}{RF}                                                                              & \multicolumn{1}{l|}{2.63}  & \multicolumn{1}{l|}{35.42} & \multicolumn{1}{l|}{2.76}  & \multicolumn{1}{l|}{40.28} & \multicolumn{1}{l|}{2.74}  & \multicolumn{1}{l|}{38.19} & \multicolumn{1}{l|}{0.66}  & \multicolumn{1}{l|}{11.11} & \multicolumn{1}{l|}{0.41}  & \multicolumn{1}{l|}{5.21}  & \multicolumn{1}{l|}{0.34}  & \multicolumn{1}{l|}{5.21}  & \multicolumn{1}{l|}{6.67}  & \multicolumn{1}{l|}{12.85} & \multicolumn{1}{l|}{5.89}  & \multicolumn{1}{l|}{10.42} & \multicolumn{1}{l|}{5.71}  & \multicolumn{1}{l|}{10.42} & \multicolumn{1}{l|}{32.18} & \multicolumn{1}{l|}{30.56} & \multicolumn{1}{l|}{32.14} & \multicolumn{1}{l|}{31.6}  & \multicolumn{1}{l|}{32.02} & 31.25 \\ \hline
\multicolumn{1}{|l|}{MLP}                                                                             & \multicolumn{1}{l|}{3.35}  & \multicolumn{1}{l|}{46.39} & \multicolumn{1}{l|}{3.31}  & \multicolumn{1}{l|}{47.19} & \multicolumn{1}{l|}{3.45}  & \multicolumn{1}{l|}{47.19} & \multicolumn{1}{l|}{1.65}  & \multicolumn{1}{l|}{24.3}  & \multicolumn{1}{l|}{1.59}  & \multicolumn{1}{l|}{21.89} & \multicolumn{1}{l|}{1.42}  & \multicolumn{1}{l|}{21.49} & \multicolumn{1}{l|}{8.32}  & \multicolumn{1}{l|}{5.22}  & \multicolumn{1}{l|}{7.63}  & \multicolumn{1}{l|}{4.42}  & \multicolumn{1}{l|}{7.15}  & \multicolumn{1}{l|}{4.62}  & \multicolumn{1}{l|}{23.81} & \multicolumn{1}{l|}{13.86} & \multicolumn{1}{l|}{22.91} & \multicolumn{1}{l|}{13.05} & \multicolumn{1}{l|}{22.84} & 12.45 \\ \hline
\multicolumn{1}{|l|}{ADA}                                                                             & \multicolumn{1}{l|}{1.79}  & \multicolumn{1}{l|}{22.39} & \multicolumn{1}{l|}{1.51}  & \multicolumn{1}{l|}{20.88} & \multicolumn{1}{l|}{1.42}  & \multicolumn{1}{l|}{20.08} & \multicolumn{1}{l|}{0.14}  & \multicolumn{1}{l|}{2.9}   & \multicolumn{1}{l|}{0.16}  & \multicolumn{1}{l|}{3.47}  & \multicolumn{1}{l|}{0.19}  & \multicolumn{1}{l|}{3.28}  & \multicolumn{1}{l|}{2.07}  & \multicolumn{1}{l|}{0.97}  & \multicolumn{1}{l|}{2.68}  & \multicolumn{1}{l|}{0.97}  & \multicolumn{1}{l|}{2.54}  & \multicolumn{1}{l|}{0.97}  & \multicolumn{1}{l|}{29.41} & \multicolumn{1}{l|}{12.55} & \multicolumn{1}{l|}{27.01} & \multicolumn{1}{l|}{11.97} & \multicolumn{1}{l|}{27.39} & 11.97 \\ \hline
\multicolumn{1}{|l|}{BAG}                                                                             & \multicolumn{1}{l|}{3.55}  & \multicolumn{1}{l|}{43.96} & \multicolumn{1}{l|}{3.71}  & \multicolumn{1}{l|}{40.11} & \multicolumn{1}{l|}{3.46}  & \multicolumn{1}{l|}{38.64} & \multicolumn{1}{l|}{0.76}  & \multicolumn{1}{l|}{11.9}  & \multicolumn{1}{l|}{0.44}  & \multicolumn{1}{l|}{11.54} & \multicolumn{1}{l|}{0.46}  & \multicolumn{1}{l|}{11.17} & \multicolumn{1}{l|}{1.45}  & \multicolumn{1}{l|}{0.73}  & \multicolumn{1}{l|}{1.29}  & \multicolumn{1}{l|}{0.92}  & \multicolumn{1}{l|}{1.34}  & \multicolumn{1}{l|}{0.92}  & \multicolumn{1}{l|}{34.21} & \multicolumn{1}{l|}{15.75} & \multicolumn{1}{l|}{33.54} & \multicolumn{1}{l|}{14.65} & \multicolumn{1}{l|}{33.43} & 14.65 \\ \hline
\multicolumn{1}{|l|}{GBC}                                                                             & \multicolumn{1}{l|}{2.56}  & \multicolumn{1}{l|}{39.91} & \multicolumn{1}{l|}{2.16}  & \multicolumn{1}{l|}{36.36} & \multicolumn{1}{l|}{2.19}  & \multicolumn{1}{l|}{35.03} & \multicolumn{1}{l|}{0.92}  & \multicolumn{1}{l|}{15.96} & \multicolumn{1}{l|}{0.92}  & \multicolumn{1}{l|}{15.3}  & \multicolumn{1}{l|}{0.78}  & \multicolumn{1}{l|}{15.3}  & \multicolumn{1}{l|}{11.05} & \multicolumn{1}{l|}{10.42} & \multicolumn{1}{l|}{11.24} & \multicolumn{1}{l|}{8.43}  & \multicolumn{1}{l|}{11.1}  & \multicolumn{1}{l|}{8.43}  & \multicolumn{1}{l|}{23.42} & \multicolumn{1}{l|}{14.86} & \multicolumn{1}{l|}{21.51} & \multicolumn{1}{l|}{13.75} & \multicolumn{1}{l|}{21.47} & 13.53 \\ \hline
\multicolumn{25}{|c|}{\textbf{Postgres}}                                                                                                                                                                                                                                                                                                                                                                                                                                                                                                                                                                                                                                                                                                                                                              \\ \hline
\multicolumn{1}{|l|}{LR}                                                                              & \multicolumn{1}{l|}{7.47}  & \multicolumn{1}{l|}{47.33} & \multicolumn{1}{l|}{7.55}  & \multicolumn{1}{l|}{48.4}  & \multicolumn{1}{l|}{7.55}  & \multicolumn{1}{l|}{47.33} & \multicolumn{1}{l|}{0.78}  & \multicolumn{1}{l|}{5.12}  & \multicolumn{1}{l|}{0.61}  & \multicolumn{1}{l|}{3.62}  & \multicolumn{1}{l|}{0.55}  & \multicolumn{1}{l|}{4.26}  & \multicolumn{1}{l|}{5.62}  & \multicolumn{1}{l|}{7.46}  & \multicolumn{1}{l|}{5.62}  & \multicolumn{1}{l|}{7.46}  & \multicolumn{1}{l|}{5.62}  & \multicolumn{1}{l|}{7.25}  & \multicolumn{1}{l|}{18.06} & \multicolumn{1}{l|}{21.54} & \multicolumn{1}{l|}{15.63} & \multicolumn{1}{l|}{18.98} & \multicolumn{1}{l|}{14.96} & 19.19 \\ \hline
\multicolumn{1}{|l|}{DT}                                                                              & \multicolumn{1}{l|}{4.36}  & \multicolumn{1}{l|}{25.42} & \multicolumn{1}{l|}{4.39}  & \multicolumn{1}{l|}{25.96} & \multicolumn{1}{l|}{3.96}  & \multicolumn{1}{l|}{24.36} & \multicolumn{1}{l|}{2.66}  & \multicolumn{1}{l|}{11.86} & \multicolumn{1}{l|}{2.39}  & \multicolumn{1}{l|}{12.95} & \multicolumn{1}{l|}{2.85}  & \multicolumn{1}{l|}{11.25} & \multicolumn{1}{l|}{11.38} & \multicolumn{1}{l|}{15.25} & \multicolumn{1}{l|}{12.39} & \multicolumn{1}{l|}{14.65} & \multicolumn{1}{l|}{11.39} & \multicolumn{1}{l|}{16.58} & \multicolumn{1}{l|}{16.71} & \multicolumn{1}{l|}{20.34} & \multicolumn{1}{l|}{15.67} & \multicolumn{1}{l|}{19.84} & \multicolumn{1}{l|}{14.78} & 21.81 \\ \hline
\multicolumn{1}{|l|}{RF}                                                                              & \multicolumn{1}{l|}{2.36}  & \multicolumn{1}{l|}{15.69} & \multicolumn{1}{l|}{2.14}  & \multicolumn{1}{l|}{14.29} & \multicolumn{1}{l|}{2.14}  & \multicolumn{1}{l|}{14.82} & \multicolumn{1}{l|}{4.29}  & \multicolumn{1}{l|}{10.56} & \multicolumn{1}{l|}{4.39}  & \multicolumn{1}{l|}{8.85}  & \multicolumn{1}{l|}{4.81}  & \multicolumn{1}{l|}{10.92} & \multicolumn{1}{l|}{14.62} & \multicolumn{1}{l|}{18.59} & \multicolumn{1}{l|}{12.39} & \multicolumn{1}{l|}{17.89} & \multicolumn{1}{l|}{15.62} & \multicolumn{1}{l|}{20.85} & \multicolumn{1}{l|}{17.35} & \multicolumn{1}{l|}{20.39} & \multicolumn{1}{l|}{17.21} & \multicolumn{1}{l|}{21.52} & \multicolumn{1}{l|}{18.54} & 22.63 \\ \hline
\multicolumn{1}{|l|}{MLP}                                                                             & \multicolumn{1}{l|}{7.41}  & \multicolumn{1}{l|}{44.44} & \multicolumn{1}{l|}{7.41}  & \multicolumn{1}{l|}{66.67} & \multicolumn{1}{l|}{6.17}  & \multicolumn{1}{l|}{55.56} & \multicolumn{1}{l|}{0.0}   & \multicolumn{1}{l|}{22.22} & \multicolumn{1}{l|}{0.0}   & \multicolumn{1}{l|}{0.0}   & \multicolumn{1}{l|}{0.0}   & \multicolumn{1}{l|}{11.11} & \multicolumn{1}{l|}{6.17}  & \multicolumn{1}{l|}{0.0}   & \multicolumn{1}{l|}{4.94}  & \multicolumn{1}{l|}{0.0}   & \multicolumn{1}{l|}{4.94}  & \multicolumn{1}{l|}{0.0}   & \multicolumn{1}{l|}{13.58} & \multicolumn{1}{l|}{22.2}  & \multicolumn{1}{l|}{11.11} & \multicolumn{1}{l|}{22.22} & \multicolumn{1}{l|}{9.88}  & 22.22 \\ \hline
\multicolumn{1}{|l|}{ADA}                                                                             & \multicolumn{1}{l|}{3.51}  & \multicolumn{1}{l|}{21.45} & \multicolumn{1}{l|}{3.51}  & \multicolumn{1}{l|}{22.14} & \multicolumn{1}{l|}{3.6}   & \multicolumn{1}{l|}{22.28} & \multicolumn{1}{l|}{1.04}  & \multicolumn{1}{l|}{3.96}  & \multicolumn{1}{l|}{1.04}  & \multicolumn{1}{l|}{3.73}  & \multicolumn{1}{l|}{1.04}  & \multicolumn{1}{l|}{3.5}   & \multicolumn{1}{l|}{15.51} & \multicolumn{1}{l|}{22.84} & \multicolumn{1}{l|}{16.2}  & \multicolumn{1}{l|}{23.31} & \multicolumn{1}{l|}{16.94} & \multicolumn{1}{l|}{24.71} & \multicolumn{1}{l|}{8.92}  & \multicolumn{1}{l|}{12.35} & \multicolumn{1}{l|}{8.35}  & \multicolumn{1}{l|}{11.89} & \multicolumn{1}{l|}{7.1}   & 10.26 \\ \hline
\multicolumn{1}{|l|}{BAG}                                                                             & \multicolumn{1}{l|}{6.36}  & \multicolumn{1}{l|}{48.15} & \multicolumn{1}{l|}{6.94}  & \multicolumn{1}{l|}{51.85} & \multicolumn{1}{l|}{6.36}  & \multicolumn{1}{l|}{55.56} & \multicolumn{1}{l|}{1.16}  & \multicolumn{1}{l|}{14.81} & \multicolumn{1}{l|}{0.58}  & \multicolumn{1}{l|}{7.41}  & \multicolumn{1}{l|}{0.58}  & \multicolumn{1}{l|}{3.7}   & \multicolumn{1}{l|}{2.31}  & \multicolumn{1}{l|}{3.7}   & \multicolumn{1}{l|}{1.73}  & \multicolumn{1}{l|}{3.7}   & \multicolumn{1}{l|}{1.16}  & \multicolumn{1}{l|}{3.7}   & \multicolumn{1}{l|}{19.08} & \multicolumn{1}{l|}{22.22} & \multicolumn{1}{l|}{19.08} & \multicolumn{1}{l|}{22.22} & \multicolumn{1}{l|}{18.5}  & 22.22 \\ \hline
\multicolumn{1}{|l|}{GBC}                                                                             & \multicolumn{1}{l|}{5.28}  & \multicolumn{1}{l|}{27.93} & \multicolumn{1}{l|}{4.12}  & \multicolumn{1}{l|}{23.73} & \multicolumn{1}{l|}{3.87}  & \multicolumn{1}{l|}{22.03} & \multicolumn{1}{l|}{1.97}  & \multicolumn{1}{l|}{12.47} & \multicolumn{1}{l|}{2.66}  & \multicolumn{1}{l|}{10.17} & \multicolumn{1}{l|}{2.66}  & \multicolumn{1}{l|}{10.17} & \multicolumn{1}{l|}{9.88}  & \multicolumn{1}{l|}{13.47} & \multicolumn{1}{l|}{10.9}  & \multicolumn{1}{l|}{15.25} & \multicolumn{1}{l|}{10.41} & \multicolumn{1}{l|}{16.95} & \multicolumn{1}{l|}{15.56} & \multicolumn{1}{l|}{24.69} & \multicolumn{1}{l|}{14.04} & \multicolumn{1}{l|}{20.34} & \multicolumn{1}{l|}{13.08} & 18.64 \\ \hline
\multicolumn{25}{|c|}{\textbf{CLCDSA}}                                                                                                                                                                                                                                                                                                                                                                                                                                                                                                                                                                                                                                                                                                                                                                   \\ \hline
\multicolumn{1}{|l|}{LR}                                                                              & \multicolumn{1}{l|}{24.59} & \multicolumn{1}{l|}{40.89} & \multicolumn{1}{l|}{24.37} & \multicolumn{1}{l|}{41.03} & \multicolumn{1}{l|}{24.22} & \multicolumn{1}{l|}{40.89} & \multicolumn{1}{l|}{1.96}  & \multicolumn{1}{l|}{3.51}  & \multicolumn{1}{l|}{1.83}  & \multicolumn{1}{l|}{2.97}  & \multicolumn{1}{l|}{1.88}  & \multicolumn{1}{l|}{2.83}  & \multicolumn{1}{l|}{2.16}  & \multicolumn{1}{l|}{7.15}  & \multicolumn{1}{l|}{2.16}  & \multicolumn{1}{l|}{7.69}  & \multicolumn{1}{l|}{2.2}   & \multicolumn{1}{l|}{7.69}  & \multicolumn{1}{l|}{9.89}  & \multicolumn{1}{l|}{34.68} & \multicolumn{1}{l|}{9.47}  & \multicolumn{1}{l|}{33.87} & \multicolumn{1}{l|}{9.41}  & 33.87 \\ \hline
\multicolumn{1}{|l|}{DT}                                                                              & \multicolumn{1}{l|}{9.28}  & \multicolumn{1}{l|}{13.3}  & \multicolumn{1}{l|}{9.0}   & \multicolumn{1}{l|}{13.07} & \multicolumn{1}{l|}{9.15}  & \multicolumn{1}{l|}{13.07} & \multicolumn{1}{l|}{2.24}  & \multicolumn{1}{l|}{3.44}  & \multicolumn{1}{l|}{2.18}  & \multicolumn{1}{l|}{3.67}  & \multicolumn{1}{l|}{2.13}  & \multicolumn{1}{l|}{3.67}  & \multicolumn{1}{l|}{1.87}  & \multicolumn{1}{l|}{13.3}  & \multicolumn{1}{l|}{1.8}   & \multicolumn{1}{l|}{12.61} & \multicolumn{1}{l|}{1.8}   & \multicolumn{1}{l|}{12.61} & \multicolumn{1}{l|}{3.98}  & \multicolumn{1}{l|}{17.2}  & \multicolumn{1}{l|}{4.09}  & \multicolumn{1}{l|}{17.89} & \multicolumn{1}{l|}{4.07}  & 17.89 \\ \hline
\multicolumn{1}{|l|}{RF}                                                                              & \multicolumn{1}{l|}{23.85} & \multicolumn{1}{l|}{29.86} & \multicolumn{1}{l|}{24.22} & \multicolumn{1}{l|}{29.51} & \multicolumn{1}{l|}{24.36} & \multicolumn{1}{l|}{28.82} & \multicolumn{1}{l|}{1.23}  & \multicolumn{1}{l|}{9.38}  & \multicolumn{1}{l|}{0.92}  & \multicolumn{1}{l|}{9.03}  & \multicolumn{1}{l|}{0.85}  & \multicolumn{1}{l|}{8.33}  & \multicolumn{1}{l|}{2.48}  & \multicolumn{1}{l|}{11.11} & \multicolumn{1}{l|}{2.27}  & \multicolumn{1}{l|}{9.38}  & \multicolumn{1}{l|}{2.2}   & \multicolumn{1}{l|}{9.03}  & \multicolumn{1}{l|}{10.86} & \multicolumn{1}{l|}{29.86} & \multicolumn{1}{l|}{10.92} & \multicolumn{1}{l|}{30.9}  & \multicolumn{1}{l|}{11.0}  & 31.25 \\ \hline
\multicolumn{1}{|l|}{MLP}                                                                             & \multicolumn{1}{l|}{17.98} & \multicolumn{1}{l|}{19.96} & \multicolumn{1}{l|}{16.49} & \multicolumn{1}{l|}{19.58} & \multicolumn{1}{l|}{15.87} & \multicolumn{1}{l|}{18.63} & \multicolumn{1}{l|}{9.25}  & \multicolumn{1}{l|}{10.46} & \multicolumn{1}{l|}{10.44} & \multicolumn{1}{l|}{10.65} & \multicolumn{1}{l|}{11.0}  & \multicolumn{1}{l|}{11.6}  & \multicolumn{1}{l|}{4.35}  & \multicolumn{1}{l|}{22.24} & \multicolumn{1}{l|}{4.39}  & \multicolumn{1}{l|}{21.29} & \multicolumn{1}{l|}{4.56}  & \multicolumn{1}{l|}{19.96} & \multicolumn{1}{l|}{9.04}  & \multicolumn{1}{l|}{34.41} & \multicolumn{1}{l|}{8.69}  & \multicolumn{1}{l|}{34.41} & \multicolumn{1}{l|}{8.47}  & 35.55 \\ \hline
\multicolumn{1}{|l|}{ADA}                                                                             & \multicolumn{1}{l|}{15.23} & \multicolumn{1}{l|}{13.92} & \multicolumn{1}{l|}{13.06} & \multicolumn{1}{l|}{11.91} & \multicolumn{1}{l|}{13.31} & \multicolumn{1}{l|}{12.2}  & \multicolumn{1}{l|}{1.72}  & \multicolumn{1}{l|}{3.01}  & \multicolumn{1}{l|}{8.87}  & \multicolumn{1}{l|}{13.06} & \multicolumn{1}{l|}{8.83}  & \multicolumn{1}{l|}{13.06} & \multicolumn{1}{l|}{1.12}  & \multicolumn{1}{l|}{5.31}  & \multicolumn{1}{l|}{2.34}  & \multicolumn{1}{l|}{10.47} & \multicolumn{1}{l|}{2.34}  & \multicolumn{1}{l|}{10.47} & \multicolumn{1}{l|}{6.33}  & \multicolumn{1}{l|}{22.96} & \multicolumn{1}{l|}{4.98}  & \multicolumn{1}{l|}{18.94} & \multicolumn{1}{l|}{4.98}  & 18.51 \\ \hline
\multicolumn{1}{|l|}{BAG}                                                                             & \multicolumn{1}{l|}{24.07} & \multicolumn{1}{l|}{24.52} & \multicolumn{1}{l|}{24.05} & \multicolumn{1}{l|}{24.17} & \multicolumn{1}{l|}{23.98} & \multicolumn{1}{l|}{24.17} & \multicolumn{1}{l|}{1.45}  & \multicolumn{1}{l|}{1.74}  & \multicolumn{1}{l|}{1.2}   & \multicolumn{1}{l|}{2.26}  & \multicolumn{1}{l|}{1.33}  & \multicolumn{1}{l|}{2.09}  & \multicolumn{1}{l|}{1.88}  & \multicolumn{1}{l|}{10.26} & \multicolumn{1}{l|}{1.52}  & \multicolumn{1}{l|}{9.22}  & \multicolumn{1}{l|}{1.38}  & \multicolumn{1}{l|}{8.52}  & \multicolumn{1}{l|}{10.93} & \multicolumn{1}{l|}{39.65} & \multicolumn{1}{l|}{11.11} & \multicolumn{1}{l|}{41.22} & \multicolumn{1}{l|}{11.22} & 41.74 \\ \hline
\multicolumn{1}{|l|}{GBC}                                                                             & \multicolumn{1}{l|}{17.44} & \multicolumn{1}{l|}{16.03} & \multicolumn{1}{l|}{17.44} & \multicolumn{1}{l|}{16.03} & \multicolumn{1}{l|}{17.44} & \multicolumn{1}{l|}{16.03} & \multicolumn{1}{l|}{6.98}  & \multicolumn{1}{l|}{10.77} & \multicolumn{1}{l|}{6.98}  & \multicolumn{1}{l|}{10.77} & \multicolumn{1}{l|}{6.98}  & \multicolumn{1}{l|}{10.77} & \multicolumn{1}{l|}{5.12}  & \multicolumn{1}{l|}{22.97} & \multicolumn{1}{l|}{5.12}  & \multicolumn{1}{l|}{22.97} & \multicolumn{1}{l|}{5.12}  & \multicolumn{1}{l|}{22.97} & \multicolumn{1}{l|}{8.47}  & \multicolumn{1}{l|}{27.27} & \multicolumn{1}{l|}{8.47}  & \multicolumn{1}{l|}{27.27} & \multicolumn{1}{l|}{8.47}  & 27.27 \\ \hline
\multicolumn{25}{|c|}{\textbf{Code review}}                                                                                                                                                                                                                                                                                                                                                                                                                                                                                                                                                                                                                                                                                                                                                              \\ \hline
\multicolumn{1}{|l|}{LR}                                                                              & \multicolumn{1}{l|}{16.67} & \multicolumn{1}{l|}{35.37} & \multicolumn{1}{l|}{18.08} & \multicolumn{1}{l|}{45.12} & \multicolumn{1}{l|}{17.51} & \multicolumn{1}{l|}{45.12} & \multicolumn{1}{l|}{6.21}  & \multicolumn{1}{l|}{18.29} & \multicolumn{1}{l|}{3.11}  & \multicolumn{1}{l|}{8.54}  & \multicolumn{1}{l|}{2.82}  & \multicolumn{1}{l|}{6.1}   & \multicolumn{1}{l|}{1.98}  & \multicolumn{1}{l|}{1.22}  & \multicolumn{1}{l|}{1.98}  & \multicolumn{1}{l|}{3.66}  & \multicolumn{1}{l|}{1.98}  & \multicolumn{1}{l|}{3.66}  & \multicolumn{1}{l|}{10.45} & \multicolumn{1}{l|}{37.8}  & \multicolumn{1}{l|}{10.45} & \multicolumn{1}{l|}{35.37} & \multicolumn{1}{l|}{10.45} & 36.59 \\ \hline
\multicolumn{1}{|l|}{DT}                                                                              & \multicolumn{1}{l|}{5.43}  & \multicolumn{1}{l|}{13.89} & \multicolumn{1}{l|}{7.07}  & \multicolumn{1}{l|}{13.89} & \multicolumn{1}{l|}{6.52}  & \multicolumn{1}{l|}{13.89} & \multicolumn{1}{l|}{5.16}  & \multicolumn{1}{l|}{8.33}  & \multicolumn{1}{l|}{4.35}  & \multicolumn{1}{l|}{6.94}  & \multicolumn{1}{l|}{4.35}  & \multicolumn{1}{l|}{6.94}  & \multicolumn{1}{l|}{1.09}  & \multicolumn{1}{l|}{4.17}  & \multicolumn{1}{l|}{1.09}  & \multicolumn{1}{l|}{4.17}  & \multicolumn{1}{l|}{1.09}  & \multicolumn{1}{l|}{4.17}  & \multicolumn{1}{l|}{5.98}  & \multicolumn{1}{l|}{22.22} & \multicolumn{1}{l|}{5.98}  & \multicolumn{1}{l|}{22.22} & \multicolumn{1}{l|}{5.98}  & 20.83 \\ \hline
\multicolumn{1}{|l|}{RF}                                                                              & \multicolumn{1}{l|}{14.78} & \multicolumn{1}{l|}{40.0}  & \multicolumn{1}{l|}{14.51} & \multicolumn{1}{l|}{40.0}  & \multicolumn{1}{l|}{16.09} & \multicolumn{1}{l|}{44.62} & \multicolumn{1}{l|}{8.97}  & \multicolumn{1}{l|}{18.46} & \multicolumn{1}{l|}{8.44}  & \multicolumn{1}{l|}{15.38} & \multicolumn{1}{l|}{6.86}  & \multicolumn{1}{l|}{9.23}  & \multicolumn{1}{l|}{3.17}  & \multicolumn{1}{l|}{7.69}  & \multicolumn{1}{l|}{2.9}   & \multicolumn{1}{l|}{10.77} & \multicolumn{1}{l|}{2.64}  & \multicolumn{1}{l|}{10.77} & \multicolumn{1}{l|}{8.44}  & \multicolumn{1}{l|}{21.54} & \multicolumn{1}{l|}{8.71}  & \multicolumn{1}{l|}{18.46} & \multicolumn{1}{l|}{8.18}  & 18.46 \\ \hline
\multicolumn{1}{|l|}{MLP}                                                                             & \multicolumn{1}{l|}{11.01} & \multicolumn{1}{l|}{31.71} & \multicolumn{1}{l|}{10.71} & \multicolumn{1}{l|}{31.71} & \multicolumn{1}{l|}{10.71} & \multicolumn{1}{l|}{31.71} & \multicolumn{1}{l|}{10.12} & \multicolumn{1}{l|}{17.07} & \multicolumn{1}{l|}{10.12} & \multicolumn{1}{l|}{14.63} & \multicolumn{1}{l|}{9.82}  & \multicolumn{1}{l|}{14.63} & \multicolumn{1}{l|}{4.76}  & \multicolumn{1}{l|}{12.2}  & \multicolumn{1}{l|}{4.76}  & \multicolumn{1}{l|}{14.63} & \multicolumn{1}{l|}{4.46}  & \multicolumn{1}{l|}{10.98} & \multicolumn{1}{l|}{5.95}  & \multicolumn{1}{l|}{23.17} & \multicolumn{1}{l|}{5.95}  & \multicolumn{1}{l|}{20.73} & \multicolumn{1}{l|}{5.95}  & 21.95 \\ \hline
\multicolumn{1}{|l|}{ADA}                                                                             & \multicolumn{1}{l|}{10.85} & \multicolumn{1}{l|}{20.48} & \multicolumn{1}{l|}{9.68}  & \multicolumn{1}{l|}{18.07} & \multicolumn{1}{l|}{9.68}  & \multicolumn{1}{l|}{18.07} & \multicolumn{1}{l|}{0.29}  & \multicolumn{1}{l|}{3.61}  & \multicolumn{1}{l|}{0.29}  & \multicolumn{1}{l|}{3.61}  & \multicolumn{1}{l|}{0.29}  & \multicolumn{1}{l|}{3.61}  & \multicolumn{1}{l|}{2.05}  & \multicolumn{1}{l|}{1.2}   & \multicolumn{1}{l|}{2.05}  & \multicolumn{1}{l|}{1.2}   & \multicolumn{1}{l|}{2.05}  & \multicolumn{1}{l|}{1.2}   & \multicolumn{1}{l|}{10.85} & \multicolumn{1}{l|}{38.55} & \multicolumn{1}{l|}{10.85} & \multicolumn{1}{l|}{38.55} & \multicolumn{1}{l|}{10.85} & 38.55 \\ \hline
\multicolumn{1}{|l|}{BAG}                                                                             & \multicolumn{1}{l|}{16.44} & \multicolumn{1}{l|}{40.26} & \multicolumn{1}{l|}{15.89} & \multicolumn{1}{l|}{29.87} & \multicolumn{1}{l|}{15.07} & \multicolumn{1}{l|}{24.68} & \multicolumn{1}{l|}{9.32}  & \multicolumn{1}{l|}{32.47} & \multicolumn{1}{l|}{9.32}  & \multicolumn{1}{l|}{38.96} & \multicolumn{1}{l|}{9.86}  & \multicolumn{1}{l|}{38.96} & \multicolumn{1}{l|}{1.92}  & \multicolumn{1}{l|}{1.3}   & \multicolumn{1}{l|}{1.92}  & \multicolumn{1}{l|}{2.6}   & \multicolumn{1}{l|}{2.19}  & \multicolumn{1}{l|}{2.6}   & \multicolumn{1}{l|}{6.58}  & \multicolumn{1}{l|}{15.58} & \multicolumn{1}{l|}{6.3}   & \multicolumn{1}{l|}{14.29} & \multicolumn{1}{l|}{6.03}  & 12.99 \\ \hline
\multicolumn{1}{|l|}{GBC}                                                                             & \multicolumn{1}{l|}{8.72}  & \multicolumn{1}{l|}{17.95} & \multicolumn{1}{l|}{8.14}  & \multicolumn{1}{l|}{19.23} & \multicolumn{1}{l|}{7.56}  & \multicolumn{1}{l|}{20.51} & \multicolumn{1}{l|}{10.47} & \multicolumn{1}{l|}{23.08} & \multicolumn{1}{l|}{9.88}  & \multicolumn{1}{l|}{24.36} & \multicolumn{1}{l|}{9.88}  & \multicolumn{1}{l|}{23.08} & \multicolumn{1}{l|}{3.49}  & \multicolumn{1}{l|}{10.26} & \multicolumn{1}{l|}{3.78}  & \multicolumn{1}{l|}{8.97}  & \multicolumn{1}{l|}{3.49}  & \multicolumn{1}{l|}{8.97}  & \multicolumn{1}{l|}{8.14}  & \multicolumn{1}{l|}{14.1}  & \multicolumn{1}{l|}{7.85}  & \multicolumn{1}{l|}{14.1}  & \multicolumn{1}{l|}{7.85}  & 14.1  \\ \hline
\end{tabular}}
\label{granular_PyExp_appendix}
\end{table*}

\begin{table*}[t]
\centering
\caption{Granular-level evaluation metric values for LIME executed on the same instance across the entire Java project, Postgres, CLCDSA, and code review datasets, with varying $\alpha$ values.}
\resizebox{\linewidth}{!}{
\begin{tabular}{|lllllllllllllllllllllllll|}
\hline
\multicolumn{1}{|l|}{\multirow{4}{*}{\textbf{\begin{tabular}[c]{@{}l@{}}ML\\ \\ Model\end{tabular}}}} & \multicolumn{24}{c|}{\textbf{Java Project}}                                                                                                                                                                                                                                                                                                                                                                                                                                                                                                                                                                                                                                                                                   \\ \cline{2-25} 
\multicolumn{1}{|l|}{}                                                                                & \multicolumn{6}{c|}{\textbf{PCPD}}                                                                                                                                          & \multicolumn{6}{c|}{\textbf{PCPI}}                                                                                                                                          & \multicolumn{6}{c|}{\textbf{NCPI}}                                                                                                                                          & \multicolumn{6}{c|}{\textbf{NCPD}}                                                                                                                                          \\ \cline{2-25} 
\multicolumn{1}{|l|}{}                                                                                & \multicolumn{2}{c|}{$\alpha=1$}                         & \multicolumn{2}{c|}{$\alpha=2$}                         & \multicolumn{2}{c|}{$\alpha=3$}                         & \multicolumn{2}{c|}{$\alpha=1$}                         & \multicolumn{2}{c|}{$\alpha=2$}                         & \multicolumn{2}{c|}{$\alpha=3$}                         & \multicolumn{2}{c|}{$\alpha=1$}                         & \multicolumn{2}{c|}{$\alpha=2$}                         & \multicolumn{2}{c|}{$\alpha=3$}                         & \multicolumn{2}{c|}{$\alpha=1$}                         & \multicolumn{2}{c|}{$\alpha=2$}                         & \multicolumn{2}{c|}{$\alpha=3$}                         \\ \cline{2-25} 
\multicolumn{1}{|l|}{}                                                                                & \multicolumn{1}{c|}{Cort.} & \multicolumn{1}{c|}{Wrng.} & \multicolumn{1}{c|}{Cort.} & \multicolumn{1}{c|}{Wrng.} & \multicolumn{1}{c|}{Cort.} & \multicolumn{1}{c|}{Wrng.} & \multicolumn{1}{c|}{Cort.} & \multicolumn{1}{c|}{Wrng.} & \multicolumn{1}{c|}{Cort.} & \multicolumn{1}{c|}{Wrng.} & \multicolumn{1}{c|}{Cort.} & \multicolumn{1}{c|}{Wrng.} & \multicolumn{1}{c|}{Cort.} & \multicolumn{1}{c|}{Wrng.} & \multicolumn{1}{c|}{Cort.} & \multicolumn{1}{c|}{Wrng.} & \multicolumn{1}{c|}{Cort.} & \multicolumn{1}{c|}{Wrng.} & \multicolumn{1}{c|}{Cort.} & \multicolumn{1}{c|}{Wrng.} & \multicolumn{1}{c|}{Cort.} & \multicolumn{1}{c|}{Wrng.} & \multicolumn{1}{c|}{Cort.} & \multicolumn{1}{c|}{Wrng.} \\ \hline
\multicolumn{1}{|l|}{LR}                                                                              & \multicolumn{1}{l|}{10.64} & \multicolumn{1}{l|}{0.31}  & \multicolumn{1}{l|}{10.64} & \multicolumn{1}{l|}{0.79}  & \multicolumn{1}{l|}{10.64} & \multicolumn{1}{l|}{0.94}  & \multicolumn{1}{l|}{8.83}  & \multicolumn{1}{l|}{66.46} & \multicolumn{1}{l|}{9.53}  & \multicolumn{1}{l|}{47.4}  & \multicolumn{1}{l|}{10.01} & \multicolumn{1}{l|}{45.04} & \multicolumn{1}{l|}{0.0}   & \multicolumn{1}{l|}{19.37} & \multicolumn{1}{l|}{0.0}   & \multicolumn{1}{l|}{19.37} & \multicolumn{1}{l|}{0.0}   & \multicolumn{1}{l|}{19.37} & \multicolumn{1}{l|}{29.59} & \multicolumn{1}{l|}{11.02} & \multicolumn{1}{l|}{45.93} & \multicolumn{1}{l|}{18.9}  & \multicolumn{1}{l|}{45.98} & 19.21                      \\ \hline
\multicolumn{1}{|l|}{DT}                                                                              & \multicolumn{1}{l|}{11.41} & \multicolumn{1}{l|}{34.17} & \multicolumn{1}{l|}{11.46} & \multicolumn{1}{l|}{33.59} & \multicolumn{1}{l|}{11.06} & \multicolumn{1}{l|}{28.16} & \multicolumn{1}{l|}{2.9}   & \multicolumn{1}{l|}{19.61} & \multicolumn{1}{l|}{2.85}  & \multicolumn{1}{l|}{20.39} & \multicolumn{1}{l|}{3.2}   & \multicolumn{1}{l|}{23.5}  & \multicolumn{1}{l|}{30.22} & \multicolumn{1}{l|}{21.17} & \multicolumn{1}{l|}{39.27} & \multicolumn{1}{l|}{21.36} & \multicolumn{1}{l|}{40.17} & \multicolumn{1}{l|}{20.78} & \multicolumn{1}{l|}{16.26} & \multicolumn{1}{l|}{6.8}   & \multicolumn{1}{l|}{15.66} & \multicolumn{1}{l|}{6.8}   & \multicolumn{1}{l|}{14.91} & 7.77                       \\ \hline
\multicolumn{1}{|l|}{RF}                                                                              & \multicolumn{1}{l|}{8.33}  & \multicolumn{1}{l|}{33.55} & \multicolumn{1}{l|}{8.33}  & \multicolumn{1}{l|}{42.76} & \multicolumn{1}{l|}{8.28}  & \multicolumn{1}{l|}{45.07} & \multicolumn{1}{l|}{1.99}  & \multicolumn{1}{l|}{23.68} & \multicolumn{1}{l|}{1.63}  & \multicolumn{1}{l|}{9.21}  & \multicolumn{1}{l|}{1.45}  & \multicolumn{1}{l|}{4.93}  & \multicolumn{1}{l|}{73.67} & \multicolumn{1}{l|}{42.43} & \multicolumn{1}{l|}{77.1}  & \multicolumn{1}{l|}{42.43} & \multicolumn{1}{l|}{79.23} & \multicolumn{1}{l|}{42.43} & \multicolumn{1}{l|}{16.88} & \multicolumn{1}{l|}{22.04} & \multicolumn{1}{l|}{17.29} & \multicolumn{1}{l|}{26.32} & \multicolumn{1}{l|}{16.83} & 26.64                      \\ \hline
\multicolumn{1}{|l|}{MLP}                                                                             & \multicolumn{1}{l|}{11.84} & \multicolumn{1}{l|}{8.25}  & \multicolumn{1}{l|}{11.89} & \multicolumn{1}{l|}{10.36} & \multicolumn{1}{l|}{11.84} & \multicolumn{1}{l|}{11.32} & \multicolumn{1}{l|}{9.43}  & \multicolumn{1}{l|}{44.53} & \multicolumn{1}{l|}{10.39} & \multicolumn{1}{l|}{23.03} & \multicolumn{1}{l|}{10.49} & \multicolumn{1}{l|}{19.58} & \multicolumn{1}{l|}{16.11} & \multicolumn{1}{l|}{25.72} & \multicolumn{1}{l|}{17.21} & \multicolumn{1}{l|}{25.91} & \multicolumn{1}{l|}{9.43}  & \multicolumn{1}{l|}{44.53} & \multicolumn{1}{l|}{32.61} & \multicolumn{1}{l|}{14.97} & \multicolumn{1}{l|}{25.24} & \multicolumn{1}{l|}{22.84} & \multicolumn{1}{l|}{32.61} & 14.97                      \\ \hline
\multicolumn{1}{|l|}{ADA}                                                                             & \multicolumn{1}{l|}{12.14} & \multicolumn{1}{l|}{0.0}   & \multicolumn{1}{l|}{12.14} & \multicolumn{1}{l|}{0.0}   & \multicolumn{1}{l|}{12.14} & \multicolumn{1}{l|}{0.0}   & \multicolumn{1}{l|}{2.73}  & \multicolumn{1}{l|}{51.58} & \multicolumn{1}{l|}{7.03}  & \multicolumn{1}{l|}{44.32} & \multicolumn{1}{l|}{7.08}  & \multicolumn{1}{l|}{44.59} & \multicolumn{1}{l|}{0.86}  & \multicolumn{1}{l|}{18.62} & \multicolumn{1}{l|}{0.05}  & \multicolumn{1}{l|}{18.62} & \multicolumn{1}{l|}{0.0}   & \multicolumn{1}{l|}{18.44} & \multicolumn{1}{l|}{29.89} & \multicolumn{1}{l|}{0.93}  & \multicolumn{1}{l|}{24.18} & \multicolumn{1}{l|}{9.87}  & \multicolumn{1}{l|}{24.28} & 9.87                       \\ \hline
\multicolumn{1}{|l|}{BAG}                                                                             & \multicolumn{1}{l|}{13.59} & \multicolumn{1}{l|}{8.74}  & \multicolumn{1}{l|}{13.59} & \multicolumn{1}{l|}{16376} & \multicolumn{1}{l|}{13.59} & \multicolumn{1}{l|}{20.4}  & \multicolumn{1}{l|}{6.31}  & \multicolumn{1}{l|}{54.28} & \multicolumn{1}{l|}{4.53}  & \multicolumn{1}{l|}{31.88} & \multicolumn{1}{l|}{4.53}  & \multicolumn{1}{l|}{29.14} & \multicolumn{1}{l|}{53.18} & \multicolumn{1}{l|}{17.3}  & \multicolumn{1}{l|}{58.07} & \multicolumn{1}{l|}{17.3}  & \multicolumn{1}{l|}{60.31} & \multicolumn{1}{l|}{17.3}  & \multicolumn{1}{l|}{11.2}  & \multicolumn{1}{l|}{5.1}   & \multicolumn{1}{l|}{10.94} & \multicolumn{1}{l|}{8.93}  & \multicolumn{1}{l|}{10.08} & 8.93                       \\ \hline
\multicolumn{1}{|l|}{GBC}                                                                             & \multicolumn{1}{l|}{11.03} & \multicolumn{1}{l|}{24.39} & \multicolumn{1}{l|}{11.03} & \multicolumn{1}{l|}{22.15} & \multicolumn{1}{l|}{11.03} & \multicolumn{1}{l|}{25.41} & \multicolumn{1}{l|}{7.62}  & \multicolumn{1}{l|}{45.33} & \multicolumn{1}{l|}{7.27}  & \multicolumn{1}{l|}{37.6}  & \multicolumn{1}{l|}{7.57}  & \multicolumn{1}{l|}{36.79} & \multicolumn{1}{l|}{5.64}  & \multicolumn{1}{l|}{30.28} & \multicolumn{1}{l|}{6.38}  & \multicolumn{1}{l|}{30.28} & \multicolumn{1}{l|}{7.62}  & \multicolumn{1}{l|}{29.47} & \multicolumn{1}{l|}{45.5}  & \multicolumn{1}{l|}{15.24} & \multicolumn{1}{l|}{37.93} & \multicolumn{1}{l|}{15.24} & \multicolumn{1}{l|}{37.93} & 16.26                      \\ \hline
\multicolumn{25}{|c|}{\textbf{Postgres}}                                                                                                                                                                                                                                                                                                                                                                                                                                                                                                                                                                                                                                                                                                                                                                                      \\ \hline
\multicolumn{1}{|l|}{LR}                                                                              & \multicolumn{1}{l|}{19.91} & \multicolumn{1}{l|}{11.83} & \multicolumn{1}{l|}{19.91} & \multicolumn{1}{l|}{15.56} & \multicolumn{1}{l|}{19.91} & \multicolumn{1}{l|}{11.83} & \multicolumn{1}{l|}{16.58} & \multicolumn{1}{l|}{25.52} & \multicolumn{1}{l|}{14.6}  & \multicolumn{1}{l|}{23.44} & \multicolumn{1}{l|}{14.79} & \multicolumn{1}{l|}{22.61} & \multicolumn{1}{l|}{16.58} & \multicolumn{1}{l|}{25.52} & \multicolumn{1}{l|}{0.45}  & \multicolumn{1}{l|}{43.36} & \multicolumn{1}{l|}{0.06}  & \multicolumn{1}{l|}{43.36} & \multicolumn{1}{l|}{0.19}  & \multicolumn{1}{l|}{43.15} & \multicolumn{1}{l|}{68.31} & \multicolumn{1}{l|}{39.21} & \multicolumn{1}{l|}{68.63} & 43.36                      \\ \hline
\multicolumn{1}{|l|}{DT}                                                                              & \multicolumn{1}{l|}{18.05} & \multicolumn{1}{l|}{41.39} & \multicolumn{1}{l|}{17.54} & \multicolumn{1}{l|}{36.97} & \multicolumn{1}{l|}{17.98} & \multicolumn{1}{l|}{39.5}  & \multicolumn{1}{l|}{8.35}  & \multicolumn{1}{l|}{10.5}  & \multicolumn{1}{l|}{8.8}   & \multicolumn{1}{l|}{11.55} & \multicolumn{1}{l|}{8.48}  & \multicolumn{1}{l|}{11.55} & \multicolumn{1}{l|}{13.97} & \multicolumn{1}{l|}{27.31} & \multicolumn{1}{l|}{11.1}  & \multicolumn{1}{l|}{29.62} & \multicolumn{1}{l|}{8.67}  & \multicolumn{1}{l|}{26.05} & \multicolumn{1}{l|}{28.25} & \multicolumn{1}{l|}{16.39} & \multicolumn{1}{l|}{28.44} & \multicolumn{1}{l|}{14.5}  & \multicolumn{1}{l|}{29.09} & 14.92                      \\ \hline
\multicolumn{1}{|l|}{RF}                                                                              & \multicolumn{1}{l|}{18.4}  & \multicolumn{1}{l|}{27.45} & \multicolumn{1}{l|}{18.4}  & \multicolumn{1}{l|}{26.97} & \multicolumn{1}{l|}{18.4}  & \multicolumn{1}{l|}{27.21} & \multicolumn{1}{l|}{2.03}  & \multicolumn{1}{l|}{11.22} & \multicolumn{1}{l|}{1.6}   & \multicolumn{1}{l|}{10.74} & \multicolumn{1}{l|}{1.23}  & \multicolumn{1}{l|}{8.59}  & \multicolumn{1}{l|}{68.62} & \multicolumn{1}{l|}{55.13} & \multicolumn{1}{l|}{67.32} & \multicolumn{1}{l|}{55.13} & \multicolumn{1}{l|}{66.52} & \multicolumn{1}{l|}{54.89} & \multicolumn{1}{l|}{37.54} & \multicolumn{1}{l|}{22.2}  & \multicolumn{1}{l|}{36.86} & \multicolumn{1}{l|}{26.25} & \multicolumn{1}{l|}{35.94} & 24.82                      \\ \hline
\multicolumn{1}{|l|}{MLP}                                                                             & \multicolumn{1}{l|}{21.74} & \multicolumn{1}{l|}{8.7}   & \multicolumn{1}{l|}{21.74} & \multicolumn{1}{l|}{7.88}  & \multicolumn{1}{l|}{21.74} & \multicolumn{1}{l|}{9.85}  & \multicolumn{1}{l|}{18.33} & \multicolumn{1}{l|}{38.26} & \multicolumn{1}{l|}{17.98} & \multicolumn{1}{l|}{37.11} & \multicolumn{1}{l|}{18.33} & \multicolumn{1}{l|}{37.44} & \multicolumn{1}{l|}{5.71}  & \multicolumn{1}{l|}{38.59} & \multicolumn{1}{l|}{8.01}  & \multicolumn{1}{l|}{38.59} & \multicolumn{1}{l|}{9.69}  & \multicolumn{1}{l|}{38.59} & \multicolumn{1}{l|}{61.6}  & \multicolumn{1}{l|}{31.2}  & \multicolumn{1}{l|}{61.05} & \multicolumn{1}{l|}{32.84} & \multicolumn{1}{l|}{57.77} & 33.33                      \\ \hline
\multicolumn{1}{|l|}{ADA}                                                                             & \multicolumn{1}{l|}{19.72} & \multicolumn{1}{l|}{11.47} & \multicolumn{1}{l|}{19.79} & \multicolumn{1}{l|}{15.58} & \multicolumn{1}{l|}{19.85} & \multicolumn{1}{l|}{18.83} & \multicolumn{1}{l|}{7.77}  & \multicolumn{1}{l|}{23.38} & \multicolumn{1}{l|}{7.65}  & \multicolumn{1}{l|}{21.0}  & \multicolumn{1}{l|}{7.65}  & \multicolumn{1}{l|}{23.59} & \multicolumn{1}{l|}{1.07}  & \multicolumn{1}{l|}{47.4}  & \multicolumn{1}{l|}{0.76}  & \multicolumn{1}{l|}{47.62} & \multicolumn{1}{l|}{0.57}  & \multicolumn{1}{l|}{47.62} & \multicolumn{1}{l|}{49.37} & \multicolumn{1}{l|}{15.58} & \multicolumn{1}{l|}{56.32} & \multicolumn{1}{l|}{15.8}  & \multicolumn{1}{l|}{56.01} & 15.8                       \\ \hline
\multicolumn{1}{|l|}{BAG}                                                                             & \multicolumn{1}{l|}{21.59} & \multicolumn{1}{l|}{36.59} & \multicolumn{1}{l|}{21.59} & \multicolumn{1}{l|}{42.57} & \multicolumn{1}{l|}{21.59} & \multicolumn{1}{l|}{45.45} & \multicolumn{1}{l|}{7.28}  & \multicolumn{1}{l|}{13.53} & \multicolumn{1}{l|}{5.02}  & \multicolumn{1}{l|}{7354}  & \multicolumn{1}{l|}{4.27}  & \multicolumn{1}{l|}{7.32}  & \multicolumn{1}{l|}{16.82} & \multicolumn{1}{l|}{37.25} & \multicolumn{1}{l|}{20.21} & \multicolumn{1}{l|}{37.25} & \multicolumn{1}{l|}{23.23} & \multicolumn{1}{l|}{37.03} & \multicolumn{1}{l|}{45.51} & \multicolumn{1}{l|}{9.31}  & \multicolumn{1}{l|}{37.85} & \multicolumn{1}{l|}{8.2}   & \multicolumn{1}{l|}{34.84} & 7.98                       \\ \hline
\multicolumn{1}{|l|}{GBC}                                                                             & \multicolumn{1}{l|}{18.67} & \multicolumn{1}{l|}{24.35} & \multicolumn{1}{l|}{18.8}  & \multicolumn{1}{l|}{21.98} & \multicolumn{1}{l|}{18.73} & \multicolumn{1}{l|}{24.35} & \multicolumn{1}{l|}{11.27} & \multicolumn{1}{l|}{22.41} & \multicolumn{1}{l|}{10.57} & \multicolumn{1}{l|}{19.83} & \multicolumn{1}{l|}{10.76} & \multicolumn{1}{l|}{20.26} & \multicolumn{1}{l|}{41.01} & \multicolumn{1}{l|}{45.69} & \multicolumn{1}{l|}{36.84} & \multicolumn{1}{l|}{46.77} & \multicolumn{1}{l|}{36.33} & \multicolumn{1}{l|}{46.12} & \multicolumn{1}{l|}{49.11} & \multicolumn{1}{l|}{21.55} & \multicolumn{1}{l|}{50.38} & \multicolumn{1}{l|}{22.2}  & \multicolumn{1}{l|}{51.56} & 20.91                      \\ \hline
\multicolumn{25}{|c|}{\textbf{CLCDSA}}                                                                                                                                                                                                                                                                                                                                                                                                                                                                                                                                                                                                                                                                                                                                                                                        \\ \hline
\multicolumn{1}{|l|}{LR}                                                                              & \multicolumn{1}{l|}{65.66} & \multicolumn{1}{l|}{0.0}   & \multicolumn{1}{l|}{65.66} & \multicolumn{1}{l|}{0.54}  & \multicolumn{1}{l|}{65.66} & \multicolumn{1}{l|}{0.27}  & \multicolumn{1}{l|}{25.81} & \multicolumn{1}{l|}{35.44} & \multicolumn{1}{l|}{32.3}  & \multicolumn{1}{l|}{33.29} & \multicolumn{1}{l|}{33.53} & \multicolumn{1}{l|}{33.42} & \multicolumn{1}{l|}{0.0}   & \multicolumn{1}{l|}{50.74} & \multicolumn{1}{l|}{0.0}   & \multicolumn{1}{l|}{50.6}  & \multicolumn{1}{l|}{0.0}   & \multicolumn{1}{l|}{50.74} & \multicolumn{1}{l|}{29.24} & \multicolumn{1}{l|}{33.96} & \multicolumn{1}{l|}{27.84} & \multicolumn{1}{l|}{36.51} & \multicolumn{1}{l|}{27.33} & 36.38                      \\ \hline
\multicolumn{1}{|l|}{DT}                                                                              & \multicolumn{1}{l|}{59.11} & \multicolumn{1}{l|}{21.23} & \multicolumn{1}{l|}{59.41} & \multicolumn{1}{l|}{21.46} & \multicolumn{1}{l|}{59.56} & \multicolumn{1}{l|}{22.6}  & \multicolumn{1}{l|}{7.17}  & \multicolumn{1}{l|}{11.19} & \multicolumn{1}{l|}{7.1}   & \multicolumn{1}{l|}{10.05} & \multicolumn{1}{l|}{7.13}  & \multicolumn{1}{l|}{9.36}  & \multicolumn{1}{l|}{15.85} & \multicolumn{1}{l|}{46.58} & \multicolumn{1}{l|}{12.77} & \multicolumn{1}{l|}{44.52} & \multicolumn{1}{l|}{13.18} & \multicolumn{1}{l|}{44.98} & \multicolumn{1}{l|}{8.11}  & \multicolumn{1}{l|}{30.14} & \multicolumn{1}{l|}{8.34}  & \multicolumn{1}{l|}{30.59} & \multicolumn{1}{l|}{8.6}   & 30.82                      \\ \hline
\multicolumn{1}{|l|}{RF}                                                                              & \multicolumn{1}{l|}{52.25} & \multicolumn{1}{l|}{35.07} & \multicolumn{1}{l|}{62.25} & \multicolumn{1}{l|}{31.94} & \multicolumn{1}{l|}{62.25} & \multicolumn{1}{l|}{32.99} & \multicolumn{1}{l|}{6.93}  & \multicolumn{1}{l|}{10.07} & \multicolumn{1}{l|}{5.37}  & \multicolumn{1}{l|}{8.68}  & \multicolumn{1}{l|}{5.12}  & \multicolumn{1}{l|}{8.33}  & \multicolumn{1}{l|}{28.87} & \multicolumn{1}{l|}{48.96} & \multicolumn{1}{l|}{28.4}  & \multicolumn{1}{l|}{49.31} & \multicolumn{1}{l|}{28.79} & \multicolumn{1}{l|}{50.35} & \multicolumn{1}{l|}{9.42}  & \multicolumn{1}{l|}{28.12} & \multicolumn{1}{l|}{8.5}   & \multicolumn{1}{l|}{27.78} & \multicolumn{1}{l|}{7.75}  & 26.39                      \\ \hline
\multicolumn{1}{|l|}{MLP}                                                                             & \multicolumn{1}{l|}{62.94} & \multicolumn{1}{l|}{14.1}  & \multicolumn{1}{l|}{63.02} & \multicolumn{1}{l|}{10.34} & \multicolumn{1}{l|}{62.98} & \multicolumn{1}{l|}{9.96}  & \multicolumn{1}{l|}{24.25} & \multicolumn{1}{l|}{15.79} & \multicolumn{1}{l|}{29.93} & \multicolumn{1}{l|}{17.67} & \multicolumn{1}{l|}{33.01} & \multicolumn{1}{l|}{17.86} & \multicolumn{1}{l|}{23.39} & \multicolumn{1}{l|}{58.27} & \multicolumn{1}{l|}{17.56} & \multicolumn{1}{l|}{62.59} & \multicolumn{1}{l|}{15.96} & \multicolumn{1}{l|}{63.72} & \multicolumn{1}{l|}{25.22} & \multicolumn{1}{l|}{45.49} & \multicolumn{1}{l|}{26.66} & \multicolumn{1}{l|}{45.3}  & \multicolumn{1}{l|}{27.56} & 44.92                      \\ \hline
\multicolumn{1}{|l|}{ADA}                                                                             & \multicolumn{1}{l|}{61.29} & \multicolumn{1}{l|}{0.97}  & \multicolumn{1}{l|}{61.29} & \multicolumn{1}{l|}{1.53}  & \multicolumn{1}{l|}{61.29} & \multicolumn{1}{l|}{1.95}  & \multicolumn{1}{l|}{16.58} & \multicolumn{1}{l|}{21.14} & \multicolumn{1}{l|}{16.62} & \multicolumn{1}{l|}{21.7}  & \multicolumn{1}{l|}{16.58} & \multicolumn{1}{l|}{21.84} & \multicolumn{1}{l|}{0.13}  & \multicolumn{1}{l|}{67.32} & \multicolumn{1}{l|}{0.59}  & \multicolumn{1}{l|}{67.18} & \multicolumn{1}{l|}{0.5}   & \multicolumn{1}{l|}{67.18} & \multicolumn{1}{l|}{21.2}  & \multicolumn{1}{l|}{11.96} & \multicolumn{1}{l|}{23.55} & \multicolumn{1}{l|}{21.7}  & \multicolumn{1}{l|}{23.64} & 22.11                      \\ \hline
\multicolumn{1}{|l|}{BAG}                                                                             & \multicolumn{1}{l|}{60.48} & \multicolumn{1}{l|}{16.61} & \multicolumn{1}{l|}{60.29} & \multicolumn{1}{l|}{15.22} & \multicolumn{1}{l|}{60.29} & \multicolumn{1}{l|}{16.09} & \multicolumn{1}{l|}{14.82} & \multicolumn{1}{l|}{6.57}  & \multicolumn{1}{l|}{12.72} & \multicolumn{1}{l|}{7.09}  & \multicolumn{1}{l|}{13.0}  & \multicolumn{1}{l|}{7.61}  & \multicolumn{1}{l|}{24.49} & \multicolumn{1}{l|}{69.72} & \multicolumn{1}{l|}{29.65} & \multicolumn{1}{l|}{70.42} & \multicolumn{1}{l|}{30.36} & \multicolumn{1}{l|}{69.9}  & \multicolumn{1}{l|}{13.99} & \multicolumn{1}{l|}{54.84} & \multicolumn{1}{l|}{11.06} & \multicolumn{1}{l|}{33.39} & \multicolumn{1}{l|}{10.54} & 30.45                      \\ \hline
\multicolumn{1}{|l|}{GBC}                                                                             & \multicolumn{1}{l|}{60.93} & \multicolumn{1}{l|}{20.67} & \multicolumn{1}{l|}{61.04} & \multicolumn{1}{l|}{20.19} & \multicolumn{1}{l|}{61.16} & \multicolumn{1}{l|}{20.19} & \multicolumn{1}{l|}{19.81} & \multicolumn{1}{l|}{18.05} & \multicolumn{1}{l|}{19.78} & \multicolumn{1}{l|}{17.58} & \multicolumn{1}{l|}{19.7}  & \multicolumn{1}{l|}{17.81} & \multicolumn{1}{l|}{25.86} & \multicolumn{1}{l|}{44.18} & \multicolumn{1}{l|}{18.92} & \multicolumn{1}{l|}{44.18} & \multicolumn{1}{l|}{18.25} & \multicolumn{1}{l|}{43.23} & \multicolumn{1}{l|}{18.32} & \multicolumn{1}{l|}{45.37} & \multicolumn{1}{l|}{17.84} & \multicolumn{1}{l|}{46.56} & \multicolumn{1}{l|}{17.72} & 46.56                      \\ \hline
\multicolumn{25}{|c|}{\textbf{Code review}}                                                                                                                                                                                                                                                                                                                                                                                                                                                                                                                                                                                                                                                                                                                                                                                   \\ \hline
\multicolumn{1}{|l|}{LR}                                                                              & \multicolumn{1}{l|}{62.41} & \multicolumn{1}{l|}{2.41}  & \multicolumn{1}{l|}{61.7}  & \multicolumn{1}{l|}{7.23}  & \multicolumn{1}{l|}{61.7}  & \multicolumn{1}{l|}{7.23}  & \multicolumn{1}{l|}{48.23} & \multicolumn{1}{l|}{34.94} & \multicolumn{1}{l|}{59.57} & \multicolumn{1}{l|}{31.33} & \multicolumn{1}{l|}{62.41} & \multicolumn{1}{l|}{33.73} & \multicolumn{1}{l|}{4.26}  & \multicolumn{1}{l|}{44.58} & \multicolumn{1}{l|}{4.26}  & \multicolumn{1}{l|}{44.58} & \multicolumn{1}{l|}{3.55}  & \multicolumn{1}{l|}{44.58} & \multicolumn{1}{l|}{24.82} & \multicolumn{1}{l|}{42.17} & \multicolumn{1}{l|}{21.99} & \multicolumn{1}{l|}{38.35} & \multicolumn{1}{l|}{19.86} & 36.14                      \\ \hline
\multicolumn{1}{|l|}{DT}                                                                              & \multicolumn{1}{l|}{46.67} & \multicolumn{1}{l|}{21.62} & \multicolumn{1}{l|}{49.33} & \multicolumn{1}{l|}{21.62} & \multicolumn{1}{l|}{49.33} & \multicolumn{1}{l|}{21.62} & \multicolumn{1}{l|}{8.0}   & \multicolumn{1}{l|}{8.11}  & \multicolumn{1}{l|}{9.33}  & \multicolumn{1}{l|}{16.22} & \multicolumn{1}{l|}{9.33}  & \multicolumn{1}{l|}{17.57} & \multicolumn{1}{l|}{13.33} & \multicolumn{1}{l|}{45.95} & \multicolumn{1}{l|}{12.67} & \multicolumn{1}{l|}{48.65} & \multicolumn{1}{l|}{10.67} & \multicolumn{1}{l|}{47.3}  & \multicolumn{1}{l|}{6.67}  & \multicolumn{1}{l|}{12.16} & \multicolumn{1}{l|}{7.33}  & \multicolumn{1}{l|}{12.16} & \multicolumn{1}{l|}{7.33}  & 13.51                      \\ \hline
\multicolumn{1}{|l|}{RF}                                                                              & \multicolumn{1}{l|}{63.29} & \multicolumn{1}{l|}{27.27} & \multicolumn{1}{l|}{61.39} & \multicolumn{1}{l|}{28.79} & \multicolumn{1}{l|}{62.66} & \multicolumn{1}{l|}{34.85} & \multicolumn{1}{l|}{28.48} & \multicolumn{1}{l|}{28.79} & \multicolumn{1}{l|}{29.11} & \multicolumn{1}{l|}{22.73} & \multicolumn{1}{l|}{27.85} & \multicolumn{1}{l|}{18.18} & \multicolumn{1}{l|}{15.82} & \multicolumn{1}{l|}{36.36} & \multicolumn{1}{l|}{25.95} & \multicolumn{1}{l|}{36.36} & \multicolumn{1}{l|}{27.22} & \multicolumn{1}{l|}{36.36} & \multicolumn{1}{l|}{25.95} & \multicolumn{1}{l|}{21.21} & \multicolumn{1}{l|}{22.15} & \multicolumn{1}{l|}{21.21} & \multicolumn{1}{l|}{17.72} & 22.73                      \\ \hline
\multicolumn{1}{|l|}{MLP}                                                                             & \multicolumn{1}{l|}{64.18} & \multicolumn{1}{l|}{17.78} & \multicolumn{1}{l|}{67.91} & \multicolumn{1}{l|}{23.33} & \multicolumn{1}{l|}{67.91} & \multicolumn{1}{l|}{20.0}  & \multicolumn{1}{l|}{41.79} & \multicolumn{1}{l|}{38.89} & \multicolumn{1}{l|}{45.52} & \multicolumn{1}{l|}{33.33} & \multicolumn{1}{l|}{38.06} & \multicolumn{1}{l|}{32.22} & \multicolumn{1}{l|}{15.67} & \multicolumn{1}{l|}{40.0}  & \multicolumn{1}{l|}{9.7}   & \multicolumn{1}{l|}{38.89} & \multicolumn{1}{l|}{8.21}  & \multicolumn{1}{l|}{38.89} & \multicolumn{1}{l|}{26.87} & \multicolumn{1}{l|}{33.33} & \multicolumn{1}{l|}{23.88} & \multicolumn{1}{l|}{33.33} & \multicolumn{1}{l|}{25.37} & 34.44                      \\ \hline
\multicolumn{1}{|l|}{ADA}                                                                             & \multicolumn{1}{l|}{40.0}  & \multicolumn{1}{l|}{16.67} & \multicolumn{1}{l|}{32.86} & \multicolumn{1}{l|}{16367} & \multicolumn{1}{l|}{36.43} & \multicolumn{1}{l|}{17.86} & \multicolumn{1}{l|}{32.86} & \multicolumn{1}{l|}{15.48} & \multicolumn{1}{l|}{33.57} & \multicolumn{1}{l|}{14.29} & \multicolumn{1}{l|}{34.29} & \multicolumn{1}{l|}{20.24} & \multicolumn{1}{l|}{4.29}  & \multicolumn{1}{l|}{44.05} & \multicolumn{1}{l|}{4.29}  & \multicolumn{1}{l|}{45.24} & \multicolumn{1}{l|}{4.29}  & \multicolumn{1}{l|}{44.05} & \multicolumn{1}{l|}{4.29}  & \multicolumn{1}{l|}{28.57} & \multicolumn{1}{l|}{4.29}  & \multicolumn{1}{l|}{28.57} & \multicolumn{1}{l|}{16.43} & 28.57                      \\ \hline
\multicolumn{1}{|l|}{BAG}                                                                             & \multicolumn{1}{l|}{73.79} & \multicolumn{1}{l|}{22.78} & \multicolumn{1}{l|}{71.72} & \multicolumn{1}{l|}{15.19} & \multicolumn{1}{l|}{73.1}  & \multicolumn{1}{l|}{15.19} & \multicolumn{1}{l|}{35.86} & \multicolumn{1}{l|}{48.1}  & \multicolumn{1}{l|}{55.86} & \multicolumn{1}{l|}{51.9}  & \multicolumn{1}{l|}{42.76} & \multicolumn{1}{l|}{51.9}  & \multicolumn{1}{l|}{12.41} & \multicolumn{1}{l|}{16.46} & \multicolumn{1}{l|}{17.24} & \multicolumn{1}{l|}{16.46} & \multicolumn{1}{l|}{17.93} & \multicolumn{1}{l|}{6.33}  & \multicolumn{1}{l|}{11.72} & \multicolumn{1}{l|}{8.86}  & \multicolumn{1}{l|}{7.59}  & \multicolumn{1}{l|}{6.33}  & \multicolumn{1}{l|}{7.59}  & 6.33                       \\ \hline
\multicolumn{1}{|l|}{GBC}                                                                             & \multicolumn{1}{l|}{62.68} & \multicolumn{1}{l|}{10.98} & \multicolumn{1}{l|}{61.27} & \multicolumn{1}{l|}{10.98} & \multicolumn{1}{l|}{62.68} & \multicolumn{1}{l|}{12.2}  & \multicolumn{1}{l|}{48.59} & \multicolumn{1}{l|}{47.56} & \multicolumn{1}{l|}{26.76} & \multicolumn{1}{l|}{51.22} & \multicolumn{1}{l|}{23.94} & \multicolumn{1}{l|}{51.22} & \multicolumn{1}{l|}{5.63}  & \multicolumn{1}{l|}{39.02} & \multicolumn{1}{l|}{11.27} & \multicolumn{1}{l|}{36.59} & \multicolumn{1}{l|}{6.34}  & \multicolumn{1}{l|}{39.02} & \multicolumn{1}{l|}{21.13} & \multicolumn{1}{l|}{8.54}  & \multicolumn{1}{l|}{21.83} & \multicolumn{1}{l|}{7.32}  & \multicolumn{1}{l|}{22.54} & 7.32                       \\ \hline
\end{tabular}}
\label{granular_LIME_appendix}
\end{table*}






\end{document}